\documentclass[10pt,aps,
pre,
superscriptaddress,
 amsmath,amssymb,
floatfix,
longbibliography
]{revtex4-2}
\usepackage{array}[=2016-10-06]

\usepackage[utf8]{inputenc}
\usepackage[T1]{fontenc}
\usepackage{color,soul}
\setulcolor{red}
\sethlcolor{yellow}

\usepackage{CJK}
\usepackage{multirow}
\usepackage{tabularx}
\usepackage{graphicx}
\DeclareGraphicsExtensions{.jpg,.png}
\usepackage{capt-of}
\usepackage[colorlinks=true, citecolor=blue, linkcolor=blue]{hyperref}
\usepackage{gensymb}

\begin{document}

\title{Self-limited stacking of non-Euclidean colloidal shells: From saddles to caps}

\author{Kyle T. Sullivan}
\affiliation{Department of Physics, University of Massachusetts, Amherst, Massachusetts 01003, USA}

\author{Mark J. Stevens}
\affiliation{Center for Integrated Nanotechnologies, Sandia National Laboratories, Albuquerque, NM 87185, USA}

\author{Gregory M. Grason}
\affiliation{Department of Polymer Science and Engineering, University of Massachusetts, Amherst, Massachusetts 01003, USA}

\date{\today}

\begin{abstract} 
The presence of geometric frustration in self-assemblies, stemming from a shape misfit between identical subunits, can lead to the spontaneous formation of finite-sized structures known as self-limitation.  Recently proposed models of curved, colloidal shells, dubbed ``curvamers'', that form one-dimensional stacks have shown curvature-induced frustration to be a promising class of particle designs for engineering self-limiting assembly.  However, existing models of curvamer assembly considered only cylindrically curved shells where elastic costs are derived purely from bending.  Here, we study the self-limiting behavior of a more general class of curvature-frustrated colloidal shells, extending shapes to include non-Euclidean geometries such as spherical caps and saddles whose deformations can also incur stretching costs.  We introduce continuum mechanical and discrete, coarse-grained models of shell stacks to study how qualitatively different intra-assembly strain gradients required for each geometry affect the buildup of super-extensive elastic costs which set the ``program'' of self-limiting behavior.  We find the accumulation of elastic energies in non-Euclidean shell stacks is sensitive to shell thickness via a dimensionless quantity related to the F\"oppl-von K\'arm\'an number that characterizes the relative strength of stretching to bending.  In particular, thin shells penalize stretching more causing faster elastic energy growth resulting in the suppression of self-limitation with sizes generally largest for cylindrical shells, smallest for spherical caps and intermediate for saddles.  We additionally find that while thin spherical caps experience the strongest elastic penalties, the rapidly increasing rate of elastic penalties in stacks of saddles lead to a robust size-selection as interparticle binding strength grows as well as relatively smaller dispersity in self-limiting stack size in comparison to cylindrical and spherical curvamers. These results provide critical guidance for the design of frustrated self-limitation in engineered colloidal systems.
\end{abstract}

\maketitle

\section{Introduction}

Self-assembly, the spontaneous organization of subunits into larger, coherent structures, has long been considered a promising tool for designing functional materials that ``build themselves'' at the nano- and micron-scale where top-down fabrication methods are difficult or impossible \cite{li-patchycolloids-2020, hamley-nanotechreview-2003,grason_teaching_2025}.  Many examples of self-assembly can be found in nature and living organisms where, often, finite and controlled size of the assembled structure plays a critical role in their functionality.  Some include protein shells that encase virus genetic material \cite{zandi-virusgrowth-2020,zlotnick-virusassembly-2011}, photonic nanostructures that produce vibrant wavelengths of light in bird feathers\cite{prum-nanostructurefeathers-2009,dufresne-nanostructurefeatherassembly-2009,saranathan-featherxrays-2012}, and multi-filament bundles that provide structural support in cell cytoskeletons\cite{popp-cellfilaments-2012}, aid in clot formation\cite{weisel-fibrin-2007} and convert mechanical sound vibrations to electrochemical signals for hearing\cite{rosario-stereocilia-2025,krey-stereocilia-2023}.  However, obtaining precise control of structure size in synthetic systems has proven difficult to achieve experimentally where most assemblies result in unlimited growth with macroscopically large aggregates of subunits that are thermodynamically driven to grow due to cohesive interactions with no characteristic size\cite{hagan_sla_2021}.  

\begin{figure*}[h]
    \centering
    \includegraphics[width=1\textwidth]{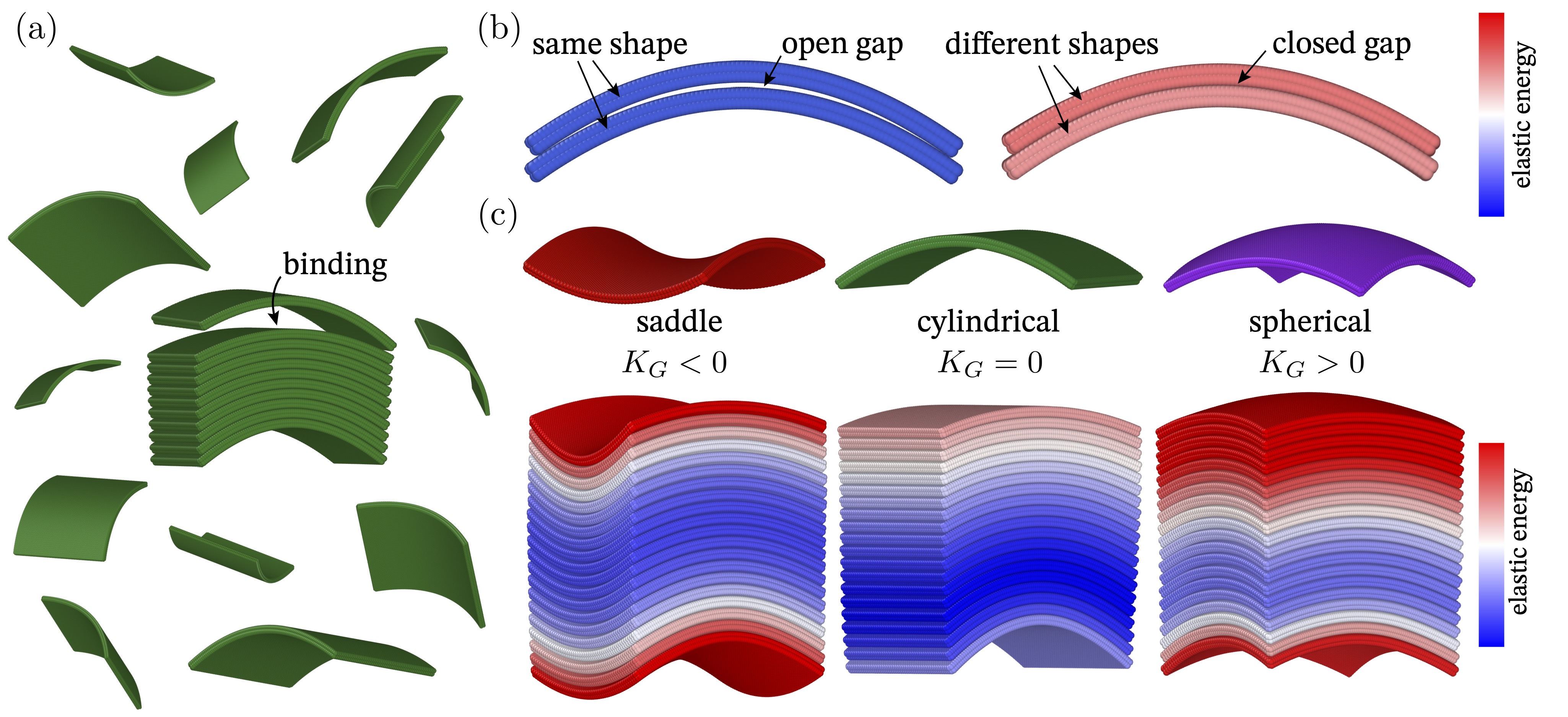}
    \caption{ (a) An illustration of the assembly process for frustrated, cylindrical shells forming one-dimensional stacks. (b) Curved shells are frustrated because attractive interactions want to close the gap between shell surfaces that occurs when shells are in their preferred shape (left) but doing so requires changing curvature incurring elastic costs (right). (c) Examples of stacking assemblies for the three shell shapes that this article will focus on colored according to average shell elastic energy.  Blue corresponds to low elastic costs while red indicates very high elastic costs.  Euclidean shells (cylinders) with zero preferred Gaussian curvature ($K_{G}=0$) exhibit relatively low elastic stresses exclusively from bending deformations while non-Euclidean shell stacks such as saddles ($K_{G}<0$) and spherical caps ($K_{G}>0$) exhibit qualitative different elastic strain gradients due to difference in stacking geometry as well as additional stretching elastic costs.}
    \label{fig:intro}
\end{figure*}

One pathway to control assembly size that has emerged in recent years is {\it geometrically-frustrated assembly} (GFA), a paradigm of self-assembly that utilizes shape misfits between identical, soft subunit particles\cite{grason-perspective-2016, hagan_sla_2021}.  As cohesive attractions cause subunits to assemble into aggregates, they are forced to deform  generating stresses within the structure --- associated with elastic strain gradients --- the energetic costs of which continue to grow with each additional subunit. This accumulation of super-extensive frustration costs in GFA has been shown to lead to assemblies with multiple morphologies\cite{schneider_shapes_2005, armon_shape_2014, hall-chiralbundles-2016, kohler_stress_2016, hall-wedges-2023, wang-polarwj-2025, leroy_collective_2023}, topological defects\cite{bruss_noneuclidean_2012, bruss_topological_2013, li-crystalcaps-2019, hackney-spins-2023}, and under the right conditions, {\it self-limiting assembly} (SLA) where the competition between elasticity and cohesive interactions results in an emergent, thermodynamically favorable size scale\cite{aggeli-chiralrods-2001, grason-twistedbundles-2020,  hagan_sla_2021, serafin-polyhedra-2021}.  Experimental systems of frustrated assemblies exhibiting frustration driven accumulation of elastic strain have so far included 1D chains of incommensurate ``polybrick'' subunits\cite{berengut-polybricks-2020} as well as width limited 2D crystals of colloids on spherical templates~\cite{meng-curvedcrystals-2014} or geometrically incompatible twisting ribbons ~\cite{zhang-ribbons-2019, serafin_frustrated_2021} . The prospect of achieving self-limitation through GFA more generally is nevertheless appealing as the self-limiting size is thermodynamically selected and so will appear spontaneously once the system reaches thermal equilibrium, and that it can do so using a single species of subunit particle.  Additionally, if the strength of attractions relative to deformability can be adjusted, a new self-limiting size can be selected without the need of a new subunit design.  Thus, GFA has the potential to be a useful mechanism to ``program'' and tune size control of self-assembled structures via the controlled misfit, flexibility and interactions of particles.

Beyond studies of GFA based on more abstract statistical mechanical frameworks \cite{meiri-gfa-2021,hackney-spins-2023,hackney-phasetransitions-2025, ortiz-tavarez_statistical_2025}, recently, many discrete particle models have been introduced to assess the accessible range of GFA systems by simulating the ground state competition between binding and deformation in terms of local properties of misfitting particles (e.g. shape, flexibility and interactions) \cite{Spivack-puzzlemers-2022,wang-polarwj-2025,wang-polybricks-2024,hall-wedges-2023,lenz_polygon_17, leroy_collective_2023, botond-tubules-2022}.  

One particularly promising system showing a large variable range of self-limiting size is the assembly of curved, colloidal shells (a.k.a. ``curvamers'') into one-dimensional stacks\cite{tanjeem-curvamers-2022, sullivan-curvamers-2024}, illustrated schematically in Fig. \ref{fig:intro}a.  As illustrated in Fig. \ref{fig:intro}b, finite-thickness curved shells of the same shape exhibit a gap between their surfaces when stacked without deformation.  Attractive surface-to-surface interactions can force gap closure by deforming shell shapes --- a motif referred to as ``curvature-focused" stacking --- introducing an elastic penalty to adhesive binding.  Initial models of curvamer assembly considered cylindrically-curved shell shapes\cite{tanjeem-curvamers-2022, sullivan-curvamers-2024}, in which stacking assemblies were predicted to achieve a self-limiting size range up to 100 shells tall, provided the interaction range is short enough to avoid unfavorable, gap-opening between shells.  Notably, by considering only Euclidean shell geometries, this prior curvamer model incorporated a limited scope of elastic distortions, namely only bending deformations, upon stack assembly. 

In this article, we study the stacking assembly of non-Euclidean shells, where we show that assembly necessarily requires deformations of Gaussian curvature and therefore qualitatively and quantitatively alters the thermodynamics of elastic energy accumulation with size (see Fig. \ref{fig:intro}c).  Specifically, changes to a shell's Gaussian curvature are associated with metric distortions on the mid-surface of the shell in the form of stretching\cite{doCarmo2016Chapter4}.  An illustrative example of this effect is how, in the process of projecting the positively curved surface of the Earth to a flat page, Greenland often appears to be roughly the same size as Africa when, in fact, Africa is approximately $14$ times larger by area.  Crucially, stretching modes of elastic deformation tend to be far more costly than bending, especially for thin shells, with the ratio of stretching to bending penalties proportional to $Y/B \sim t^{-2}$ where $Y$ is the two-dimensional Young's modulus,  $B$ is the bending modulus and $t$ is the thickness of the shell. Though not yet studied in the context of SLA, the complex elasticity of non-Euclidean surfaces has been shown to play an important role producing a wide variety of soft matter phenomena including the buckling transitions of spherical shells\cite{aggarwal-viralshellelasticity-2012,singh-viralcapsidshells-2020,kosmrlj-sphericalshellstatmech-2017}, and the wrinkling and snapping elastic instabilities of ribbons, plates and shells that occur in response to swelling \cite{ciarlet-leafripples-2009,holmes-noneuclideanshells-2017,holmes-flytrapinstabilities-2018,holmes-confinedswelling-2019,holmes-stimulishelltheory-2024}, mechanical stimuli\cite{hayward-shellsnapping-2015,davidovitch-wrinkledsheets-2016, roback-dynamicalmodesnoneuclidean-2026}, and incompatible geometries\cite{efrati-unconstrainednoneuclideanplates-2009,efrati-noneuclideanplatebuckling-2009,efrati-noneuclideanplatetheory-2010,davidovitch-confinedsheets-2016, santangelo-frustratedribbon-2025}.  

We find that the non-Euclidean shells generally accumulate elastic penalties with stack size faster than the cylindrical shells, the rate greatly depending on the strength of stretching penalties relative to bending which is measured by a dimensionless {\it F\"oppl-von K\'arm\'an number}~\cite{seung-defectelasticity-1988, lidmar-viralshells-2003} $\eta \sim w^4\kappa_0^2/t^2$, where $\kappa_0$ characterizes the magnitude of shell curvature and $w$ is the wide, lateral shell width.  We show that the self-limiting behavior is nearly identical in the limit of small $\eta$, where bending dominates, while in the thin-shell limit (large $\eta$), additional stretching dependent costs for non-Euclidean shells tend to suppress the self-limiting stack size relative to cylindrical shells. Notably, the nature of elastic energy accumulation with size depends on the {\it sign} of the preferred Gaussian curvature, $K_{G0}$, due to differences in the geometry of parallel stacking spherical and saddle-like surfaces (cf Fig. \ref{fig:intro}.  We find that a key consequence of this is that hyperbolic, saddle-shaped shells exhibit an especially flat dependence of stack size on inter-curvamer binding, as well as a smaller range of finite-temperature size fluctuations in stack size, features that may be attractive for engineered self-limiting assemblies.  We conclude with a simplified analysis of the gap-opening frustration escape mode that occurs for finite ranges of attraction and find that  spherical cap stacks becoming gap-opened at significantly smaller sizes than stacks of saddles or cylindrical shells, reducing the potentially accessible range of self-limiting assembly for $K_G^{(0)}>0$ curvamers.  

The remainder of this article is organized as follows.  In Sec. \ref{section:models} we first introduce the geometry of curved shell stacking, and then describe  the continuum mechanical model of frustrated shell stacking, followed by the coarse-grained, discrete shell model.  In Sec. \ref{section:sla} we present the predictions of the continuum theory and results from the energy-minimized discrete model on the self-limiting ground states of shell stacks with different geometries and describe the dependence of self-limiting assembly on preferred Gaussian curvature and the ratio of stretching to bending cost.  Next, in Sec. \ref{section:fluctuations}, we model and compare relative fluctuations of stack size around the self-limiting size for the distinct curvamer shapes.  Finally, we analyze the transition from gap-closed to gap-opened stacking in Sec. \ref{section:gaps} before concluding with a discussion in Sec. \ref{section:discussion}.

\section{Models}
\label{section:models}

In this section, we develop an analytic continuum model and a coarse-grained numerical model of elastic shell stacking where a shell's ``bottom'' surface preferentially binds to another shell's ``top'' surface aligned along a common axis perpendicular to the center of each shell's midsurface.  As first reported in ref. \cite{tanjeem-curvamers-2022}, this stacking configuration is necessary to propagate elastic stresses over long ranges which enable self-limiting behavior and can be induced through the use of several patchy attractive zones along shell surfaces.  Below we first discuss the geometry of frustrated shell stacking and how it induces the curvature strains that ultimately generate self-limiting elastic costs in Sec. \ref{section:geometry}.  Then we summarize a theory of shallow shell elasticity in Sec. \ref{section:shellelasticity} before applying it to derive the equilibrium equations of state for elastic shell stacks in Sec. \ref{section:shellstacktheory}.  Similar to the continuum theory in ref. \cite{tanjeem-curvamers-2022}, we restrict our model to the case of curvature-focused stacking where shell surfaces adhesively bind to each other without gaps. Such stacking can be induced through the use of short-ranged, stiff attractive interactions, however an analysis of this assumption will be presented later in the results section through a scaling argument for the transition to gap-opened stacking.  Then in Sec. \ref{section:cgmodel} we introduce a coarse-grained bead-spring model of discrete shell stacking with finite-ranged attraction that relaxes many of the assumptions of the continuum theory. 

\begin{figure*}[h]
    \centering
    \includegraphics[width=0.95\textwidth]{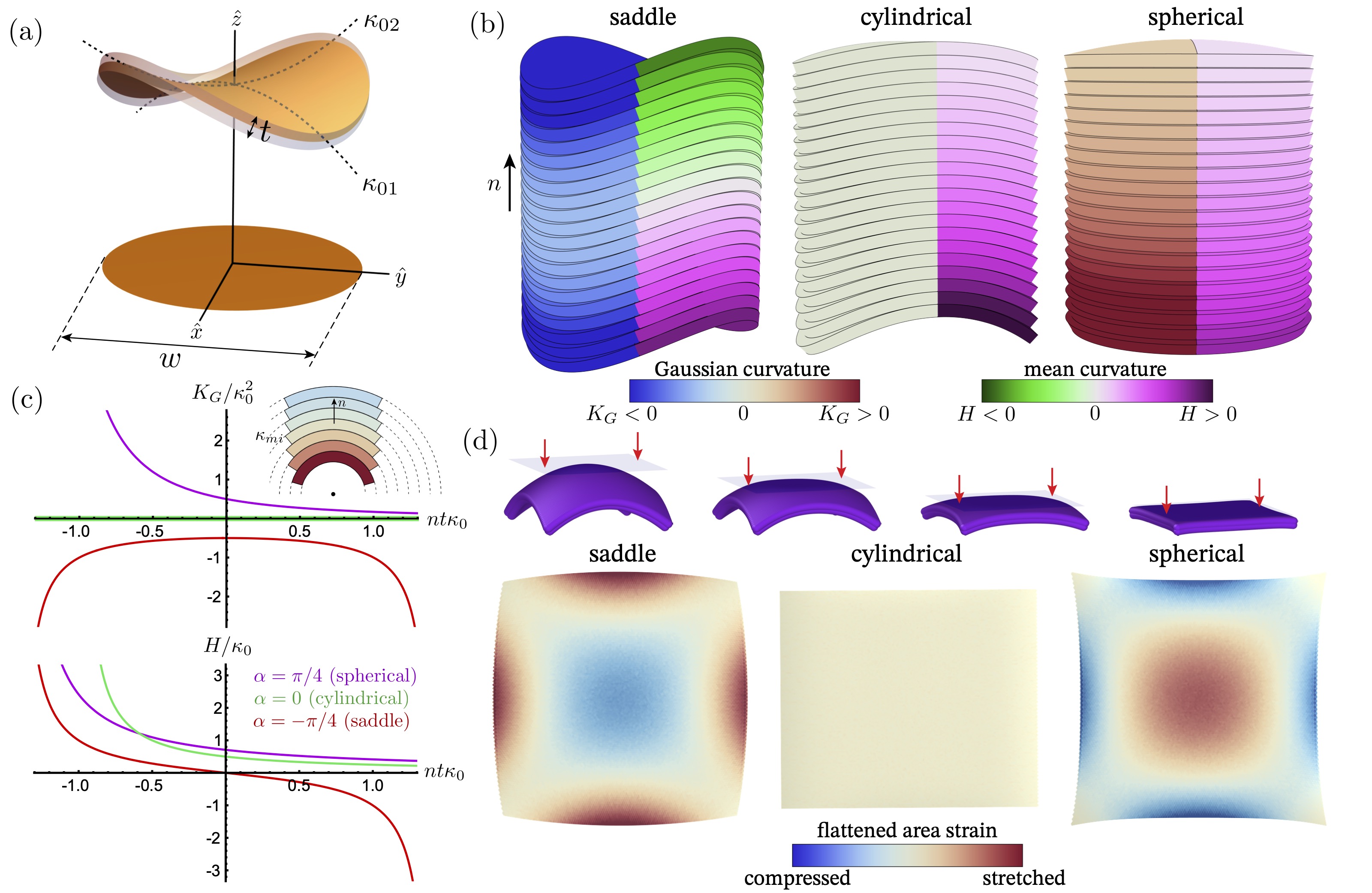}
    \caption{Geometry of shell stacks. (a) The design of a curved shell in the continuum model with circular footprint of diameter $w$, thickness $t$, and preferred principal curvatures $\kappa_{01}$, $\kappa_{02}$.  (b) Curvature-focused, gap-closed stacking requires gradients of Gaussian curvature (left half of each stack) for non-Euclidean shells while cylindrical shells maintain preferred zero Gaussian curvature.  Gradients in mean curvature (right half of stacks) are present for all shell shapes and generate elastic bending penalties. (c) Plots of the in-stack profiles of Gaussian (top) and mean (bottom) curvature for the stacks shown in (b).  (d) The dilational, area strain of the mid-surface of coarse-grained shells after being flattened.  In general, variations away from preferred Gaussian curvature require in-plane stretching meaning the gradients of Gaussian curvature seen in (b) will generate stretching costs.}
    \label{fig:stackgeometry}
\end{figure*}

\subsection{Geometry of curved shell stacking}
\label{section:geometry}

We first introduce the basic geometry of shells, and most importantly, the kinematics of stacking parallel curved layers, which depends crucially on the absence or presence of Gaussian curvature.

Shells are assumed to be thin, quasi-two-dimensional objects with uniform thickness, $t$, and are thus fully described by their mid-surface (see Fig. \ref{fig:stackgeometry}a).  For the continuum model, we will additionally assume that shells have a circular footprint of diameter $w$ in their long dimension and are shallow, meaning the radius of curvature in any direction is large compared to the size of the shell.  The shape of a shell is then determined by the out-of-plane displacement $h(\vec{x})$ of a point $\vec{x}$ in the $xy$-plane to the mid-surface of the shell (i.e. assuming the Monge parameterization).  We describe shells as generalized paraboloids, parameterized by their mid-point principal curvatures, $\kappa_{1}$ and $\kappa_{2}$,  along the $x$- and $y$-axes, respectively,
\begin{equation}
    \label{eqn:xyheightfunction}
    h(\vec{x}) = h_0-\frac{1}{2}\kappa_1~x^2 - \frac{1}{2}\kappa_2~y^2\text{.}
\end{equation}
We further assume the lateral width of shells, $w$, is sufficiently small that shells remain shallow (i.e. $|\kappa_{1,2} w| \ll 1$) such that we may neglect variations of curvature over the shell surface.  In this approximation the mean and Gaussian curvature are respectively
\begin{equation}
    H = \frac{1}{2}(\kappa_1 + \kappa_2)\text{;}  \ \  K_G = \kappa_1\kappa_2\text{.}
\end{equation} 
We denote a shell's stress-free, preferred shape by curvature quantities  with a subscript zero and we additionally define a characteristic curvature scale, $\kappa_0 = \sqrt{\kappa_{01}^2+\kappa_{02}^2}$ that is proportional to the root-mean-square (RMS) of the preferred curvatures averaged along different local directions in the shell.  Finally, we introduce a {\it shape angle}, $\alpha$, to parameterize preferred shell curvatures,
\begin{align}
    &\kappa_{01} = \kappa_0\cos\alpha\text{,}\\
    &\kappa_{02} = \kappa_0 \sin\alpha\text{,}
\end{align}
with $\alpha = -\pi/4\text{, } 0 \text{, } \pi/4$ corresponding to saddle, cylindrical and spherical cap shells, respectively.  In this way, $\kappa_0$ quantifies the ``degree'' or amount of curvature (i.e. RMS curvature) the shell possesses while $\alpha
$ determines how that curvature is distributed among the two axes on the shell.

Our model considers curvamer binding via attractive interactions between bottom and top surfaces of adjacent neighbor shells, such that when interactions are sufficiently short-ranged and stiff, shell surfaces adhere completely, with the mid-surfaces then forming uniformly spaced layers.  This configuration of stacked so-called parallel surfaces~\cite{hyde_beyond_1997}, which we refer to as ``curvature-focused'' stacking, is the central geometrical mechanism at play in this class of GFA and is a motif featured prominently in smectic phases of liquid crystals\cite{sethna_spheric_1982,didonna-bluephases-2002,didonna-bluephases-2003,kamien-focalconics-2023}.  Consequently, the radii of curvature $r_i(z)$ of a shell a distance $z$ away from the middle in a curvature-focused stack must be related to the radii of curvature of the middle shell $r_{mi}$ according to $r_i(z) = r_{mi} + z$~\cite{hyde_beyond_1997}.
This requires the mean and Gaussian curvatures to vary in a stack as
\begin{equation}
\label{eq: H(z)}
    H(z) = \frac{H_m+ zK_{Gm}}{1+2zH_m+z^2K_{Gm}}\text{,}
\end{equation}
and 
\begin{equation}
\label{eq: KG(z)}
    K_G(z) = \frac{K_{Gm}}{1+2zH_m+z^2K_{Gm}}\text{,}
\end{equation}
for a middle shell (at $z=0$) with mean curvature $H_m$ and Gaussian curvature $K_{Gm}$.  As illustrated in Fig. \ref{fig:stackgeometry}b-c, changes in mean and Gaussian curvatures increase the further one travels from the center of the stack, with the exception of Euclidean shell stacks ($K_{Gm}=0$) whose Gaussian curvature vanishes everywhere.  That is, Gaussian curvature of parallel stacking is only uniform for Euclidean shell stacks, whereas parallel stacking of non-Euclidean shells necessarily introduces gradients of $K_G$ in the stack.

Critically, deviations away from the preferred Gaussian curvature of a shell generate elastic stretching penalties while changes in mean curvature are the source of elastic bending penalties.  This fact, demonstrated visually in Fig. \ref{fig:stackgeometry}d by flattening shells using the coarse-grained model and analyzing mid-surface area strains (i.e. metric distortions of the mid-surface), will be explicitly modeled in Sec. \ref{section:shellelasticity}.  Based on this, we will show that non-Euclidean shell stacks are subject to larger elastic costs in general compared to stacks of Euclidean shells which only feature penalties for bending distortions, which become relatively less costly as shells become thinner.  

We note further that the profile of Gaussian curvature in non-Euclidean shell stacks is qualitatively different depending on the sign of $K_{Gm}$.  Stacking around a saddle shell ($K_{Gm}<0$, $H_m = 0$) produces a symmetric Gaussian curvature profile, while stacking around a spherical cap ($K_{Gm} > 0$, $H_m > 0$) is asymmetric --- shells in the top half of a stack become flatter while those in the bottom half become more curved overall.  Notably, variation of $K_G$ grows {\it quadratically} with $z$ for saddles, while it is linear for spherical shells.  Thus, we may anticipate the nature of elastic energy accumulation within shell stacks (and therefore self-limiting behavior) should differ depending on the sign of $K_{G0}$, with stacks of negatively curved shells likely exhibiting a more strongly non-linear accumulation of stretching costs with stack size than positively curved shells.

\subsection{Continuum model of shell deformations}
\label{section:shellelasticity}

\begin{figure*}
    \centering
    \includegraphics[width=1\linewidth]{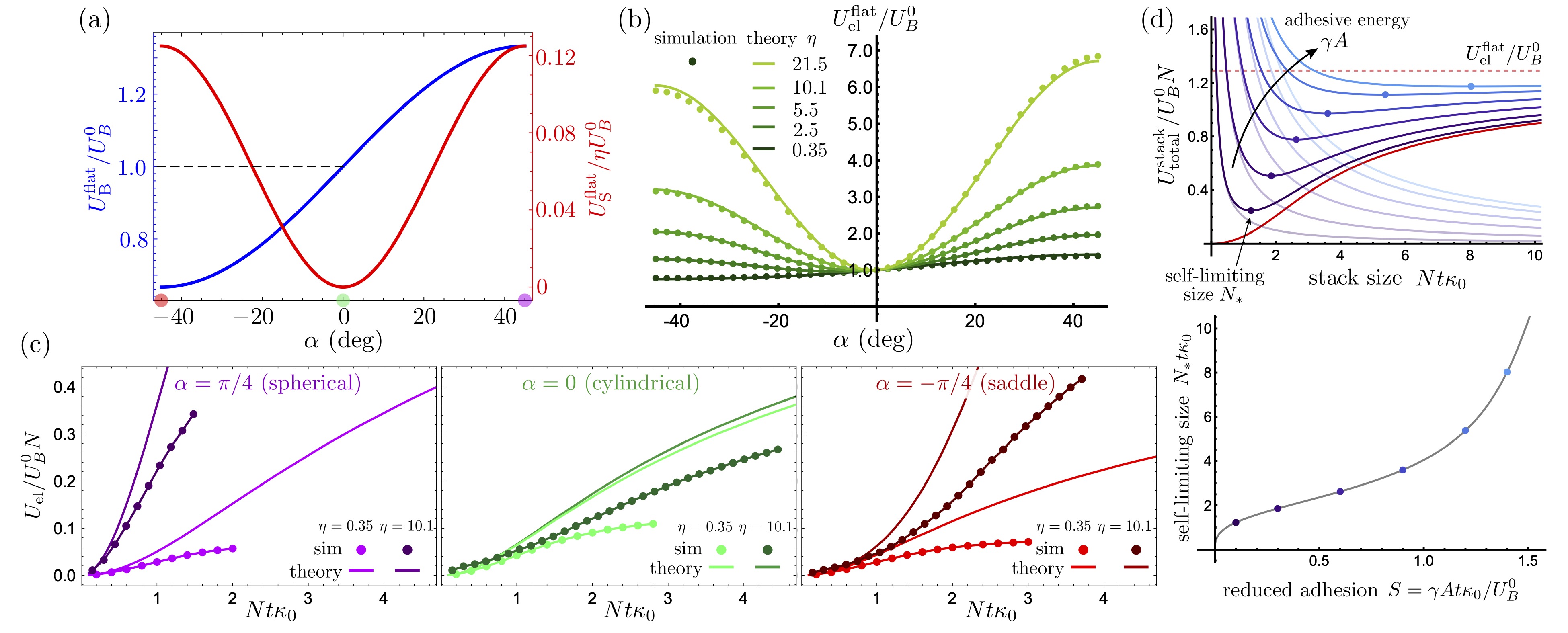}
    \caption{(a) The energy to flatten a shell, separated into bending (blue) and stretching costs (red), for different preferred shapes $\alpha$, where $\kappa_{01} = \kappa_0\cos{\alpha}$ and $\kappa_{02} = \kappa_0\sin{\alpha}$, and Poisson's ratio $\nu = 0.3$.  Energies are normalized by $U_B^0$ which is the cost to flatten a cylinder with the same preferred RMS curvature $\kappa_0 = \sqrt{\kappa_{01}^2+\kappa_{02}^2}$. (b) The total elastic energy of flattened shells.  The values of $\eta$ for the continuum model (solid lines) are determined by fitting to the coarse-grained, simulation data (dots). Increasing $\eta$ results in stronger stretching penalties.  (c) The accumulation of elastic costs as stack size increases for shells with different preferred Gaussian curvatures and thicknesses. (d) The self-limiting assembly size minimizes the per subunit free energy (top).  Increasing adhesive strength increases self-limiting size for the same shell design, thus the accumulation of per shell elastic costs set the ``program'' for self-limiting behavior (bottom).  Coarse-grained shells shown here have horizontal spring constants $k_h = 1.0\, \gamma A / d^2$ in (b) and $k_h = 0.5\, \gamma A / d^2$ in (c).}
    \label{fig:models_continuum}
\end{figure*}

Here, we briefly summarize the elastic model we use for shell deformations, known as Marguerre-von K\'arm\'an (MvK) theory \cite{ciarlet-mvkshells-1986} which describes the geometrically non-linear coupling of in-plane (i.e. metric) distortions with out-of-plane (bending) deflections for thin, 2D elastic membranes whose mid-surface is curved in the stress-free configuration.  In this study we exploit the solutions of Seffen~\cite{seffen-bistableshells-2006}, which describe the case of orthotropic, uniformly curved, shallow shells with elliptically shaped footprints and provide explicit expressions for elastic stretching and bending energies in terms of deviations away from preferred shell curvatures.

As reviewed in more detail in Appendix \ref{section:appendix-elasticity}, the core assumption of shallow shell elasticity is that the strain, $\bar\epsilon_{ij}$, experienced in a deformed shell can be approximated as the strain, $\epsilon_{ij}^0$, of a flat plate deformed to the preferred shape of the shell (the ``pre-strain'') subtracted from the strain, $\epsilon_{ij}$ in deforming a planar plate to the current configuration of the shell,
\begin{equation}
    \bar\varepsilon_{ij} = \varepsilon_{ij} - \varepsilon_{ij}^0\text{ .}
\end{equation}
In general, the strain is $\varepsilon_{ij} \simeq \frac{1}{2}\left( \partial_iu_j + \partial_ju_i +  \partial_ih\partial_jh  \right)$
where $u_i$ are the $i,j = x,y$ components of the displacements and $h$ is the displacement in the $\hat{z}$ direction, normal to the reference plane, and neglecting second-order strain contributions $\partial_iu_k\partial_ju_k$ under the assumption that rotations in-plane are small.  Similar to the variational approach to the F\"oppl-von K\'arm\'an (FvK) equations\cite{LandauLifshitzElasticityChapter2} for a flat plate\cite{seung-defectelasticity-1988}, the in-plane stretching free energy is $F_{s} = \frac{1}{2}\int\left(2\mu \bar{\varepsilon}^2_{ij} + \lambda\bar{\varepsilon}^2_{kk}\right)\text{d}A$ with 2D Lam\'{e} coefficients $\mu$ and $\lambda$.  Minimization of in-plane components ($u_i$) of the displacement subject to a given, shallow paraboloidal shape, eq. (\ref{eqn:xyheightfunction}), yields a condition for mechanical equilibrium, $\partial_i \sigma_{ij}=0$, and a compatibility condition relating in-plane to out-of-plane displacements (see Appendix \ref{section:appendix-elasticity}) 
\begin{equation}
    \nabla^2_{\perp} \sigma_{ii} =-Y\Delta K_{G}\text{,}
\end{equation} where $\Delta K_{G} = K_G-K_{G0}$ is the deviation away from preferred Gaussian curvature and $Y=4\mu (\lambda+\mu)/(\lambda+2 \mu)$ is the 2D Young's modulus of the shell.  This equation captures mechanical and energetic consequences of Gauss's well-known theorem proving that changes in the Gaussian curvature necessarily require metric distortion~\cite{doCarmo2016Chapter4}, i.e. changes of distance of material points along the mid-surface of the shell.  As changes in Gaussian curvature require second-derivatives of stress, under appropriate boundary conditions the in-plane stresses grow as $\sigma \approx -Y \Delta K_Gw^2$, notably growing with {\it wide} lateral dimension of the shell $w$.  Critically, according to eq. (\ref{eq: KG(z)}), non-Euclidean shells necessarily acquire these in-plane stretching deformations upon parallel stacking, as $\Delta K_G \neq 0$.

Following Seffen~\cite{seffen-bistableshells-2006}, we compute the elastic energy of paraboloidal shells in terms of the vector of surface curvature $\vec{\kappa} = (\kappa_1, \kappa_2)$ and its differences from the preferred shell curvatures $\vec{\kappa}_0 = \kappa_0 ( \cos \alpha, \sin \alpha)$.  This includes the elastic energy of bending distortions

\begin{equation}
\label{eqn:1shellBend}
    \frac{U^{\rm shell}_{\rm B}(\vec{\kappa},\vec{\kappa}_0)}{A} = \frac{B_\parallel }{2} \Big[\left(\kappa_1 - \kappa_{01}\right)^2 +  \left(\kappa_2 - \kappa_{02}\right)^2\Big]  + B_\perp  \left(\kappa_1 - \kappa_{01}\right)\left(\kappa_2 - \kappa_{02}\right) \text{,}
\end{equation}
where $A = \pi w^2/ 4$ is the shell area and $B_\parallel$ are $B_\perp$ are bending moduli, as well as elastic energy of in-plane strains, or ``stretching",
\begin{equation}
    \label{eqn:1shellStretch}
   \frac{U^{\rm shell}_{\rm S}(\vec{\kappa},\vec{\kappa}_0) }{A} = \frac{Y w^4 }{6144}\left(\kappa_1\kappa_2 - \kappa_{01}\kappa_{02}\right)^2\text{,}
\end{equation}
where $Y$ is the 2D Young's modulus\footnote{For a shell composed of a homogeneous elastic material of thickness $t$ with Young's modulus $E$ and Poisson ratio $\nu$, bending moduli are given by $B_\parallel = Et^3/12(1-\nu^2)$, $B_\perp = B_{\parallel}\nu$ and the 2D Young's modulus is $Y = Et$. }.  

To assess the relative strength of bending and stretching energies, it is useful to compare the elastic energies needed to flatten the shells (i.e. $\vec{\kappa} = 0$).  These are, up to numerical prefactors (dependent on $\alpha$ and $B_{\parallel}/B_\perp$), proportional to $U_B^{0}/A \equiv B \kappa_0^2/2$ and $U_S^{0}/A \equiv Y (\kappa_0 w)^4/6144$, respectively, taking $B_{\parallel} = B$.  

The ratio of these energy scales defines a key quantity, the FvK number,
\begin{equation}
\label{eqn:eta}
\eta \equiv  \frac{U_S^{0}}{U_B^{0} }= \frac{1}{3072}\frac{ Y w^4\kappa_0^2}{B }=  \frac{1-\nu^2}{256}\frac{ w^4\kappa_0^2}{t^2 }\text{,}
\end{equation}
where on the right-hand side we use the relationship between bending and 2D Young's modulus for a homogeneous elastic shell of thickness $t$ and Poisson ratio $\nu$.  Notably, the FvK number shows that the relative cost of metric versus curvature distortions increases as shells become wide and thin.  

Fig. \ref{fig:models_continuum}a plots the energy costs to flatten, both as functions of variable $\eta$ but also the explicit dependence on stress-free shell shape, through the parameter $\alpha$.  Notably the stretching energy of a deformed shell is symmetric with respect to the sign of the preferred Gaussian curvature meaning the stretching cost to flatten a shell is the same for a spherical cap as a saddle shell (and zero for a cylindrical shell, $\alpha =0$).  In contrast, assuming $\nu >0$, the reduced bending cost of a flattened shell is lowest for negatively curved, saddle shells and greatest for positively curved, spherical caps.  The difference stems from the Poisson effect with bending costs from flattening a shell becoming identical for all shapes as $\nu \rightarrow 0$.  The total flattening energy is plotted in Fig. \ref{fig:models_continuum}b for different preferred shell shapes and compared to the energy of flattened coarse-grained shells (described below), where it becomes clear that this elastic cost is dominated by stretching for $\eta \gg 1 $.

\subsection{Continuum model of cohesive stacking assembly}
\label{section:shellstacktheory}

Here, we describe a thermodynamic model of curvature-focused stacking assembly in curvamer particles.  In equilibrium, a stack comprised of $N$ shells will have two components contributing to the total free energy: the adhesive energy from attractive interactions between adjacent shells, and the elastic energy cost of deforming each shell in the stack,
\begin{equation}
    U^{\rm stack}_{\rm total}(N) = U_{\rm adh}(N) + U_{\rm el}(N)\text{.} 
\end{equation}
Denoting $\gamma$ as the adhesive strength per unit area of contact between shell surfaces, the energy of binding two shells together is then $-\gamma A$, where $A = \pi w^2/4$.  Since shells are only shallowly curved, the contact area can be approximated as the total area of the shell as in the flat case.  The total adhesive energy for $N-1$ ``bonds'' between $N$ shells in a stack is simply
\begin{equation}
    U_{\rm adh}(N) = -\gamma AN + \gamma A\text{,}
\end{equation}
where the first term is an extensively growing bulk term while the second term is a constant energetic cost associated with having exposed boundaries at the top and bottom of the stack. 

We construct the elastic energy in a stack of $N$ curvamer particles, which satisfy the constraint of perfect conformal contact between their surfaces and the constraints of parallel layer stacking summarized in Sec. \ref{section:geometry}.  Thus, the shape profile is parameterized by the shape of a single curvamer in the stack, which we choose to be the mid-point.  Labeling shells by integers $n \in [-N/2,N/2] $, the mid-point corresponds to the shell at $n = 0$, neglecting small discretization corrections in the limit of $N\gg1$.  Defining the curvature of the mid-shell as $\vec{\kappa}_m$, the curvatures at position $n$ simply follow the curvature-focusing form
\begin{equation}
\kappa_{i}(n,\vec{\kappa}_m) = \frac{\kappa_{mi}}{1+nt\kappa_{mi}} \text{.}
\end{equation}
To compute the total stack energy for a given $\vec{\kappa}_m$, we simply sum bending and stretching energies of the single shells, eqs. (\ref{eqn:1shellBend}) and (\ref{eqn:1shellStretch}), converting the sum to an integral in the limit of large $N$
\begin{equation}
        U^{\rm stack}_{\rm S,B}(N,\vec{\kappa}_0,\vec{\kappa}_{m})  = \int_{-N/2}^{N/2} U^{\rm shell}_{\rm S,B}\big(\vec{\kappa}_m(n), \vec{\kappa}_0  \big) \text{d}n\text{.}
\end{equation}
For a given stack size $N$ and preferred shell shape $\vec{\kappa}_0$, the elastic energy is computed by minimizing over the principle curvatures of the central particle at $n=0$, which is described in detail in Appendix \ref{section:appendix-stacking}, yielding
\begin{equation}
    U_{\rm el}(N) \equiv {\rm min}_{\vec{\kappa}_m} \Big[U^{\rm stack}_{\rm S}(N,\vec{\kappa}_0,\vec{\kappa}_{m}) +U^{\rm stack}_{\rm B}(N,\vec{\kappa}_0,\vec{\kappa}_{m}) \Big] .
\end{equation}
As described previously~\cite{hagan_sla_2021}, in the canonical ensemble, thermodynamic self-limitation in size derives from a minimum in the interaction (free) energy per particle of assembly.  In the context of GFA such a minimum relies on the balance of cohesive costs (at the free boundary of a finite structure) and the super-extensive growth of elastic costs of frustration with increasing assembly size.  Defining the interaction free energy per particle as $\epsilon(N) \equiv U^{\rm stack} (N)/N$, we have
\begin{equation}
\epsilon(N) = \epsilon_0 + \frac{\gamma A}{N} + \epsilon_{\rm ex}(N) \, 
\end{equation}
where $\epsilon_0 = - \gamma A$ is a constant and $\epsilon_{\rm ex}(N)\equiv U_{\rm el}(N)/N$ is the so-called {\it excess} energy density of assembly beyond the bulk and surface terms derived from local cohesive forces.  We compute the self-limiting stack size $N_*$ from the minimization of $\epsilon(N)$, which is defined by the following equation of state
\begin{equation}
\gamma A = \Big[N^2 \partial_N \epsilon_{\rm ex}(N) \Big]_{N=N_*} \ .
\end{equation}
In Appendix \ref{section:appendix-stacking}, we show that by rescaling curvature variables by preferred RMS curvature $\kappa_0$, stack sizes by $(\kappa_0 t)^{-1}$ and energies by the bending cost of flattening $U_B^{0}$, the thermodynamics of the curvature-focused stacking model of shallow shells depends only on three dimensionless quantities:  the FvK number $\eta$; the shell shape parameter $\alpha$; and a dimensionless measure of adhesion
\begin{equation}
    S \equiv  \frac{\gamma A(\kappa_0 t)}{U_B^{0} } \ .
\end{equation}
This measure of ``stickiness'' $S$ can be thought of as the ratio of the boundary cost per subunit in a stack of size equal to the preferred RMS curvature radius ($\kappa_0^{-1}$) relative to the elastic bending cost of flattening a curvamer.  Intuitively, it can be expected that the minimum $N_*$ of $\epsilon(N)$ increases with $S$ as shown schematically in Fig. \ref{fig:models_continuum}d for a single shell design. In what follows, we consider the additional variation of stack thermodynamics and self-limiting stack size with parameters that reflect the possibility of preferred non-Euclidean shapes, specifically $\alpha \neq 0$ (non-cylindrical shapes) and variable ratio of stretching to bending costs reflected in the FvK number $\eta$.

\begin{figure}
    \centering
    \includegraphics[width=0.6\textwidth]{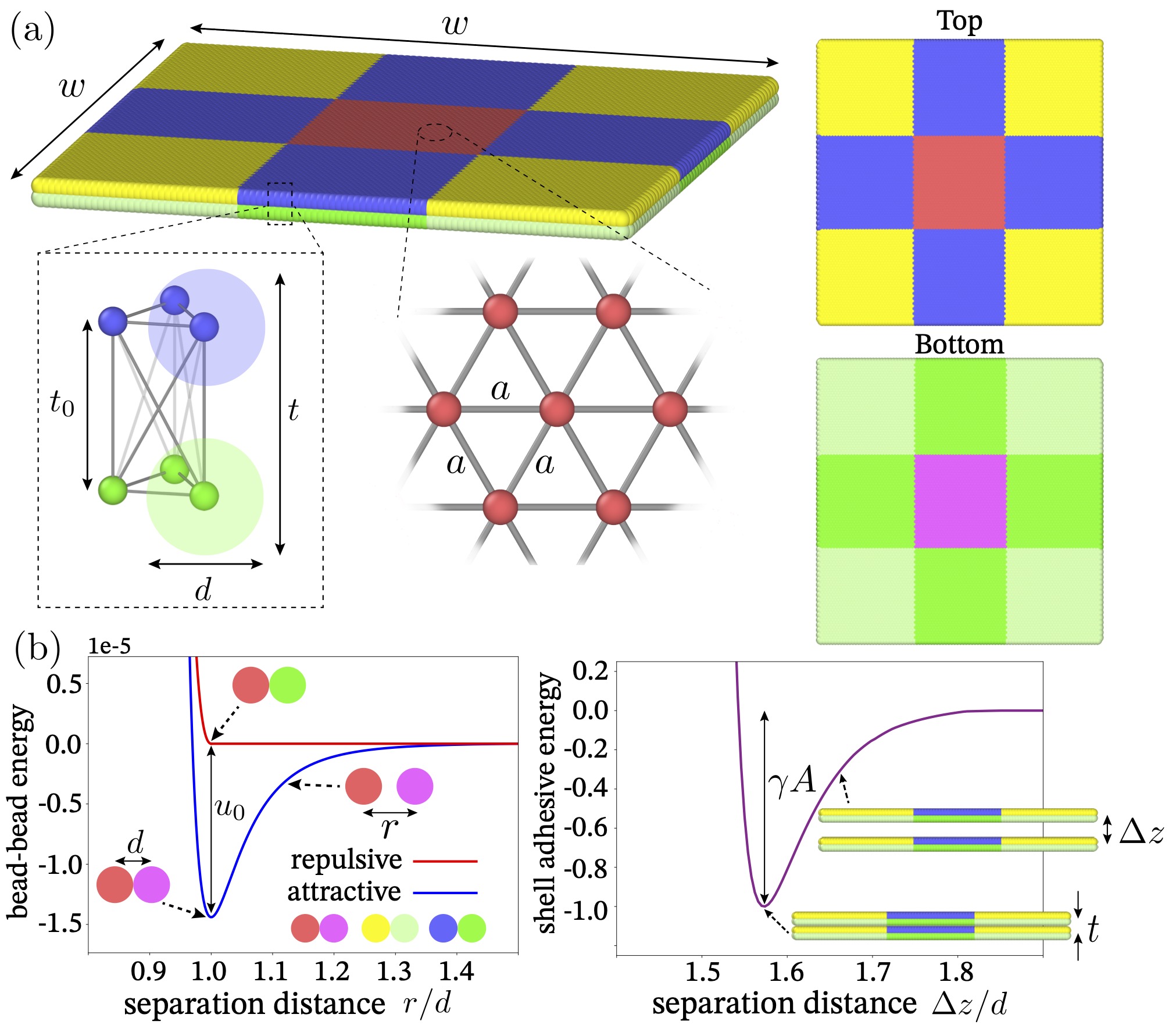}
    \caption{The coarse-grained, numerical model. (a) Discrete coarse-grained shells are formed by two layers of beads with soft-core diameter $d$ arranged in triangular lattices (lattice constant $a$) connected by harmonic springs.  Shells have a square footprint of side length $w$. There are a total of 6 different bead types (colors) which are localized to 9 different ``patchy'' zones on each layer.  (b) Lennard-Jones and Weeks-Chandler-Andersen potentials mediate attractive and repulsive bead-bead interactions (left), respectively, which give rise to overall repulsive-attractive inter-shell interactions (right).  Only interactions between three different bead type pairings (red-pink, yellow-light green, blue-dark green) are attractive, with all other pair interactions being repulsive to favor aligned stacking of shells.  Bead interaction strengths $u_0$ are set so that the total adhesive strength, $\gamma A$, between flat shells remains constant for all designs and attraction ranges. Renderings of coarse-grained shells made with \texttt{OVITO}\cite{ovito}.}
    \label{fig:cgmodel}
\end{figure}

\subsection{Coarse-grained model of non-Euclidean shell stacking assembly}
\label{section:cgmodel}

To test the continuum model and connect discrete shell properties to self-limiting assembly behavior, we perform numerical energy minimization calculations on a coarse-grained, bead-spring model of discrete, curved shell stacks using the \texttt{LAMMPS} software package\cite{LAMMPS}.  This model is a generalization of the previous 2D models of cylindrically curved shells~\cite{tanjeem-curvamers-2022, sullivan-curvamers-2024} to the case of both Euclidean and non-Euclidean shells in 3D.  Each shell consists of two square layers of beads arranged on a triangular lattice mesh that define the top ($+$) and bottom ($-$) surfaces of the structure, which are separated by a distance $t_0$ (see Fig. \ref{fig:cgmodel}a).  To construct the $+$ and $-$ meshes, we begin with a flat bilayer where the triangular lattice spacing is $a$ in both layers, and then deform the bilayer so that the mid-surface curves according to the Monge parameterization for a generalized paraboloid/hyperboloid given in eq. (\ref{eqn:xyheightfunction}), displacing the vertices of the $+$ and $-$ meshes so that they lie along the reference surface normal $\pm t_0/2$ away from the mid-surface respectively (see Appendix \ref{section:appendix-simdetails}).  This deformed configuration is used to set preferred rest-lengths of the springs within and between the mesh layers.  Three types of harmonic bonds connect the beads together to give the shell its elastic rigidity.  Those that connect beads in the same layer (``horizontal'') have spring constant $k_h$, while those that connect the same lattice point on different layers (``vertical'') have spring constant $k_v$.  Bonds that go from one layer to the other connecting adjacent lattice points (``cross'') have spring constant $k_c$ and provide shear resistance along the rectangular faces of the triangular prism unit cell. Relations between the spring constants are set so that a preferentially flat coarse-grained shell has the same effective elastic properties as an ideal plate with Poisson ratio $\nu_{xy} = 0.333$ in the plane of the plate and $\nu_z = 0.001$ out-of-plane satisfying the Kirchoff-Love hypothesis that plate thickness is unchanged under deformation (see Appendix \ref{section:appendix-simdetails} for spring constant ratio formulae).  We choose $k_h$ to be the free variable that sets the overall rigidity of the shell while the other spring constants are determined according the appropriate formula.  The rest lengths of every bond are independently set and are calculated based on the bead positions when the shell is in its preferred shape.  

We introduce ``patchy'' interactions between neighboring shells that favor aligned stacking along a common axis and suppress sliding lateral motions that interrupt self-limiting frustration propagation\cite{tanjeem-curvamers-2022}, patterning the top and bottom surface of each curvamer into nine patchy zones, as shown in Fig. (\ref{fig:cgmodel})a.  For aligned curvamer pairs, beads in each zone of one shell interact attractively with those in the complementary zones on the other via a pairwise Lennard-Jones potential
\begin{equation}
\label{eqn:ulj}
    u_{\rm LJ}(r) = 4u_0\left[ \left( \frac{\sigma}{r-\Delta}\right)^{12} - \left( \frac{\sigma}{r-\Delta}\right)^{6}\right]
\end{equation}
where $u_0$ is the bead-bead interaction strength, $\sigma$ is the attractive well width, and $\Delta$ is a shift parameter that sets the soft core diameter, $d$, of the beads.  Throughout the rest of this article, we measure lengths in the numerical model in units of $d$ by choosing $\Delta = d-2^{1/6}\sigma$ so that the bead diameter remains constant.  Additionally, interactions between beads in non-complementary zones are mediated by a Weeks-Chandler-Anderson potential 
\begin{equation}
    u_{\rm WCA}(r) = \begin{cases}
        u_{\text{LJ}}(r) + u_0\text{,} & r\leq d \\
        0\text{,} & r > d
    \end{cases}
\end{equation}
that is repulsive when beads overlap and smoothly goes to zero at $r = d$ (see Fig. \ref{fig:cgmodel}b). Beads within the same shell do not interact.  The entire shell surface is attractive when two shells are aligned correctly promoting stacking configurations that propagate frustration  stresses. The patchy interactions are necessary to strongly penalized lateral sliding motions between shells.  Fig. \ref{fig:cgmodel}b shows the repulsive and attractive nature of the total shell-shell interaction energy as a function of mid-shell separation distance when two shells are perfectly aligned. For every attractive range $\sigma$, we set the interaction strength $u_0$ such that the total adhesive energy between two flat shells is held at a constant value $-\gamma A$ when separated by their effective thickness $t \approx t_0 + d$ in the aligned orientation.  Furthermore, we measure all energies in the numerical model in units of the adhesive binding energy $\gamma A$.  

The exact coarse-grained shell model used in this article consists of $21,312$ beads in each shell using lattice constant $a = 0.316\,d$.  Additionally, each shell has a square footprint so that before being prescribed preferred curvatures, its mid-surface mesh forms a square with side length $w = 30\,d$.  To change $\eta$ we vary the structural thickness, $t_0$, and preferred RMS radius of curvature, $r_0$ (see Table \ref{table:shelldesigns} for shell designs used in this paper).  As the shell footprint is a square in the coarse-grained model and not circular, rather than use eq. (\ref{eqn:eta}) to calculate the value of $\eta$ for each coarse-grained shell design, we instead fit the continuum elastic theory from Sec. \ref{section:shellelasticity} to the flattened shell energies of coarse-grained shells to determine the corresponding theoretical value of $\eta$ (see Fig. \ref{fig:models_continuum}b).  For each of these designs, we study the self-limiting behavior of cylindrical ($\alpha = 0\degree$), spherical cap ($\alpha = 45\degree$), and saddle ($\alpha = -45\degree$) shell shapes.  For all the results shown in Sections \ref{section:sla}-\ref{section:fluctuations} shell-shell interactions are identical with microscopic interaction range $\sigma = 0.25 d$.  Throughout this study, zero temperature, energy minimization is performed on pre-built, curvature-focused stacks using a conjugate gradient descent algorithm with \texttt{LAMMPS}\cite{LAMMPS} to determine the  bead positions of the mechanically equilibrated structure and its final energy.

\section{Results}

We now consider how both shell elastic penalties and Gaussian curvature stacking profiles affect the accumulation of elastic energy and the self-limiting size selection of ground state stacking based on the continuum and coarse-grained curvamer models.
We first compare the accumulation of bending and stretching costs for different shell geometries and then study the suppression of self-limiting behavior in thin non-Euclidean shells in Section \ref{section:sla}.   In Section \ref{section:fluctuations}, we then study the relationship between differences in intra-stack accumulation of elastic energy and dispersity of selected stack size expected at finite temperature.  Last, in Section \ref{section:gaps}, we explore how the transition from frustration propagating, curvature-focused stacking to frustration escaping, gap-opened stacking changes from Euclidean to non-Euclidean curvamer shapes.

\subsection{Geometrically tunable self-limiting behavior }
\label{section:sla}

\begin{figure*}
    \centering
    \includegraphics[width=0.9\textwidth]{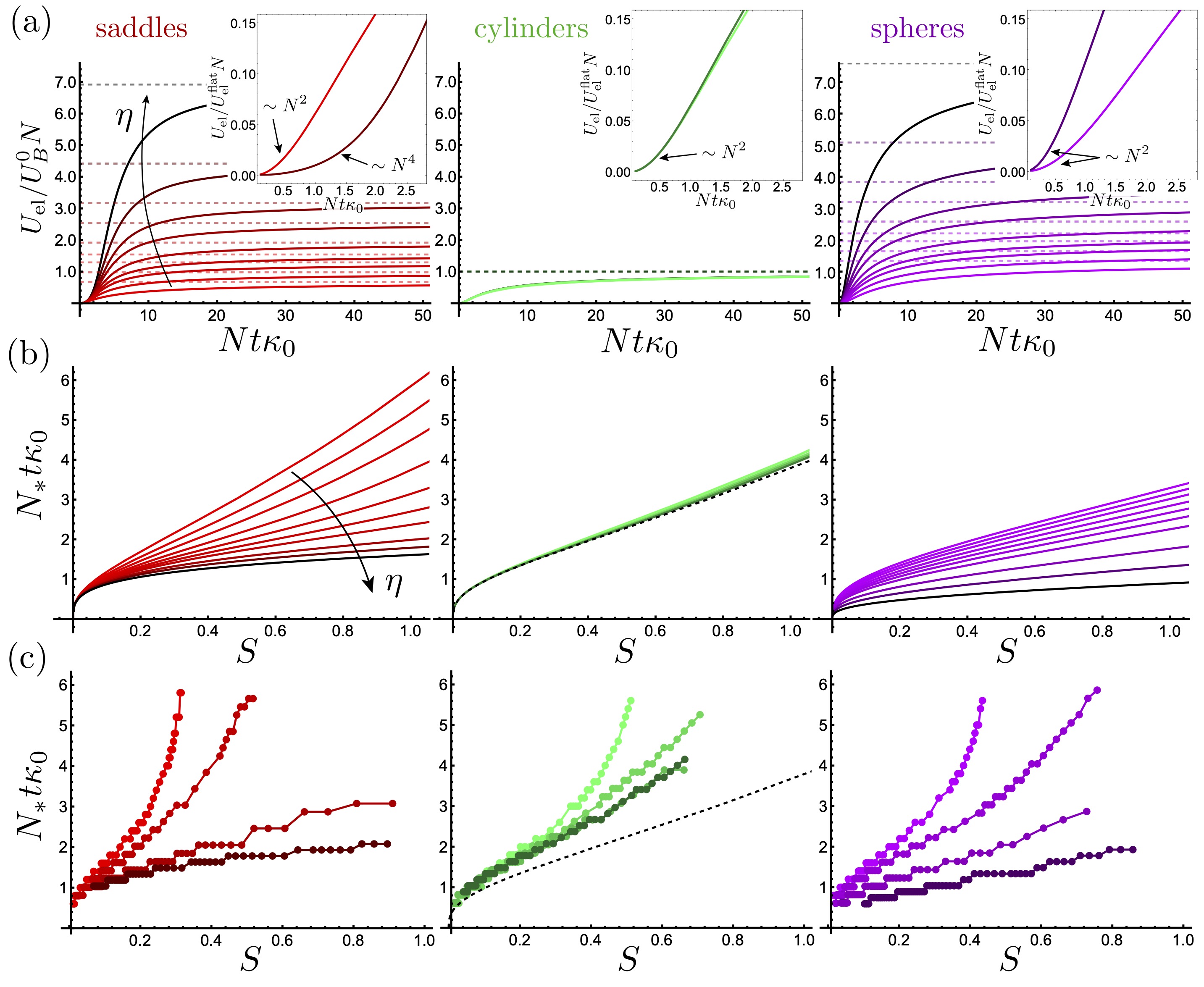}
    \caption{(a) Elastic energy densities versus stack size.  Transparent dashed lines denote shell flattening energies.  Insets show elastic energy density scalings with saddle stacks notably accumulating elastic costs as $\sim N^4$ and cylindrical and spherical cap stacks as $\sim N^2$.  Saddles and spherical caps generate stretching costs in addition to bending which are controlled by shell thickness and width through FvK number $\eta$, while cylindrical shell stacks only generate bending costs and are unaffected by $\eta$.  (b)-(c) Self-limiting size increases with reduced adhesion in all cases in the continuum theory (b) and coarse-grained model (c).  Increasing $\eta$ suppresses self-limiting size for the non-Euclidean shells due to higher elastic penalties from stretching.  The black dashed line plotted with the cylinders represents the curvature-focusing theory from ref. \cite{tanjeem-curvamers-2022}.  The data plotted in (c) are for $\eta= 0.05,\,0.35,\,2.5,\,10.1$.}
    \label{fig:results_shapes}
\end{figure*}

In Fig. \ref{fig:results_shapes}a, we plot the total elastic energy density (the excess energy density) as a function of stack size predicted by the continuum model, for an increasing sequence of FvK numbers for $0.03<\eta<25 $.  When shells are thick and $\eta$ is small the energy density curves remain similar, regardless of preferred Gaussian curvature, and saturate around the flattening energy of bending, $U_B^{0}$  (i.e. a constant elastic cost per curvamer).  When shells become thinner for larger $\eta$, there is a significant increase in the stack elastic energy density of non-Euclidean shells as stretching costs are heavily penalized relative to bending, while the elastic energy of Euclidean, cylindrical shell stacks is unaffected.  

Despite appearing similar at first glance, meaningful differences in the initial rates of elastic energy accumulation exist between positive and negative preferred Gaussian curvatures, which is made by comparing the growth of stretching costs in the limit of small stack size $N \ll N_{\kappa_0}$, where $N_{\kappa_0}\equiv( \kappa_0 t)^{-1}$ is the size when stacked shell layers have height equal to the preferred radius of curvature $r_0 = \kappa_0^{-1}$.  In this limit, stacking only weakly perturbs the shape of curvamer shells from their preferred shape, and we expect a mid-layer shape $\vec{\kappa}_m \simeq \vec{\kappa}_0$, and the stretching energy per shell is simply $U_S/N \sim Y A \langle |\Delta K_G|^2 \rangle w^4$, where $\Delta K_G (n) = K_G(n) -K_{G0}$ is the deviation from preferred Gaussian curvature at different positions in the stack.  Intuitively, concentric stacking of spherical shells implies Gaussian curvature changes linearly in stack for small $nt$ since the two principal curvatures focus in concert with $nt$, such that $|\Delta K_G| \propto \kappa_0^3 n t$ for spheres and therefore $U^{\rm sphere}_S/N \approx Y A \kappa_0^6 t^2 w^4 N^2$.  Notably, this quadratic increase with $N$ is the same power-law as the growth of bending energy per shell (for all shapes) $U_B/N \sim BA \kappa_0^4 t^2 N^2$, in accordance with the elastic energy of cylindrical shells \cite{tanjeem-curvamers-2022, sullivan-curvamers-2024}.  However, as shown in the parallel stacking formula eq. (\ref{eq: KG(z)}), when $H_m = 0$ the Gaussian curvature changes {\it quadratically} with stack height, since to linear order, curvature increases in one of the principal directions are exactly canceled by curvature decreases in the other direction, and $|\Delta K_G| \propto |K_{G0}|^2 (nt)^2$ to lowest order.  From this we expect a strongly anharmonic growth of the stretching energy in stacks of saddles, $U^{\rm saddle}_S/N \approx Y A \kappa_0^8 t^4 w^4 N^4$. This quartic dependence on stack size implies that, while saddle-shaped curvamers accumulate stretching costs due to their non-Euclidean shape, for small sizes $N  \lesssim N_{\kappa_0}$ the growth of that energy is much weaker than in positively-curved spherical caps, and only reaching a comparable accumulation in the large stack regime when $N  \gtrsim N_{\kappa_0}$.

In Fig. \ref{fig:results_shapes}b-c, the self-limiting stack size $N_*$ is plotted as a function of reduced adhesion $S$ for each of the three shapes.  We see qualitatively similar behavior between the continuum and coarse-grained models.  As a result of the faster accumulation of stretching elastic energy imposed by changes in Gaussian curvature, self-limiting size of non-Euclidean shells decreases significantly with the FvK number $\eta$, while the self-limiting behavior of cylindrical shells is essentially independent of $\eta$.  
It is straightforward to show that the balance of bending energy and cohesion select a preferred stack size for cylindrical shells 
\begin{equation}
\label{eqn:NBscaling}
    N_*({\rm cylinder}) \sim (\kappa_0 t)^{-1} S^{1/3}
\end{equation}
for the limiting case of small cohesion.  For small enough $\eta$, where stretching is negligible, $N_*$ follows a similar scaling for non-Euclidean shells as eq. (\ref{eqn:NBscaling}), but becomes $\eta$-dependent in the limit of thin shells. For spheres, the parabolic growth of stretching costs per curvamer implies
\begin{equation}
        N_*({\rm sphere}) \sim (\kappa_0 t)^{-1} (S/\eta)^{1/3} \ \ \ \  {\rm for} \ \eta \gg 1 .
\end{equation}
While for saddle-shaped curvamers in the thin shell limit, we expect the different power-law dependence of
\begin{equation}
        N_*({\rm saddle}) \sim (\kappa_0 t)^{-1} (S/\eta)^{1/5} \ \ \ \  {\rm for} \ \eta \gg 1 .
\end{equation}
due to the softer accumulation of stretching costs with size described above.

\begin{figure*}
    \centering
    \includegraphics[width=0.98\textwidth]{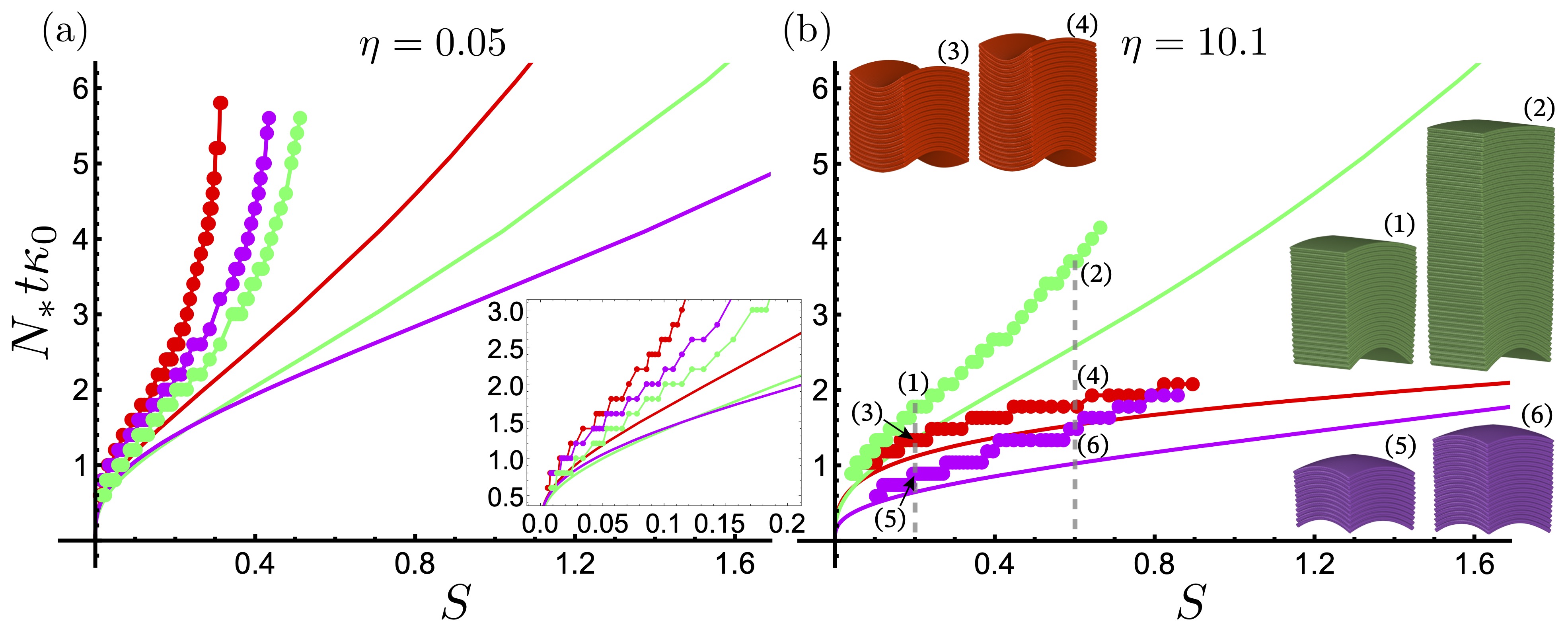}
    \caption{Comparisons of self-limiting sizes for different shapes for $\eta = 0.05$ in (a) and $\eta = 10.1$ in (b).  Stack sizes are generally ranked with $N_*({\rm saddle})>N_*({\rm sphere})>N_*({\rm cylinder})$ all being very close to each other for small $\eta$ and then transition to $N_*({\rm cylinder})>N_*({\rm saddle})>N_*({\rm sphere})$ with cylindrical shell stacks being considerably larger at high $\eta$.   Visualizing self-limiting stacks for $S=0.2$ and $S=0.6$, saddles increase from 18 to 24, spheres increase from 12 to 20, and cylinders increase from 24 to 50 shells tall.}
    \label{fig:results_sla-thickvthin}
\end{figure*}

To compare relative magnitudes of self-limiting size, we plot spheres, cylinders and saddles together in Fig. \ref{fig:results_sla-thickvthin}.  For small $\eta$ (relatively thick shells), the coarse-grained model shows self-limiting size to be only slightly different between the three shapes with $N_*({\rm saddle})>N_*({\rm sphere})>N_*({\rm cylinder})$.  Focusing in on small stacks, the continuum theory exhibits the same ranking but at larger sizes predicts that spheres and cylinders cross-over so that $N_*({\rm cylinder})>N_*({\rm sphere})$.  We speculate that this cross-over at larger sizes is not seen in the coarse-grained model due to gap-opened stacking behavior (i.e. departure from conformal binding) weakening the accumulation of frustration cost and ultimately leading to a divergence of self-limiting size (see Sec. \ref{section:gaps} below).  For thin shells with large values of $\eta$, the ranking is consistent between the two models with $N_*({\rm cylinder})>N_*({\rm saddle})>N_*({\rm sphere})$.  Additionally, the difference in self-limiting size between the shapes is noticeably greater than in the case for small $\eta$ with cylindrical shells having much larger stacks compared to the non-Euclidean shells, due to the significant increase to elastic energy required by stretching of saddles and caps.  The fact that stacks of saddle shaped particles are larger than cap shaped particles is consistent with the softer accumulation of stretching energy with stack size described above (i.e. $N^4$ compared to $N^2$).  The softer accumulation of stretching energy is also consistent with a relatively flatter dependence of $N_*$ on cohesion for saddles, relative to spherical caps and cylindrical shells in the larger $S$ range.    We highlight this fact in Fig. \ref{fig:results_sla-thickvthin}(b) where self-limiting stack sizes are compared for two values of reduced adhesion.  Tripling the adhesive strength from $S=0.2$ to $S=0.6$, stack size increases by 6 for the saddles, 8 for spherical caps, and 26 for the cylindrical shells.   This feature arguably makes saddles an attractive target for engineering self-limiting stack sizes, as the thermodynamically selected size is found to be relatively less sensitive to small changes in cohesive or elastic properties of particles.

\subsection{Size fluctuations and robust assembly}
\label{section:fluctuations}

Motivated by the gradual increase of the saddle shell self-limiting size curve discussed above and seen in Fig. \ref{fig:results_sla-thickvthin}, we analyze fluctuations around the self-limiting size with the intent to characterize which shell geometries ($\eta$ and $\alpha$) would lead to the smallest fluctuations and generally more robust self-limitation.  Near its finite-sized minimum $N_*$, the interaction free energy density can be approximated as a harmonic well with depth $-\epsilon_*$ and convexity $\epsilon_*''=\partial_N \epsilon\big|_{N_*}$ which is proportional to the energy required to change the assembly size to $N_* \pm 1$.  Ideal aggregation theory predicts that in a high-concentration, supersaturated state, particles assemble into a stacks following a roughly Gaussian distribution peaked around $N_*$ and characterized by size fluctuations around that size as\cite{hagan_sla_2021} 
\begin{equation}
\label{eqn:nflux}
    \frac{\langle \Delta N_*\rangle^{1/2}}{N_*} \simeq \frac{1}{(\beta N_*^3\epsilon_*'')^{1/2}}\text{,}
\end{equation}
where $\beta = 1/k_{\rm B} T$ is the inverse temperature.  As shown in the prior section, the minimum in $\epsilon(N)$ derives from distinct elastic penalties that grow with stack size and vary in magnitude and form depending on shell geometry (i.e. value and sign of preferred Gaussian curvature).  Here we consider how these distinct elastic effects impact the relative size-fluctuations of self-limiting stacks.

\begin{figure}
    \centering
    \includegraphics[width=0.75\textwidth]{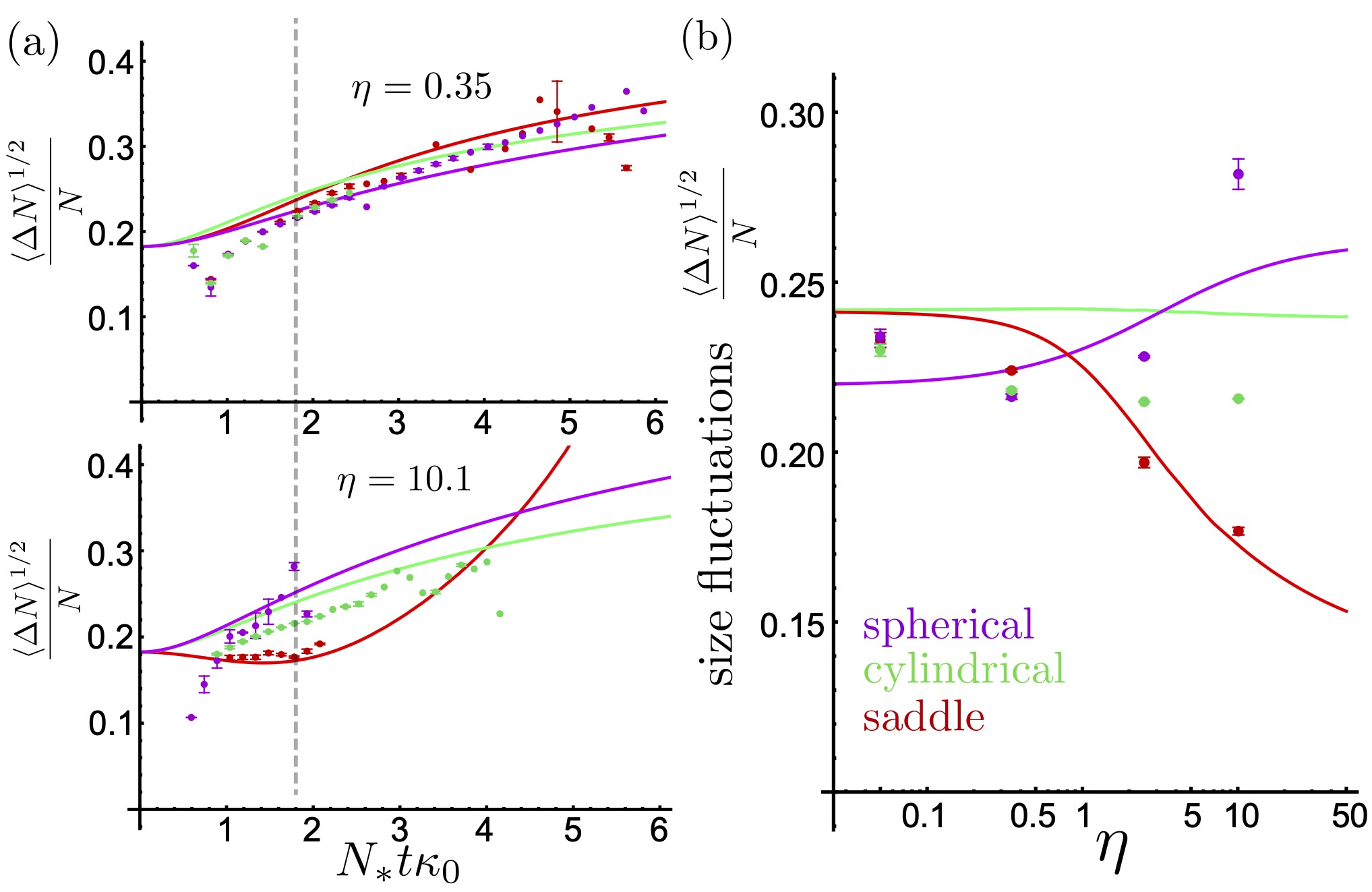}
    \caption{Fluctuations in stack size expected to be seen in thermal equilibrium relative to the self-limiting size for binding strength $E_{\rm adh} = 10k_B T$.  (a) Variations in stack size for thick shells (top) generally increase with self-limiting size and are similar for all preferred shapes.  For thin shells (bottom) there is a noticeable decrease in fluctuations for saddle shells compared to cylinders and spherical caps.  (b)  The change in relative stack size fluctuations for fixed self-limiting size ($Nt\kappa_0 = 1.8$) as shell thickness is decreased (increasing $\eta$). Thinner shells increase the cost of stretching penalties and results in size fluctuations decreasing in saddle stack as saddle are most sensitive to stretching penalties ($U_S \sim N^5$).  Error bars are the standard deviation of relative size fluctuations for shell stacks that have the same self-limiting size for different spring constants ($k_h$).}
    \label{fig:results_flux}
\end{figure}

The energetic density convexity $\epsilon_*''$ is readily obtained by taking derivatives with respect to size in the continuum theory and is approximated in the coarse-grained model by fitting the per shell energetic data around the self-limiting size with a quadratic fit.  Plotting the relative size fluctuations as a function of self-limiting size for thick shells ($\eta = 0.35$) in Fig. \ref{fig:results_flux}a, we see that all three shapes have generally the same degree of fluctuations and increase with self-limiting size, increasing from roughly 20-30\% as the mean self-limiting size $N_*$ varies from 1 to 5 times the preferred curvature scale $N_{\kappa_0}$.  When shells are thin ($\eta = 10.1$), however, the fluctuations in the saddles are suppressed, reaching a minimum of roughly 17\% before beginning to increase again for larger sizes.  We analyze this phenomenon further in Fig.\ref{fig:results_flux}b by comparing size fluctuations of stacks that exhibit the same reduced self-limiting size $N_* = 1.8 N_{\kappa_0}$, but for increasing FvK number $\eta$.  Decreasing the thickness causes the variance in saddles to decrease while the spherical caps and cylindrical shells remain large unchanged (relative size fluctuations of 20-25\%).  The smaller fluctuations imply that at thermal equilibrium, the distribution of stack sizes will have a much narrower peak at the self-limiting size for the saddles than the other shapes which would be more likely to see a broader dispersity of stack sizes resulting from the assembly process.  This fact combined with the slow increase in self-limiting size with adhesion suggests that, as long as the desired self-limiting size was within feasible range, saddles would be the optimal shape for achieving robust self-limitation that is insensitive to large uncertainties in interaction strength or shell stiffness.  

We attribute both effects to the uniquely ``soft'' and anharmonic accumulation of elastic cost in saddles (growing as $\sim N^4$ per particle) compared to cylinders and caps (growing as $\sim N^2$ per particle).  Relative to a quadratic growth in elastic cost, the quartic growth acts more like a non-linear ``switch", in effect leading to a steep and strongly nonlinear restraint on growth once stacks exceed the natural curvature scale $N \approx N_{\kappa_0}$.  Thus, for saddles the energy minimum is both relatively locked into this particular size scale and it is relatively deeper than the corresponding minima that result form the quadratic energy growth of cylinders and caps.   

\subsection{Gap-opened stacking and frustration escape}
\label{section:gaps}

\begin{figure*}
    \centering
    \includegraphics[width=1\textwidth]{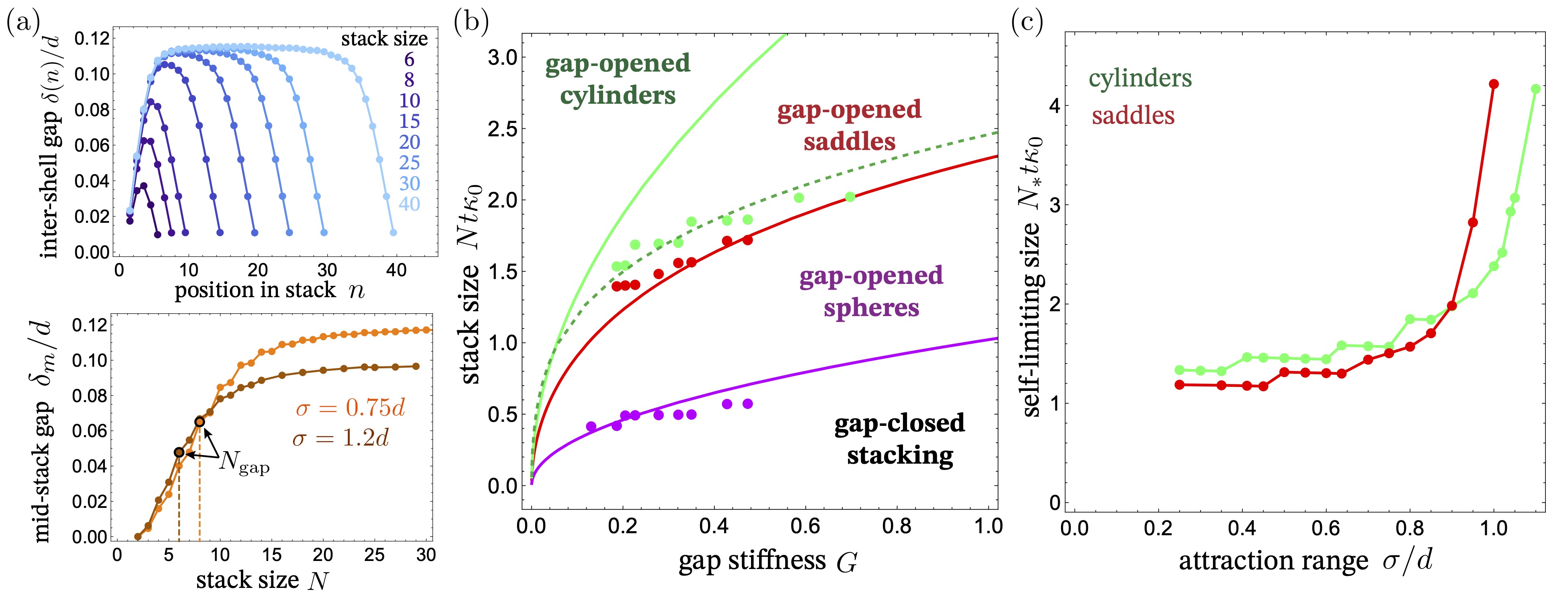}
    \caption{(a) For stacking assemblies with finite-ranged attractions, inter-shell spacing is maximal in the middle of the stack (top) which increases with stack size and saturates as shells become uniformly curved and spaced (bottom).  The stack size $N_{\rm gap}$ marks the transition from curvature-focused, gap-closed stacking to uniformly curved, gap-opened stacking.  Data shown for spherical cap stacks with attraction range $\sigma = 0.25d$ (top).  (b) The transition to frustration escaping, gap-opened stacking for thin shells ($\eta = 10.1$) occurs at larger stack sizes as gap stiffness is increased but depends on preferred Gaussian curvature.  Spherical caps become gap-opened first at the smallest size, followed by saddles and then cylindrical shells at larger stack sizes.  Simulation data are shown as dots while predictions from the continuum scaling argument are shown as solid lines.  The dashed line is the detailed analytic prediction for cylindrical shells from ref. \cite{sullivan-curvamers-2024} adjusted for the new definition of gap stiffness $G$ which features a transition from $N_{\rm gap} \sim G^{1/2}$ to $N_{\rm gap} \sim G^{1/3}$ at higher values of gap stiffness. (c)  Increasing the range of attraction lowers the gap stiffness $G$ and causes self-limiting size to diverge.  Stacks of thin saddles become unlimited at slightly smaller ranges of attraction than cylinders. All data shown for $\eta = 10.1$ with spring constant $k_h = 1.6\, \gamma A / d^2$.}
    \label{fig:results_gaps}
\end{figure*}

Prior studies of cylindrical shell curvamers with finite-attraction ranges revealed a key mechanism of ``escaping frustration" facilitated by compliance of inter-particle binding\cite{tanjeem-curvamers-2022,sullivan-curvamers-2024}.  That is, for sufficiently long-range and compliant inter-surface binding, curvamers can effectively bind without some or all of the shape deformation required by conformal, curvature-focused stacking.  This behavior is marked by a transition from curvature-focused, gap-closed stacking to a uniformly-shaped, gap-opened stacking motif at a threshold stack size $N_{\rm gap}$.  For $N \gtrsim N_{\rm gap}$, curvamers tend toward uniformly stacked shapes, effective evading the superextensive growth of elastic energy and marking a transition to unlimited stack size above a critical cohesive strength.  For cylindrical shells, the gap open threshold $N_{\rm gap}$ was shown to depend on a dimensionless ratio of inter-subunit interaction stiffness to intra-subunit bending stiffness, with compliant interactions and stiffer shells causing the transition to occur at smaller sizes and shrinking the effective range self-limitation~\cite{sullivan-curvamers-2024}.  

Here we consider this gap-opening behavior, and its effect to limit the range of accessible self-limiting assembly for non-Euclidean curvamer geometries, notably to understand the effects of their enhanced elastic penalties for curvature-focused stacking at large $\eta$.   
To explore the nature of the gap-opening transition in non-Euclidean shell stacks, we consider a simple scaling estimate based on the continuum theory and compare to  coarse-grained calculations using a variety of longer ranged attractions.  For the coarse-grained model, we will restrict our analysis to relatively thin shells with $\eta = 10.1$ for one spring stiffness ($k_h = 1.6\, \gamma A / d^2$) while varying the microscopic attraction range $\sigma$ from $0.25d$ to $1.2d$.  

As modeled previously for cylindrical shells~\cite{sullivan-curvamers-2024}, in the limit of infinitely large stacks, we expect shells to become uniformly shaped and open up gaps between their surfaces that relieve elastic costs required by curvature focusing.  In this ``gap-opened'' stacking motif, the per subunit stack energy is a compromise between the elastic costs of slightly misshapen shells and the adhesive cost of partial inter-shell gaps (i.e. whose magnitude varies over the surface of inter-shell contact).  We construct an upper bound to this energy density by assuming all shells in an infinitely large stack maintain their preferred shapes, and estimate the adhesive cost of non-uniform adhesion between curvamer surfaces.  Approximating the adhesive interaction strength per area between shell surfaces to be harmonic, $u(\delta) = -\gamma + \gamma'' \delta^2/2$, where $\delta$ is the surface-surface separation and $\gamma'' = \partial_{\delta}^2 u\big|_{\delta = 0}$ is the stiffness of interactions (i.e. convexity of attractive well), we find the adhesive energy cost associated with gap-opening to be (see Appendix \ref{section:appendix-gaps})
\begin{equation}
    \frac{U_{\rm gap}(\vec{\kappa}_{0})}{A} \simeq \frac{\gamma''t^2w^4\kappa_0^4}{3072}\left(1+\frac{1}{2}\cos4\alpha \right)\text{.}
\end{equation}
From this, we define a dimensionless measure of inter-shell stiffness to intra-shell (bending) stiffness 
\begin{equation}
    G =  \frac{\gamma''t^2w^4\kappa_0^2}{1536B}\text{,}
\end{equation}
which we refer to as the ``gap stiffness'' for simplicity, and can be related to an effective range of attraction through $\sigma_{\rm eff} = \sqrt{\gamma/\gamma''} \sim \sqrt{S/G}$.  

The transition size $N_{\rm gap}$ from gap-closed to gap-opened stacking in the continuum model is estimated by calculating when the small sizes elastic energy accumulation (see Sec. \ref{section:sla}) crosses over the gap-opened energy $U_{\rm gap}$.  For thick shells ($\eta \ll1$), bending dominates and all shapes transition to gap-opened stacking as $N_{\rm gap}\sim (\kappa_0 t)^{-1}G^{1/2}$.  This matches the small-$G$ scaling results of ref. \cite{sullivan-curvamers-2024}, which also performed a more thorough analysis of gap-opening behavior in cylindrical shell stacks and showed that $N_{\rm gap}\sim (\kappa_0 t)^{-1}G^{1/3}$ when interactions are stiff and uniformly curved shells deviate away from their preferred curvature.  For thin shells ($\eta \gg1$), cylindrical and spherical cap stacks both maintain a quadratic dependence on stack size for their elastic energy but the addition of expensive stretching penalties introduces a factor of $\eta$ so that spherical cap stacks become gap-opened as $N_{\rm gap}\sim (\kappa_0 t)^{-1} \eta^{-1/2}G^{1/2}$.  Finally, stacks of  stretching dominant, thin saddles become gap-opened as $N_{\rm gap}\sim (\kappa_0 t)^{-1}\eta^{-1/4}G^{1/4}$.

For the coarse-grained model, the distance between adjacent shell surfaces in a stack is measured at the center of the stack, where it is generically maximal.  As shown in Fig. \ref{fig:results_gaps}a, this mid-stack gap size, $\delta_m$, grows with stack size and eventually saturates as the stack becomes infinitely tall.  We then define $N_{\rm gap}$ to be the stack size for which the mid-stack gap is half of the gap size associated with an infinitely tall stack $\delta_m(N_{\rm gap}) =\delta_\infty/2$.  In practice, we take $\delta_\infty$ to be the mid-stack gap associated with the largest stack size that was successfully minimized and must be calculated for each unique combination of spring constant $k_h$, interaction range $\sigma$, shape $\alpha$ and shell geometry $\eta$.  Mapping the  attractive range $\sigma$ to interaction stiffness $\gamma''$ is done by fitting a parabola to the minimum of the flat plate-plate interaction well (see Appendix \ref{section:appendix-simdetails}).  

As shown in Fig. \ref{fig:results_gaps}b, the gap opening transition size for the continuum scaling argument and coarse-grained model generally agree well and increase with gap-stiffness $G$.  For the thin shells analyzed ($\eta = 10.1$), spherical cap stacks become gap-opened at significantly smaller sizes with $N_{\rm gap}^{\rm cylinder}>N_{\rm gap}^{\rm saddle}>N_{\rm gap}^{\rm sphere}$.  Furthermore, as can be seen in Fig. \ref{fig:results_gaps}c, increasing the range of attraction (thus making interactions more compliant and decreasing $G$) causes the self-limiting size for both saddle and cylindrical shells to increase and eventually diverge.  The self-limiting size of the saddle stacks, however, begins to diverge at a smaller range of attraction than the cylindrical stacks, owing to the fact that $N_{\rm gap}^{\rm cylinder}>N_{\rm gap}^{\rm saddle}$ for equivalent $G$.  This gives evidence that the non-Euclidean curvamer geometries are more prone to gap opening, relative to Euclidean ones, due to prohibitively larger costs of thin-shell stretching relative to the smaller costs of thin shell bending.

Finally, we note the absence of the spherical cap stacks from the self-limiting size analysis in Fig. \ref{fig:results_gaps}c.  For these parameters, stacks of spherical caps were unstable to break into sub-stacks during energy minimization, especially for shorter ranges of attraction.  For the largest ranges of attraction sampled which yielded coherent, unbroken spherical cap stacks (e.g. those seen in Fig. \ref{fig:results_gaps}b), no finite-sized minimum was detected in the range of stack sizes calculated (up to 60 shells), indicating that self-limitation had likely diverged at a much smaller attraction range than either saddles or cylinders.  

\section{Discussion and conclusions} 
\label{section:discussion}

In this paper, we have demonstrated that Gaussian curvature plays a key role in shaping the frustrated self-assembly of stacking colloidal shell particles.  Notably, the self-limitation deriving from the build up of cumulative elastic costs of frustration in non-Euclidean shell particle stacks depends critically on shell geometry via the FvK number $\eta$, a quantity that has also been shown to control the helicoid and spiral morphologies of twisted ribbons\cite{ghafouri-ribbons-2005}.  As demonstrated with our continuum mechanical and discrete, coarse-grained models, increasing $\eta$ strengthens the penalty for stretching deformations resulting in stacks accumulating higher elastic stresses which favor smaller self-limiting sizes.    The preferred Gaussian curvature of curvamer shells also affects the rate of elastic energy accumulation upon stacking assembly, most notably for thin shells (large $\eta$) where positively curved-caps exhibit a relative rapid growth of elastic cost $U_{\rm el}/N \sim (N/N_{\kappa_0})^2$ while stacking saddle shells leads to a soft accumulation of frustration cost $U_{\rm el}/N\sim (N/N_{\kappa_0})^4$, for small sizes ($N<N_{\kappa_0}$).  The effect of higher elastic costs is to suppress self-limiting size so that in general the ranking of self-limiting stack sizes is $N_{\rm cylinder} > N_{\rm saddle} > N_{\rm sphere}$. In contrast, cylindrical shell stacks do not experience stretching and elastic energy growth and self-limiting size is unaffected by $\eta$.  A direct consequence of the anharmonic growth of elastic costs for saddle shell stacks is that the window for self-limitation is pushed to larger values of $S$ as stack size changes very slowly with adhesive strength $N \sim (S/\eta)^{1/5}$ and fluctuations around the self-limiting size are suppressed relative to those exhibited by cylindrical and cap shaped particles.  In addition, we have shown that the gap-opened frustration escape mode seen with cylindrical shells also affects non-Euclidean shells so that the effective upper bound of self-limiting behavior depends on the gap stiffness $G$ scaling as $N_{\rm gap}^{\rm saddle}\sim \eta^{-1/4}G^{1/4}$ for thin saddle shells and $N_{\rm gap}^{\rm sphere}\sim \eta^{-1/2}G^{1/2}$ for thin spherical cap shells.  Thus, stiffer adhesive interactions with shorter ranges of attraction promote self-limiting assembly behavior up to larger stack sizes.    

Taken together, these results suggest that preferred Gaussian curvature of colloidal shells may be a useful design feature for engineering the nature and impacts of frustration accumulation in self-assembling systems.  Specifically, minimal saddle shapes (i.e. $H_0 = 0$ and $K_{G0} < 0$) offer two distinct advantages over the previously studied cylindrical shapes.  Foremost, the selected self-limiting stack size $N_*$ is shown to be more stable over a larger range of cohesion for saddles in comparison to cylindrical shells, suggesting that optimal stacks may be less sensitive to perturbations in environmental conditions that alter inter-particle binding.  Beyond this, the nature of stacking for minimal saddle shapes is {\it apolar}, leading to stacks that are symmetric top to bottom, while for any shell shape with $H_0 \neq 0$, like cylinders and caps, stacks become {\it polar} in structure, with shells adopting relatively flatter or more curved shapes on one side of the stack or the other.

To assess the potential implications for the sensitivity to Gaussian curvature in  stacking self-assembly, we consider several possible experimental realizations of colloidal curvamers: photo-lithographically fabricated polymeric microshells \cite{tanjeem-shapechanging-2022,kuenstler-curvedhydrogels-2020,jeon-trilayergels-2020,na-greyscaleshapes-2016}, ligand-funtionalized nanoplatelets \cite{Jana_NanoPlatlets_2017, guillemeney_curvature_2022, abecassis-nanoplatelet_curvature-2024,mourchid-thermoresponsive_nanoplatelets-2025,cadmium-nanoplatelets-2008}, and curved shell particles derived from DNA origami \cite{dietz-foldingdna-2009,Han-dnacurvature-2011}.  We estimate the range of FvK number $\eta$ available to each of these systems by looking at typical values of width and thickness and take the preferred radius of curvature to be on the same scale as the width $r_0 \sim w$ to satisfy the shallow shell condition, in which case the FvK number is $\eta \approx 0.003 (w/t)^2$.  Polymeric shells typically have their large dimension in the range $w\approx 5-15\mu\text{m}$ and thicknesses on the order $t\approx 0.1-0.5\mu\text{m}$, giving a range of $\eta \approx 0.3-70$.  Nanoplatelets typically have $w\approx 5-100\text{nm}$ and $t\approx 1-2\text{nm}$ and thus have a range of $\eta \approx 0.02-30$.  Finally, DNA origami subunits have lateral extents $w\approx 50-200\text{nm}$ and thicknesses $t\approx 5-15\text{nm}$ meaning the approximate range of FvK number is $\eta \approx 0.03-5$. Taken together these estimates suggest that in each of these systems, modestly large values of $\eta$ (of order $\sim 10^1$) will be accessible, such that the presence of stretching costs for non-Euclidean shell geometries is sufficient to alter the stacking thermodynamics relative to the Euclidean case, but that costs are not necessarily so large as to suppress finite-size stacking assembly altogether.

Beyond the direct experimental realizations, a number of open questions regarding frustrated assembly of curvamers still remain.  First, the analysis on self-limiting size here (as well as in nearly all other extant GFA models) was performed with mechanically equilibrated stacks at zero temperature, and thus do not account for entropic contributions to the free energy at finite-temperatures that may alter the optimal stacking size.  While translational entropy causes the distribution of stack sizes in thermal equilibrium to depend weakly on temperature\cite{hagan_sla_2021}, other forms of entropy such as rotations and vibrational modes may create an effective repulsion over long ranges within a stack that weaken binding free energies and result in the per shell free energy minimum to be shifted to smaller sizes.  Another feature which is likely to impact the feasible range of self-limiting assembly is the ability of weak attractions between otherwise finite stacks of curvamers to promote ``super-stacks" of weakly-bound ``sub-stacks", which are ultimately unstable to unlimited growth at low enough temperatures~\cite{wang-polybricks-2024}.  How the specific shape (i.e. Euclidean versus non-Euclidean) influences the ability of imperfectly fitting curvamers to generate sufficient binding to promote this ``super-stack'' condensation remains to be studied.  

Additionally, studies on tubule forming assemblies assembly\cite{Cheng_SoftMatter_2012,videbaekk-economical-2024} have shown that subunit geometry and interactions can lead to a wide variety of off-target structures.  Since our analysis only applies to systems in thermal equilibrium, questions regarding how assembly dynamics evolve starting from an out-of-equilibrium state, potential kinetic traps that stall the assembly process, and the structures that ultimately form remain unanswered.   Furthermore, as short-ranged attractions are necessary to avoid gap-opened frustration escape, there will likely be an elastic energy barrier associated with shell binding as highly curved shells with large aspect ratios will have to deform before their attractive surfaces can ``feel'' each other and overcome their shape misfit gap.  It remains to be seen what defines the regime of feasible self-assembly where it is possible to both have access to a wide range of self-limiting sizes and reasonable assembly timescales.  

Finally, as in nearly all prior models of GFA, here we have only considered stacks of identical curvamer shells.  To date, questions about the role of dispersity in the shape frustration between subunits, e.g. different shell shapes and curvatures, remain unstudied.  It remains to be understood if systems of mixed shape frustration permit the propagation of self-limiting frustration costs, what new morphological structures emerge, and if the stoichiometry of subunit shape can be leveraged as an addition control parameter to tune self-limiting assembly size.  These questions point to the need of further studies beyond ground state thermodynamics of single-species GFA, including finite-temperature dynamical simulations and models of mixed-frustration.

\section*{Author contributions}
The research described in this manuscript was designed and carried out by KTS, MJS and GMG. The manuscript was written by KTS, MJS and GMG.

\section*{Data availability}

Data for this manuscript include numerical energy minimization codes for discrete coarse-grained shell assemblies, as well as example parameter inputs for generating the data analyzed in the plots. Additionally, these include Mathematica notebooks and numerical data used to generate plots and schematics in the manuscript. These data are available at \url{https://hdl.handle.net/20.500.14394/59645}, Scholarworks repository, hosted by UMass Amherst.

\section*{Conflicts of interest}
There are no conflicts to declare.

\section*{Appendices}
\appendix

\section{Simulation details of the coarse-grained model}
\label{section:appendix-simdetails}

The geometry of the coarse-grained shallow shell model is restricted to shells that form approximately uniformly curved paraboloids.  Thus, the fictitious mid-surface layer of the shell is described by the Monge gauge, 
\begin{equation}
    \vec{R}(\vec{x}_M) = \left( x_M^i,\,y_M^i,\,-\frac{1}{2}\kappa_{0x}{x_M^i}^2 -\frac{1}{2}\kappa_{0y}{y_M^i}^2-\kappa_{0xy}{x_M^i}{y_M^i}\right)
\end{equation}
where $x_M^i$ and $y_M^i$ are the coordinates of a flat, triangular mesh with lattice constant $a$, and takes the shape of a square with side length $w$.  The positions of the beads that make up the coarse-grained shell are then found by projecting the mid-surface lattice points along the normal up or down by half the structural thickness $t_0/2$
\begin{align}
    \vec{R}^{\pm}(\vec{x}_M) &= \vec{R}(\vec{x}_M) \pm \frac{t_0}{2}\hat{N}_{\rm surf}(\vec{x}_M)
\end{align}
where the surface normal is 
\begin{equation}
    \hat{N}_{\rm surf}(\vec{x}_M) = \frac{1}{N_{\rm surf}}\left( \kappa_{0x}x_M^i + \kappa_{0xy}y_M^i,\, \kappa_{0y}y_M^i + \kappa_{0xy}x_M^i,\,1\right)
\end{equation}
with normalization
\begin{equation}
    N_{\rm surf} = \sqrt{1 + (\kappa_{0x}x_M^i + \kappa_{0xy}y_M^i)^2 + (\kappa_{0y}y_M^i + \kappa_{0xy}x_M^i)^2}\text{ .}
\end{equation}  

Harmonic springs, or ``bonds'', then connect neighboring beads to give the shell its preferred shape and structural rigidity.  The rest length of each bond is calculated by measuring the Euclidean distance from one bead to its neighbor when in the preferred shape of the shell with curvatures $\kappa_{0x}=\kappa_{01}$, $\kappa_{0y}=\kappa_{02}$, and $\kappa_{0xy} = 0$.  Bonds that connect adjacent beads in the same layer (top or bottom) are labeled ``horizontal'' and have spring constant $k_h$.  Bonds that connect beads in separate layers but which correspond to the same original mid-surface mesh point $\vec{x}_M$ are ``vertical'' with spring constant $k_v$.  Meanwhile, bonds that connect beads on separate layers but correspond to different beads on the mid-surface mesh are ``cross'' bonds with spring constant $k_c$.  Relations between the three spring constants are set by examining stretching deformations to the triangular prism unit cell of a flat, coarse-grained shell.  Applying stretching strains $\epsilon_{xx}$, $\epsilon_{yy}$, $\epsilon_{zz}$, the energy of the (periodic) unit cell is calculated and minimized to obtain the Poisson's ratios $\nu_{xy}$ and $\nu_z$ corresponding to in-plane stretching and out-of-plane stretching.  This set of Poisson's ratio equations yields the following spring constants,
\begin{equation}
    \frac{k_c}{k_h} = \frac{(1-3\nu_{xy})(1+\alpha_r^2)}{4\nu_{z}\alpha_r^2+3\nu_{xy}-1}\text{,}
\end{equation}
\begin{equation}
    \frac{k_v}{k_h} = \frac{3(1-3\nu_{xy})(1-\nu_{xy}-2\nu_{z}\alpha_r^2)}{2\nu_{z}(4\nu_{z}\alpha_r^2+3\nu_{xy}-1)}\text{,}
\end{equation}
where $\alpha_r = t_0/a$ is the aspect ratio of the unit cell.

\begin{table}
\caption{Coarse-grained shell designs}
\begin{tabularx}{\columnwidth}{  >{\centering\arraybackslash}X  >{\centering\arraybackslash}X
>{\centering\arraybackslash}X
>{\centering\arraybackslash}X
 }
\hline \hline 
  $t_0$ & $r_0$  & $\eta$ (fitted)  \\ 
 \hline
$4.434$ & $54.053$ &  $0.05$\\
$2.35$ & $32.893$  & $0.35$\\
$1.2$ & $21.221$  & $2.5$\\
$0.8$ & $21.221$  & $5.5$\\
$0.6$ & $21.221$  & $10.1$\\
$0.6$ & $15.015$  & $21.5$\\
 \hline \hline
\end{tabularx}
\label{table:shelldesigns}
\end{table}

\begin{figure}
    \centering
    \includegraphics[width=0.6\textwidth]{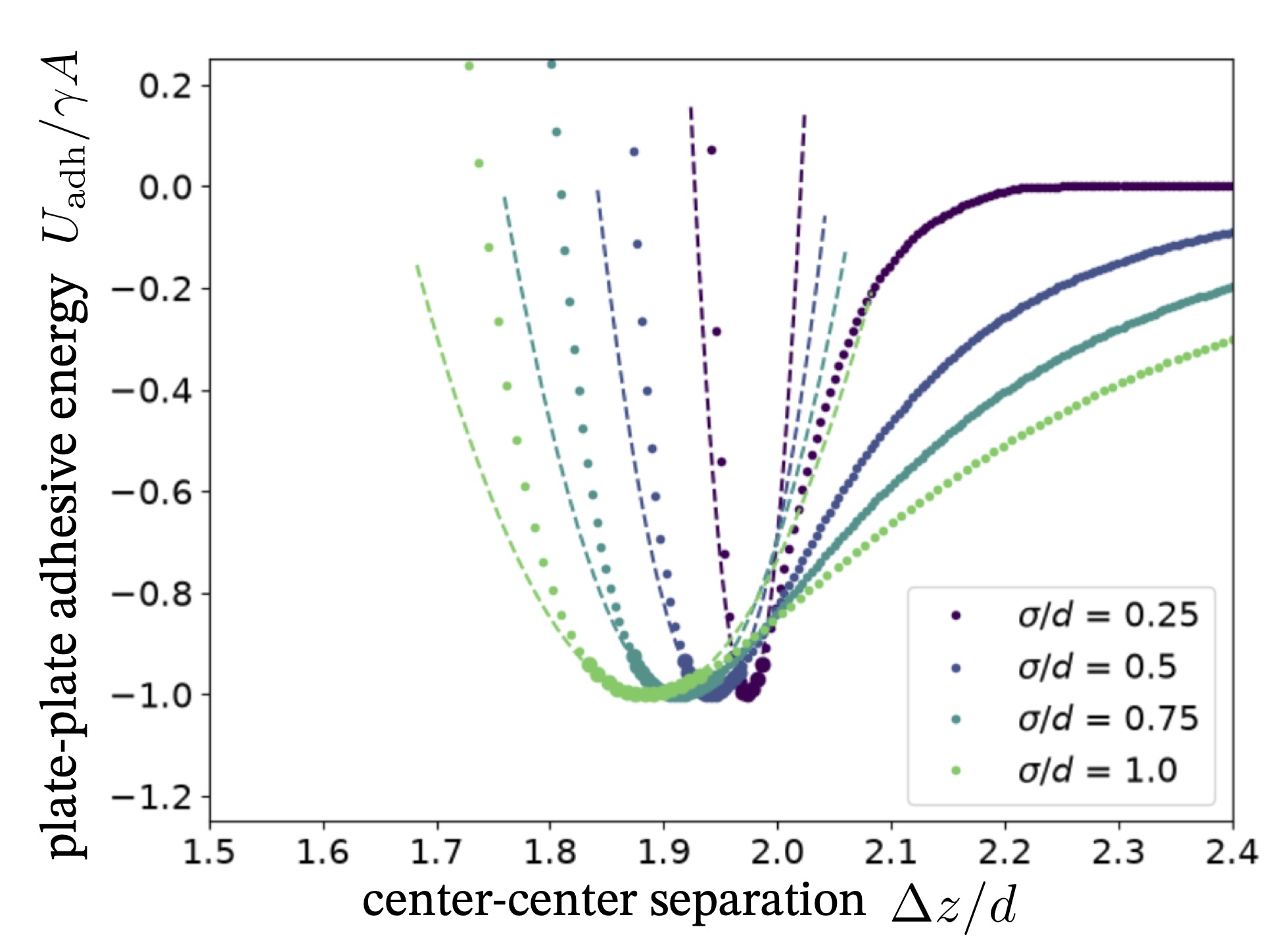}
    \caption{The adhesive interaction energy between two flat coarse-grained shells as a function of separation distance.  A quadratic fit (dashed line) was performed on the points within $0.05\sigma$ of the well minimum (fitted points denoted with large dots) to obtain the adhesive stiffness $\gamma''$.  Energies calculated for a shell with structural thickness $t_0 = d$.}
    \label{fig:appendix_plateadhesion}
\end{figure}

Mapping the coarse-grained parameters to those of the continuum model involves only a few simple calculations.  We assume the adhesive energy per unit area $\gamma$ is approximately independent of shell curvature and calculate the interaction energy between two flat shells at various separation distances (see Fig. \ref{fig:appendix_plateadhesion}).  Finding the minimum of this interaction energy gives both the adhesive strength $\gamma A$ and effective shell thickness $t$.  Performing a quadratic fit on the adhesive energies in the range $t\pm0.05\sigma$ allows us to approximate the interaction energy as harmonic and therefore calculate $\gamma''$.  To find the characteristic bending energy $U_B^0$ for a shell design with shape $\alpha$, we flatten a shell corresponding to the same design $t$, $w$, and $\kappa_0$ but with cylindrical shape ($\alpha = 0$).  Performing energy minimization over the in-plane $x$ and $y$ bead degrees of freedom then returns the flattened bending energy according to eqn. (\ref{eqn:1shellBend}).  The reduced adhesive energy, or stickiness, $S = t\kappa_0 \gamma A / U_B^0$ is then obtained.  In this paper, we set the bead-bead interaction strength, $u_0$, of the Lennard-Jones potential of eqn. (\ref{eqn:ulj}) such that the minimum adhesive energy between two flat plates $\gamma A$ is constant, so $S$ is varied by adjusting the coarse-grained shell spring constants (via $k_h$).

\begin{figure}
    \centering
    \includegraphics[width=0.6\textwidth]{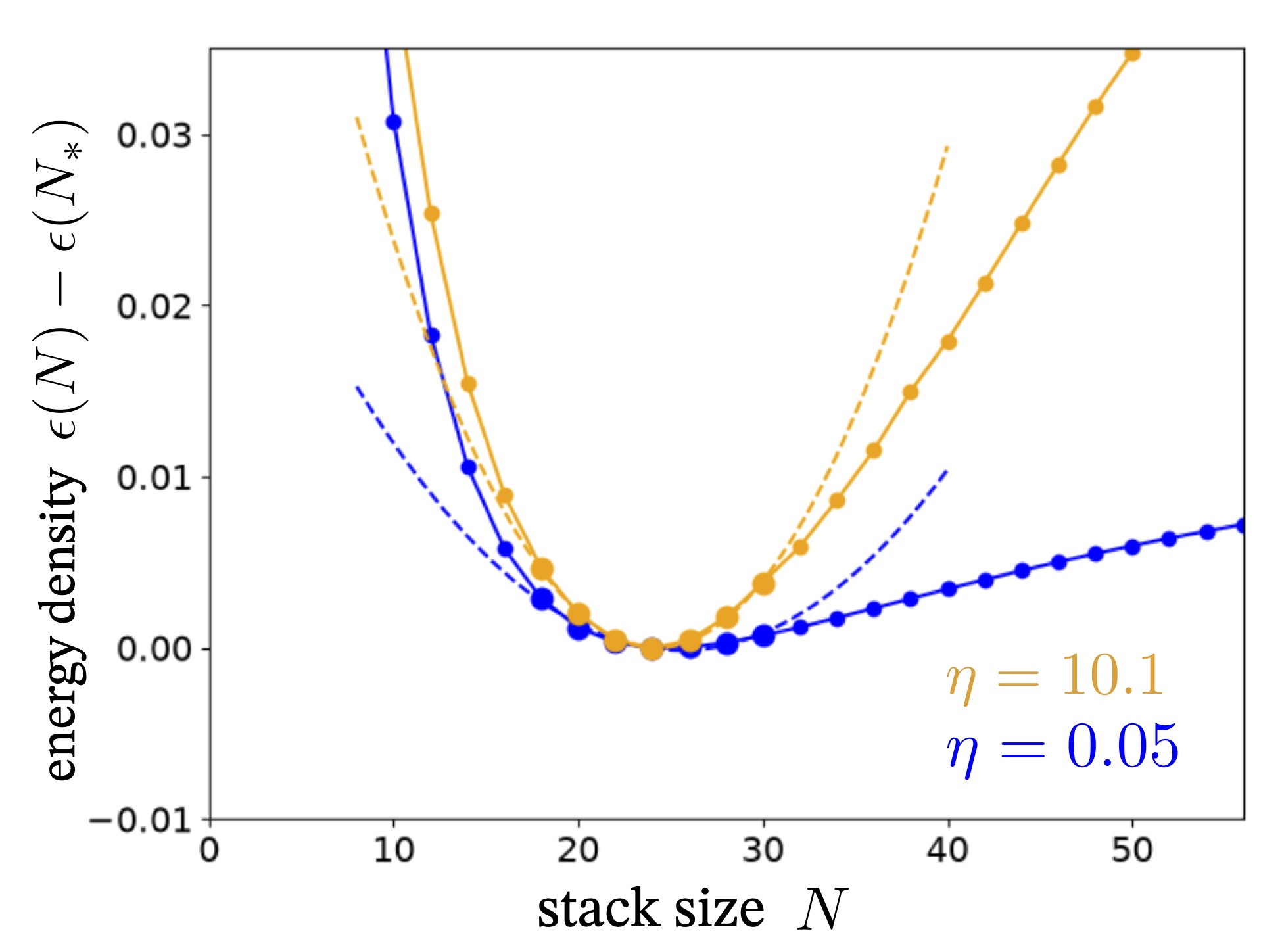}
    \caption{The self-limiting size $N_*$ is calculated as the minimum of the stack energy per subunit $\epsilon(N) = U^{\rm stack }(N)/N$.  The energy density convexity $\epsilon''_*$ is calculated by performing a quadratic fit (dashed lines) around $N_* \pm 6$ (bold points).  Thin saddle stacks (orange) have higher convexity stemming from per subunit stretching costs accumulating as $\sim N^4$ resulting in lower size fluctuations compared to stacks of thick saddles (blue) for the same self-limiting size ($N_*=24$).  Thin saddle energies shown were calculated for $k_h = 0.4 \gamma A/d^2$ while thick saddles shown correspond to $k_h = 0.18 \gamma A / d^2$.}    
    \label{fig:appendix_nflux}
\end{figure}

We use the following procedure to determine the self-limiting behavior for one shell design and shape.   To find the equilibrium configuration and associated energy of a stack, we initialize $N$ coarse-grained shells in a  vertical stack, uniformly spaced a distance $t$ apart from their mid-surfaces, with the principal curvatures of each shell changing through the stack so that they are curvature-focused and maintain gap-free surface-surface contact.  Zero temperature, energy minimization is then performed on the bead positions using a conjugate gradient descent algorithm with \texttt{LAMMPS}\cite{LAMMPS} and the final energy, associated with a local minimum in the energetic landscape, is recorded.  As shown in Fig. \ref{fig:appendix_nflux}, this is done for stacks up to 60 shells tall and the stack size among these with the lowest equilibrium energy per subunit is the self-limiting size for this value of $S$ (one choice of $k_h$).  We begin with a small value of $k_h$ so that shells are flexible and maintain a stacked configuration without breaking up into smaller sub-stacks, a phenomenon also reported in ref. \cite{tanjeem-curvamers-2022} for stiff shells.  Once these stacks are equilibrated, we increase the value of $k_h$ a small amount corresponding to a new $S$, and perform energy minimization on the previously minimized configuration.  Repeating this process gives the self-limiting size curve for this shell design ($\eta$) and shape ($\alpha$). We perform a quadratic fit (dashed lines in Fig. \ref{fig:appendix_nflux}) of the energy density around $N_* \pm 6$ yielding convexity $\epsilon''_*$ and allowing for the calculation of relative size fluctuations expected at finite temperatures according to eqn. (\ref{eqn:nflux}).

\section{Parameterization of shallow shell shape}
\label{section:appendix-shape}
The mid-surface of a uniformly thick, shallow shell can be parameterized according to
\begin{equation}
    \vec{R}(\vec{x}) = \big(x+u_x(\vec{x})\big)\hat{x} + \big(y+u_y(\vec{x})\big)\hat{y} + h(\vec{x})\hat{z}\text{,}
\end{equation}
where $\vec{u}(\vec{x})$ and $h(\vec{x})$ are the in-plane and out-of-plane displacements of a point $\vec{x}$ on a flat plane to the mid-surface $\vec{R}(\vec{x})$.  The shape of the shell is then prescribed by the out-of-plane displacement function $h(\vec{x})$.  In this work, we will make the additional simplification that curvature does not vary spatially on the surface restricting shell shapes to approximately uniformly curved paraboloids/hyperboloids, 
\begin{equation}
\label{eqn:outofplanedisp}
    h(\vec{x}) = h_0-\frac{1}{2}\kappa_{x}x^2-\frac{1}{2}\kappa_y y^2 - \kappa_{xy}xy\text{,}
\end{equation}
with curvatures $\kappa_{ij} = -\partial_i\partial_jh(\vec{x})$.  

In the differential geometry of surfaces, it is often convenient to speak in terms the principal (maximal/minimal) curvatures $\kappa_1$ and $\kappa_2$ of the surface which are obtained by diagonalizing the curvature tensor $\kappa_{ij}$ and are related to the curvatures in eqn. (\ref{eqn:outofplanedisp}) by
\begin{align}
    &\kappa_{x} = \kappa_{1}\cos^2{\theta} + \kappa_{2}\sin^2{\theta}\text{,} \nonumber \\
    &\kappa_{y} = \kappa_{1}\sin^2{\theta} + \kappa_{2}\cos^2{\theta} \text{,} \\
    &\kappa_{xy} = (\kappa_{1}-\kappa_{2})\sin{\theta}\cos{\theta} \text{,}\nonumber 
\end{align}
where the principal axes form an orthonormal basis on the surface with the first principal axis rotated counter-clockwise away from the $x$-axis by an angle $\theta$.  In a shell's stress-free, preferred shape, these will be denoted $\kappa_{01}$, $\kappa_{02}$, taking $\theta_0 = 0$ so they align with the material frame $x$ and $y$ axes.  The curvature tensor, being related to the second fundamental form, can then be used to calculate several invariant quantities associated with the shell shape: the Gaussian curvature
\begin{equation}
    K_G = \kappa_1\kappa_2\text{,}
\end{equation}
from the determinant, and the mean curvature
\begin{equation}
    H = \frac{1}{2}(\kappa_1 + \kappa_2)\text{,}
\end{equation}
from the trace.  A third invariant quantity is the Frobenius norm $||\kappa_{ij} || = \sqrt{\text{Tr}(\kappa_{ij}^2})$, which applied to the preferred shell shape gives
\begin{equation}
    \kappa_0 = \sqrt{\kappa_{01}^2+\kappa_{02}^2} \ \text{,}
\end{equation}
a useful characteristic curvature proportional to the root mean square of the principal curvatures that we will use to rescale curvatures to dimensionless variables 
\begin{equation}
    \bar{\kappa}_i \equiv \kappa_i/\kappa_0\text{.}
\end{equation}
This Pythagorean relation also suggests a new way to parameterize preferred shell shape using a shape angle $\alpha$,
\begin{align}
    &\kappa_{01} = \kappa_0\cos\alpha\text{,}\\
    &\kappa_{02} = \kappa_0 \sin\alpha\text{,}
\end{align}
with $\alpha = -\pi/2\text{, } 0 \text{, } \pi/2$ corresponding to saddle, cylindrical and spherical shells, respectively.  Additionally, in its stress free state we take the in-plane displacements $\vec{u}_i$ to be zero so that the shallow shell is simply described with a Monge gauge parameterization according to the height function $h(\vec{x})$ in eqn. (\ref{eqn:xyheightfunction}).

\section{Continuum shallow shell elasticity}
\label{section:appendix-elasticity}
As mentioned in Sec. \ref{section:shellelasticity} the strain of a deformed shallow shell is 
\begin{equation}
    \bar\epsilon_{ij} = \epsilon_{ij} - \epsilon_{ij}^0\text{,}
\end{equation}
where $\epsilon_{ij}^0$, of a flat plate deformed to the preferred shape of the shell (the ``pre-strain'') subtracted from the strain, $\epsilon_{ij}$ in deforming a flat plate to the current configuration of the shell.

In general, strain is given by
\begin{equation}
    \epsilon_{ij} = \frac{1}{2}\left( \partial_iu_j + \partial_ju_i + \partial_iu_k\partial_ju_k + \partial_ih\partial_jh  \right) \text{,}
\end{equation}
where $u_i$ are the components of the in-plane displacements and $h$ is the out-of-plane displacement. As is often done, the in-plane displacements are assumed to be small so the second order term can be neglected while the quadratic out-of-plane deflection term is kept as there are no lower order terms in $h$.  

By writing the elastic free energy functional $F_{s} = \frac{1}{2}\int\left(2\mu \bar{\varepsilon}^2_{ij} + \lambda\bar{\varepsilon}^2_{kk}\right)\text{d}A$ with 2D Lam\'{e} coefficients $\mu$ and $\lambda$ in terms of the shell stress $\bar\epsilon_{ij}$ and taking variations with respect to $u_{i}$ we find that equilibrium is satisfied for $\partial_i \sigma_{ij} = 0$ which implies that the stress tensor can be written in terms of any Airy stress function $\chi(\vec{x})$,
\begin{equation}
    \sigma_{ij} = \epsilon_{ik}\epsilon_{jl}\partial_{k}\partial_{l}\chi(\vec{x})\text{.}
\end{equation}
Enforcing that $\chi$ correctly relates the strains to the displacement field yields the compatibility condition
\begin{equation}
    \frac{1}{Y}\nabla^4\chi(\vec{x}) = -\Delta K_G(\vec{x})\text{,}
\end{equation}
where $Y$ is the 2D Young's modulus and $\Delta K_G = K_G^0 - K_G$ is the Gaussian curvature difference of the preferred and deformed states.  As shown by K. Seffen\cite{seffen-bistableshells-2006}, by assuming shells are uniformly curved ($\Delta K_G(\vec{x})=\Delta K_G$ everywhere) and thus are paraboloids/hyperboloids, and using force free boundary conditions along the outer edges of the shell (no external loads), the Airy stress function reduces to a fourth order polynomial whose coefficients can be solved exactly,
\begin{equation}
    \chi(x,y) = Q\left[\frac{1}{4}(x^4 +y^4) + \frac{1}{2}x^2y^2-\frac{1}{8}w^2(x^2+y^2)\right]\text{,}
\end{equation}
where
\begin{equation}
    Q = -\frac{1}{16}Yw^4\Delta K_G\text{.}
\end{equation}
The resulting stress tensor has components
\begin{align}
    \sigma_{xx} &= Q\left(x^2 + 3y^2 - \frac{w^2}{4}\right)\text{,}\\ 
    \sigma_{yy} &= Q\left(3x^2 + y^2 - \frac{w^2}{4}\right)\text{,}\\
    \sigma_{xy} &= -2Qxy\text{.}
\end{align}
Writing the stretching free energy in terms of stresses,
\begin{equation}
    F_{s} = \frac{1}{2Y}\int\left[(1+\nu) \sigma^2_{ij} - \nu\sigma^2_{kk}\right]\text{d}A\text{,}
\end{equation}
and integrating over the entire circular footprint  then gives eqn. (\ref{eqn:1shellStretch}).  This result matches that of Seffen\cite{seffen-bistableshells-2006} for isotropic shallow shells with circular footprints of diameter $w$.  Likewise, we utilize the form of the elastic shell bending energy derived by Seffen which simplifies to eqn. (\ref{eqn:1shellBend}) in the isotropic, circular shell limit.

Normalizing by the characteristic bending energy of a flattened cylindrical shell $U_B^0 = BA\kappa_0^2/2$ and rescaling curvatures by $\kappa_0$, the single shell elastic energy components are 
\begin{equation}
\label{eqn:dimlessUS}
    \frac{U^{\rm shell}_{\rm S}(\vec{\bar{\kappa}},\vec{\bar{\kappa}}_0)}{U_B^0} = \eta\left(\bar{\kappa}_1\bar{\kappa}_2 - \bar{\kappa}_{01}\bar{\kappa}_{02}\right)^2
\end{equation}
for the dimensionless stretching energy, and
\begin{equation}
\label{eqn:dimlessUB}
    \frac{U^{\rm shell}_{\rm B}(\vec{\bar{\kappa}},\vec{\bar{\kappa}}_0)}{U_B^0} = \left(\bar{\kappa}_1 - \bar{\kappa}_{01}\right)^2 +  \left(\bar{\kappa}_2 - \bar{\kappa}_{02}\right)^2  + 2\nu\left(\bar{\kappa}_1 - \bar{\kappa}_{01}\right)\left(\bar{\kappa}_2 - \bar{\kappa}_{02}\right) 
\end{equation}
for the dimensionless bending energy.

\begin{figure}
    \centering
    \includegraphics[width=0.6\textwidth]{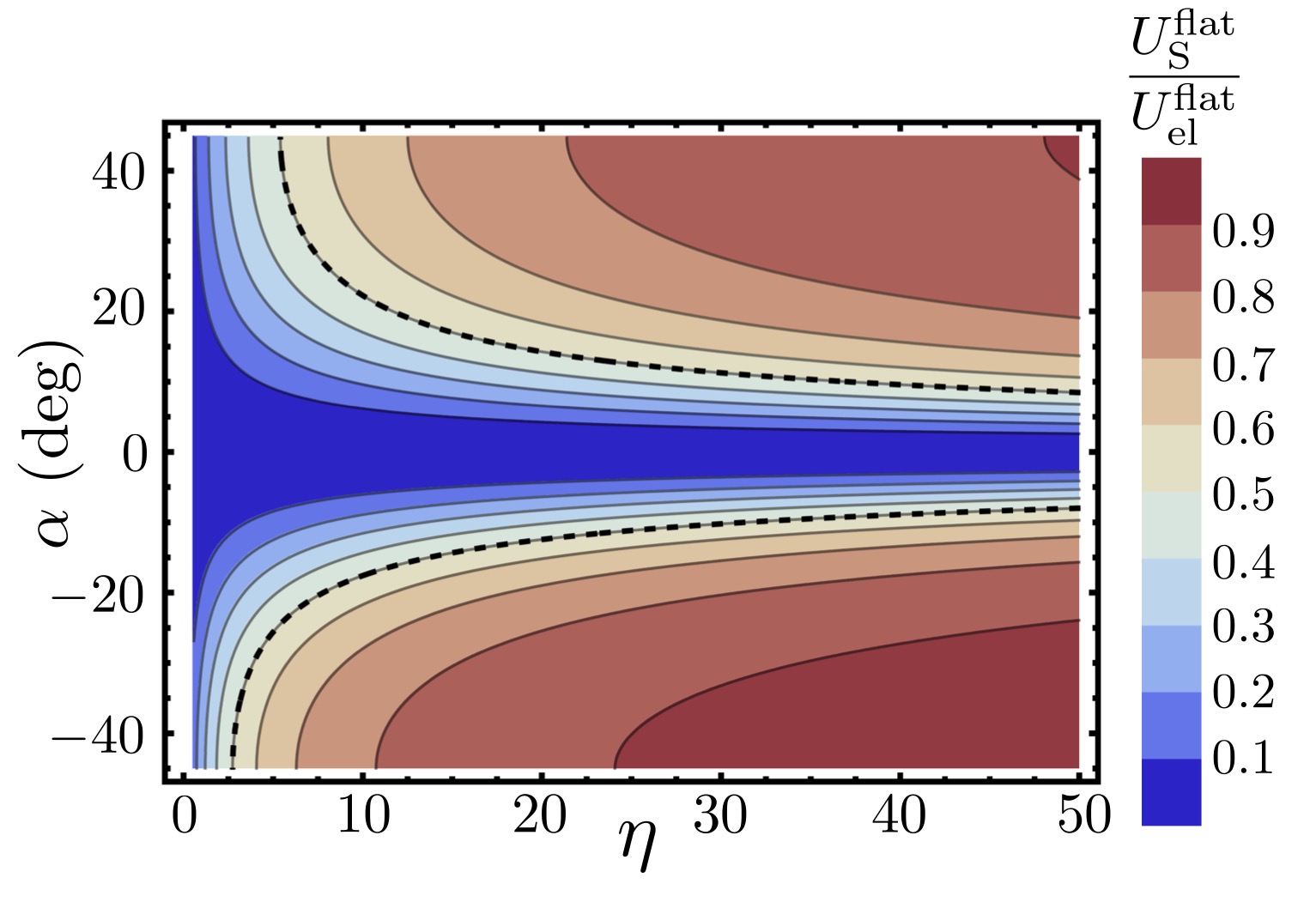}
    \caption{Phase diagram of flattened shell elastic energy for different preferred shapes ($\alpha$) and FvK numbers ($\eta$).  Blue regions denote shells whose flattened energy is dominated by bending, while red regions represent stretching dominance.  The black dashed line marks where stretching costs exactly equal bending costs.}
    \label{fig:appendix_flatshells}
\end{figure}

Fig. \ref{fig:appendix_flatshells} shows a phase diagram of single shell flattening energies for different shapes ($\alpha$) and thicknesses ($\eta$) with the black dashed lines marking the transition from regions of bending dominance (blue) to stretching dominance (red).  Notably, these transition lines asymptotically approach each other only as $\eta$ approaches infinity, leaving a large envelope of bending dominance surrounding $\alpha = 0$.  This suggests shells that are nearly cylindrical should have very similar energetics and stacking behaviors as stretching contributions become negligible within a tolerance of approximately $\alpha = 0\pm10\degree$ for shells with $\eta < 50$.  Additionally, the transition from bending to stretching dominance occurs at slightly smaller values of $\eta$ for negatively curved shells than for positively curved shells.  This can be attributed to the cross term of principal curvatures in eqn. (\ref{eqn:dimlessUB}) where negative preferred Gaussian curvature lowers the overall bending energy by an amount proportional to the shell's Poisson ratio.  Thus, while a spherical shell ($\alpha = 45\degree$) and saddle shell ($\alpha = -45\degree$) with identical $\eta$ both experience the same stretching energy when flattened, stretching makes up a larger percentage of the total elastic energy for the saddle.  As a result the transition from bending to stretching dominance occurs approximately at $\eta \approx 2.5$ for saddles and $\eta \approx 5$ for spherical caps.

\section{Energetics of continuum shell stacks}
\label{section:appendix-stacking}

Recalling the discussion of stacking geometry in Sec. \ref{section:geometry}, in order to satisfy the continuous perfect contact required for curvature-focused stacking, the radius of curvature between adjacent shells in a stack must differ only by the shell thickness.  Thus, the curvature profile in a stack $N$ shells tall is entirely determined by the shape of any one shell.  Here, we choose the middle shell in the stack $\kappa_{mi} = 1/r_{mi}$ to be our free variable so that the principal radii of curvature in a stack vary according to $r_i(n) = r_{mi} + nt$ for shells indexed by $n=-N/2,\dots,N/2$.  Normalizing by the RMS curvature $\kappa_0$, the scaled curvature profile becomes
\begin{equation}
    \bar{\kappa}_i(n,\bar{\kappa}_{mi}) =\frac{\bar{\kappa}_{mi}}{1+nt\kappa_0\bar{\kappa}_{mi}} \text{.}
\end{equation}
Defining the scaled stack size parameter $\bar{N} = Nt\kappa_0$,
which measures the total height of a stack relative to the preferred radius of curvature, we substitute the curvature profile into the single shell stretching and bending energies of eqns. (\ref{eqn:dimlessUS}) and (\ref{eqn:dimlessUB}).  Summing over all shells, we find
\begin{align}
    U_{\rm S,B}(N,\vec{\bar{\kappa}}_0,\vec{\bar{\kappa}}_{m})  &= \sum_{n=-N/2}^{N/2} U^{\rm shell}_{\rm S,B}(\bar{\kappa}_i(n,\bar{\kappa}_{mi}) ,\vec{\bar{\kappa}}_0)  \\ &\simeq \int_{-N/2}^{N/2} U^{\rm shell}_{\rm S,B}(\bar{\kappa}_i(n,\bar{\kappa}_{mi}) ,\vec{\bar{\kappa}}_0) \text{d}n\\
     &= \frac{1}{t\kappa_0}\int_{-\bar{N}/2}^{\bar{N}/2} U^{\rm shell}_{\rm S,B}(\bar{\kappa}_i(n,\bar{\kappa}_{mi}) ,\vec{\bar{\kappa}}_0) \text{d}\bar{n}
\end{align}
where the discrete sum is well-approximated by an integral over $n$ in the continuum limit $t\kappa_0\ll1$,  $N\gg1$.

The normalized stack energy per subunit is then 
\begin{align}
    \frac{U_{\rm stack}(N,\vec{\bar{\kappa}}_0,\vec{\bar{\kappa}}_{mi})}{U_B^0N} 
    &= -\frac{S}{t\kappa_0} + \frac{S}{\bar{N}}+ \frac{\epsilon_{\rm el}(\bar{N},\vec{\bar{\kappa}}_0,\vec{\bar{\kappa}}_{mi})}{U_B^0}\text{,}
\end{align}
where 
\begin{equation}
    S = \frac{\gamma A}{U_B^0}t\kappa_0
\end{equation}
is a reduced measure of the adhesive strength, or ``stickiness'' of the shells.  It can be understood as the ratio of the exposed boundary adhesive energy in a stack ($\gamma A$) to the energy to flatten every shell in the stack ($NBA\kappa_0^2/2$) for stack size $N = 1/t\kappa_0$.  The third term describes the per-subunit buildup of elastic frustration energy
\begin{equation}
    \frac{\epsilon_{\rm el}(\bar{N},\vec{\bar{\kappa}}_0,\vec{\bar{\kappa}}_{mi})}{U_B^0} = \frac{U_{\rm S}(\bar{N},\vec{\bar{\kappa}}_m,\vec{\bar{\kappa}}_0)}{U_B^0 N} + \frac{U_{\rm B}(\bar{N},\vec{\bar{\kappa}}_m,\vec{\bar{\kappa}}_0)}{U_B^0 N}\text{,}
\end{equation}
with scaled stack stretching energy per subunit
\begin{equation}
    \label{eqn:stackEstretching}
    \frac{U_{\rm S}(\bar{N},\vec{\bar{\kappa}}_m,\vec{\bar{\kappa}}_0)}{U_B^0 N} =\frac{\eta}{\bar{N}}\int_{-\bar{N}/2}^{\bar{N}/2}\big[ \bar{\kappa}_1(\bar{n},\bar{\kappa}_{m1}) \bar{\kappa}_2(\bar{n},\bar{\kappa}_{m2}) - \bar{\kappa}_{01}\bar{\kappa}_{02}\big]^2\text{d}\bar{n}
\end{equation}
and scaled stack bending energy per subunit
\begin{align}
    \frac{U_{\rm B}(\bar{N},\vec{\bar{\kappa}}_m,\vec{\bar{\kappa}}_0)}{U_B^0 N} &=\frac{1}{\bar{N}}\int_{-\bar{N}/2}^{\bar{N}/2}\bigg[ \big(\bar{\kappa}_1(\bar{n},\bar{\kappa}_{m1}) - \bar{\kappa}_{01}\big)^2 + \big(\bar{\kappa}_2(\bar{n},\bar{\kappa}_{m2}) - \bar{\kappa}_{02}\big)^2 \nonumber \\&+ 2\nu \big(\bar{\kappa}_1(\bar{n},\bar{\kappa}_{m1}) - \bar{\kappa}_{01}\big)\big(\bar{\kappa}_2(\bar{n},\bar{\kappa}_{m2}) - \bar{\kappa}_{02}\big)\bigg]\text{d}\bar{n}\text{.}
\end{align}

Stack configurations in mechanical equilibrium are obtained by solving for the middle shell principal curvatures $\bar{\kappa}_{mi}^*$ that minimize the total elastic energy 
\begin{equation}
    \label{eqn:stackeqnofstate}
    \frac{\partial}{\partial \bar{\kappa}_{mi}}\epsilon_{\rm el}(\bar{N},\bar{\kappa}_{m1},\bar{\kappa}_{m2},\vec{\bar{\kappa}}_0)\big|_{\bar{\kappa}_{mi} = \bar{\kappa}_{mi}^*} = 0\text{.}
\end{equation}  The equations of state from eqn. (\ref{eqn:stackeqnofstate}) amount to a system of coupled quadratic equations in the preferred principal curvatures 
\begin{align}
    \label{eqn:stateeqn1}&a\bar{\kappa}_{01}\bar{\kappa}_{02} + b \bar{\kappa}_{01}+c\bar{\kappa}_{02} + d = 0\text{,}\\
    \label{eqn:stateeqn2}&e\bar{\kappa}_{01}\bar{\kappa}_{02} + f \bar{\kappa}_{01}+g\bar{\kappa}_{02} + h = 0 \text{,}
\end{align}
that depend on stack size $\bar{N}$ and middle shell curvatures $\bar{\kappa}_{m1}$ and $\bar{\kappa}_{m2}$.  Note that $a$ and $d$ here are not to be confused with the lattice constant or bead diameter parameters of the coarse-grained model.
The coefficients $a,\ldots,h$ correspond to the following integral equations (taking reduced form $a=e=1$):
\begin{equation}
    b(\bar{N},\eta,\vec{\bar{\kappa}}_{m}) = \frac{1}{\eta \bar{\kappa}_{m2}}\frac{\int_{-\bar{N}/2}^{\bar{N}/2}(1+\bar{n}\bar{\kappa}_{m1})^{-2}\text{d}\bar{n}}{\int_{-\bar{N}/2}^{\bar{N}/2}(1+\bar{n}\bar{\kappa}_{m2})^{-1}(1+\bar{n}\bar{\kappa}_{m1})^{-2}\text{d}\bar{n}}\text{ ,}
\end{equation}
\begin{equation}
    c(\bar{N},\eta,\vec{\bar{\kappa}}_{m}) = \frac{\nu}{\eta \bar{\kappa}_{m2}}\frac{\int_{-\bar{N}/2}^{\bar{N}/2}(1+\bar{n}\bar{\kappa}_{m1})^{-2}\text{d}\bar{n}}{\int_{-\bar{N}/2}^{\bar{N}/2}(1+\bar{n}\bar{\kappa}_{m2})^{-1}(1+\bar{n}\bar{\kappa}_{m1})^{-2}\text{d}\bar{n}}\text{ ,}
\end{equation}
\begin{multline}
    d(\bar{N},\eta,\vec{\bar{\kappa}}_{m}) =-\frac{\bar{\kappa}_{m1}}{\eta \bar{\kappa}_{m2}}\frac{\int_{-\bar{N}/2}^{\bar{N}/2}(1+\bar{n}\bar{\kappa}_{m1})^{-3}\text{d}\bar{n}}{\int_{-\bar{N}/2}^{\bar{N}/2}(1+\bar{n}\bar{\kappa}_{m2})^{-1}(1+\bar{n}\bar{\kappa}_{m1})^{-2}\text{d}\bar{n}}\\- \bar{\kappa}_{m1}\bar{\kappa}_{m2}\frac{\int_{-\bar{N}/2}^{\bar{N}/2}(1+\bar{n}\bar{\kappa}_{m2})^{-2}(1+\bar{n}\bar{\kappa}_{m1})^{-3}\text{d}\bar{n}}{\int_{-\bar{N}/2}^{\bar{N}/2}(1+\bar{n}\bar{\kappa}_{m2})^{-1}(1+\bar{n}\bar{\kappa}_{m1})^{-2}\text{d}\bar{n}}  -\frac{\nu}{\eta}\text{ ,}
\end{multline}

\begin{equation}
    f(\bar{N},\eta,\vec{\bar{\kappa}}_{m}) = \frac{\nu}{\eta \bar{\kappa}_{m1}}\frac{\int_{-\bar{N}/2}^{\bar{N}/2}(1+\bar{n}\bar{\kappa}_{m2})^{-2}\text{d}\bar{n}}{\int_{-\bar{N}/2}^{\bar{N}/2}(1+\bar{n}\bar{\kappa}_{m1})^{-1}(1+\bar{n}\bar{\kappa}_{m2})^{-2}\text{d}\bar{n}}\text{ ,}
\end{equation}
\begin{equation}
    g(\bar{N},\eta,\vec{\bar{\kappa}}_{m}) = \frac{1}{\eta \bar{\kappa}_{m1}}\frac{\int_{-\bar{N}/2}^{\bar{N}/2}(1+\bar{n}\bar{\kappa}_{m2})^{-2}\text{d}\bar{n}}{\int_{-\bar{N}/2}^{\bar{N}/2}(1+\bar{n}\bar{\kappa}_{m1})^{-1}(1+\bar{n}\bar{\kappa}_{m2})^{-2}\text{d}\bar{n}}\text{ ,}
\end{equation}
\begin{multline}
    h(\bar{N},\eta,\vec{\bar{\kappa}}_{m}) =-\frac{\bar{\kappa}_{m2}}{\eta \bar{\kappa}_{m1}}\frac{\int_{-\bar{N}/2}^{\bar{N}/2}(1+\bar{n}\bar{\kappa}_{m2})^{-3}\text{d}\bar{n}}{\int_{-\bar{N}/2}^{\bar{N}/2}(1+\bar{n}\bar{\kappa}_{m1})^{-1}(1+\bar{n}\bar{\kappa}_{m2})^{-2}\text{d}\bar{n}}\\- \bar{\kappa}_{m1}\bar{\kappa}_{m2}\frac{\int_{-\bar{N}/2}^{\bar{N}/2}(1+\bar{n}\bar{\kappa}_{m1})^{-2}(1+\bar{n}\bar{\kappa}_{m2})^{-3}\text{d}\bar{n}}{\int_{-\bar{N}/2}^{\bar{N}/2}(1+\bar{n}\bar{\kappa}_{m1})^{-1}(1+\bar{n}\bar{\kappa}_{m2})^{-2}\text{d}\bar{n}}  -\frac{\nu}{\eta}\text{ .}
\end{multline}

Equations (\ref{eqn:stateeqn1})and (\ref{eqn:stateeqn2}) can be solved for the preferred curvatures which have solutions 
\begin{align}
    {\bar{\kappa}}^{\pm}_{01}(\bar{N},\bar{\kappa}_{m1},\bar{\kappa}_{m2}) &= \frac{1}{f-b}(d-h+bg-cf \pm \Delta^{1/2})
    \label{eqn:k01eq}\\
    \bar{\kappa}_{02}^{\pm}(\bar{N},\bar{\kappa}_{m1},\bar{\kappa}_{m2}) &= \frac{1}{c-g}(bg-cf \pm\Delta^{1/2})\label{eqn:k02eq}
\end{align}
with
\begin{align}
    \Delta=(h-d)^2&+b^2(c-g)^2+c^2(f-b)^2
    +2b(h-d)(c-g)\\&+2c(h-d)(f-b) - 2(c-g)(f-b)(bc-2d)\text{.}
\end{align}
In the case of symmetric spherical caps where $\bar{\kappa}_{01} = \bar{\kappa}_{02}$, these equations of state for mechanical equilibrium simplify to 
\begin{equation}
    \bar{\kappa}_{0i}^{\pm}(\bar{N},\bar{\kappa}_{m1},\bar{\kappa}_{m2})=-(b+c)\pm \sqrt{(b+c)^2-4d}\,\text{.}
\end{equation}

Recalling that our aim is to effectively integrate out the middle shell curvatures, we employ a numerical root finding scheme to invert eqns. (\ref{eqn:k01eq}) and (\ref{eqn:k02eq}) by solving for the roots $\bar{\kappa}_{m1}^*$, $\bar{\kappa}_{m2}^*$ that satisfy 
\begin{align}
    &\bar{\kappa}_{01}^{\pm}(\bar{N},\bar{\kappa}_{m1}^*,\bar{\kappa}_{m2}^*)=\cos{\alpha}\\
    &\bar{\kappa}_{02}^{\pm}(\bar{N},\bar{\kappa}_{m1}^*,\bar{\kappa}_{m2}^*)=\sin{\alpha}\text{.}
\end{align}
Beginning at small stack sizes where the middle shell curvatures should only deviate slightly away from the preferred values, we use Newton's method to solve for the $\bar{\kappa}_{mi}^*$ starting the search at $\bar{\kappa}_{0i}$.  Once found, the stack size is incremented a small amount and the previously found roots serve as the starting point for a new Newton's method search of the $\bar{\kappa}_{mi}$ that mechanically equilibrate the new stack size.  We start this algorithm at $\bar{N}=0.01$ and increment by $\Delta\bar{N}=0.01$ until reaching $\bar{N}=10$ after which we increment in steps of $\Delta\bar{N}=0.1$ up to $\bar{N}=100$.  This process is repeated for different preferred shapes from $\alpha = -45 \degree$ (saddles) to $45\degree$ (spherical caps) in steps of $0.5\degree$ and for $\eta = 0.025,\ 0.05,\ 0.25,\ 0.5,\ 2.5,\ 5,\ 10,\ 15,\ 20,\ 25,\ 30,\ 35,\ 40,\ 45, \
50$ with Poisson's ratio $\nu = 0.333$.  A third order interpolation is then performed over the data which can then be inverted to find $\bar{\kappa}_{m1}^*(\bar{N},\alpha,\pm)$ and $\bar{\kappa}_{m2}^*(\bar{N},\alpha,\pm)$ that correspond to the plus and minus branches of the equation of state.  Substituting these middle shell curvatures into the elastic energy density we find the ($+$) branch solutions have lower energy.  When elastic energies are presented in the main text, they will represent minimal, mechanically equilibrated solutions
\begin{equation}
    \epsilon_{\rm el}(\bar{N},\alpha,\eta) = \epsilon_{\rm el}(\bar{N},\alpha,\eta,\bar{\kappa}_{mi})\big|_{\bar{\kappa}_{mi} = \bar{\kappa}_{mi}^*(\bar{N},\alpha,+)}\text{ .}
\end{equation}

\section{Estimating the gap-opened stacking transition}
\label{section:appendix-gaps}
Let center-to-center separation between two identical shells be $\Delta z$.  Approximating top ($+$) and bottom ($-$) surfaces as paraboloids/hyperboloids and taking their principal curvatures to be $\kappa_i^{\pm}\approx \kappa_i \mp t\kappa^2_i/2$, the  surface-surface separation, $\delta$, between identically shaped shells is
\begin{equation}
    \delta(x,y) = \Delta z - t -\frac{1}{2}t\kappa_1^2 x^2 -\frac{1}{2}t\kappa_2^2 y^2\text{.}  
\end{equation}
We approximate the surface-surface adhesive energy per unit area to be harmonic, $u(\delta) = -\gamma + \gamma''\delta^2/2$, where $\gamma$ characterizes the strength of adhesive binding per area and $\gamma'' \equiv \partial^2_{\delta}u\big|_{\delta=0}$ is the ``stiffness'' of interactions or convexity of the attractive well.  The average squared surface separation distance between the shells is minimized for $\Delta z^* = t + tw^2(\kappa_1^2  +\kappa_2^2 )/32 $ yielding
\begin{equation}
    \langle \delta^2\rangle = \frac{t^2 w^4}{1024}\left(\kappa_1^4+\kappa_2^4-\frac{2}{3}\kappa_1^2\kappa_2^2 \right)\text{.} 
\end{equation}

 In the limit of infinitely large stacks, shells will become uniformly shaped and open up gaps between their surfaces that relieve elastic costs.  In this ``gap-opened'' stacking motif, the finite per subunit stack energy will be a compromise between the elastic costs of slightly misshapen shells and the adhesive cost of the inter-shell gaps.  We estimate an upper bound of this energy density by assuming all shells in an infinitely large stack maintain their preferred shapes.  In this limit, stacked shells in their preferred shapes have an average squared gap size of $\langle \delta^2\rangle\big|_{\vec{\kappa}_0} = \delta_{\infty}^2[1+\cos(4\alpha)/2]$ where the characteristic gap size (up to a shape dependent term) is
 \begin{equation}
     \delta_{\infty} = \frac{1}{16\sqrt{6}}tw^2\kappa_0^2\text{.}
 \end{equation}
 The total adhesive energy between two shells is then
\begin{align}
    \frac{U_{\rm adh}(\vec{\kappa}_0)}{A} &= -\gamma + \frac{1}{2}\gamma''\delta_{\infty}^2\left(1+\frac{1}{2}\cos4\alpha \right)\\
    &= -\gamma + \frac{\gamma''t^2w^4\kappa_0^4}{3072}\left(1+\frac{1}{2}\cos4\alpha \right)\text{,}
\end{align} 
where we identify the second term as the adhesive energetic cost due to gap-opening $U_{\rm gap}(\vec{\kappa}_0)$.  Normalizing by the characteristic bending energy of flattening, $U_B^0$, we find
\begin{equation}
    \frac{U_{\rm gap}(\vec{\kappa}_0)}{U_B^0} = G\left(1+\frac{1}{2}\cos4\alpha \right)\text{,}
\end{equation}
defining $G$ to be the ratio of inter-shell adhesive stiffness to intra-shell bending stiffness, a dimensionless measure of ``gap stiffness''
\begin{equation}
G\equiv\frac{\left(\gamma''\delta_{\infty}^2/2\right)}{U_B^0} = \frac{\gamma''t^2w^4\kappa_0^4}{1536B}\text{.}
\end{equation}

\section*{Acknowledgements}
The authors are grateful to R. Hayward, K. Fransen, T. Videbaek, L. Niu, R. Mathew, and C. Koertje for valuable discussions and input about this work. This study was supported by the US National Science Foundation through Award No. NSF-DMR 2349818 and the Brandeis Center for Bioinspired Soft Materials, an NSF MRSEC, DMR-2011846. Simulation studies and numerical calculations were performed on the UMass UNITY Cluster at the Massachusetts Green High Performance Computing Center.  This work was performed, in part, at the Center for Integrated Nanotechnologies, an Office of Science User Facility operated for the U.S. Department of Energy (DOE) Office of Science. Sandia National Laboratories is a multimission laboratory managed and operated by National Technology \& Engineering Solutions of Sandia, LLC, a wholly owned subsidiary of Honeywell International, Inc., for the U.S. DOE’s National Nuclear Security Administration under contract DE-NA-0003525. The views expressed in the article do not necessarily represent the views of the U.S. DOE or the United States Government.

\bibliographystyle{apsrev4-2}
\bibliography{main}

\begin{thebibliography}{85}%
\makeatletter
\providecommand \@ifxundefined [1]{%
 \@ifx{#1\undefined}
}%
\providecommand \@ifnum [1]{%
 \ifnum #1\expandafter \@firstoftwo
 \else \expandafter \@secondoftwo
 \fi
}%
\providecommand \@ifx [1]{%
 \ifx #1\expandafter \@firstoftwo
 \else \expandafter \@secondoftwo
 \fi
}%
\providecommand \natexlab [1]{#1}%
\providecommand \enquote  [1]{``#1''}%
\providecommand \bibnamefont  [1]{#1}%
\providecommand \bibfnamefont [1]{#1}%
\providecommand \citenamefont [1]{#1}%
\providecommand \href@noop [0]{\@secondoftwo}%
\providecommand \href [0]{\begingroup \@sanitize@url \@href}%
\providecommand \@href[1]{\@@startlink{#1}\@@href}%
\providecommand \@@href[1]{\endgroup#1\@@endlink}%
\providecommand \@sanitize@url [0]{\catcode `\\12\catcode `\$12\catcode
  `\&12\catcode `\#12\catcode `\^12\catcode `\_12\catcode `\%12\relax}%
\providecommand \@@startlink[1]{}%
\providecommand \@@endlink[0]{}%
\providecommand \url  [0]{\begingroup\@sanitize@url \@url }%
\providecommand \@url [1]{\endgroup\@href {#1}{\urlprefix }}%
\providecommand \urlprefix  [0]{URL }%
\providecommand \Eprint [0]{\href }%
\providecommand \doibase [0]{https://doi.org/}%
\providecommand \selectlanguage [0]{\@gobble}%
\providecommand \bibinfo  [0]{\@secondoftwo}%
\providecommand \bibfield  [0]{\@secondoftwo}%
\providecommand \translation [1]{[#1]}%
\providecommand \BibitemOpen [0]{}%
\providecommand \bibitemStop [0]{}%
\providecommand \bibitemNoStop [0]{.\EOS\space}%
\providecommand \EOS [0]{\spacefactor3000\relax}%
\providecommand \BibitemShut  [1]{\csname bibitem#1\endcsname}%
\let\auto@bib@innerbib\@empty
\bibitem [{\citenamefont {Li}\ \emph {et~al.}(2020)\citenamefont {Li},
  \citenamefont {Palis}, \citenamefont {M{\'e}rindol}, \citenamefont {Majimel},
  \citenamefont {Ravaine},\ and\ \citenamefont
  {Duguet}}]{li-patchycolloids-2020}%
  \BibitemOpen
  \bibfield  {author} {\bibinfo {author} {\bibfnamefont {W.}~\bibnamefont
  {Li}}, \bibinfo {author} {\bibfnamefont {H.}~\bibnamefont {Palis}}, \bibinfo
  {author} {\bibfnamefont {R.}~\bibnamefont {M{\'e}rindol}}, \bibinfo {author}
  {\bibfnamefont {J.}~\bibnamefont {Majimel}}, \bibinfo {author} {\bibfnamefont
  {S.}~\bibnamefont {Ravaine}},\ and\ \bibinfo {author} {\bibfnamefont
  {E.}~\bibnamefont {Duguet}},\ }\href {https://doi.org/10.1039/C9CS00804G}
  {\bibfield  {journal} {\bibinfo  {journal} {Chem. Soc. Rev.}\ }\textbf
  {\bibinfo {volume} {49}},\ \bibinfo {pages} {1955} (\bibinfo {year}
  {2020})}\BibitemShut {NoStop}%
\bibitem [{\citenamefont {Hamley}(2003)}]{hamley-nanotechreview-2003}%
  \BibitemOpen
  \bibfield  {author} {\bibinfo {author} {\bibfnamefont {I.~W.}\ \bibnamefont
  {Hamley}},\ }\href {https://doi.org/https://doi.org/10.1002/anie.200200546}
  {\bibfield  {journal} {\bibinfo  {journal} {Angewandte Chemie International
  Edition}\ }\textbf {\bibinfo {volume} {42}},\ \bibinfo {pages} {1692}
  (\bibinfo {year} {2003})}\  \BibitemShut {NoStop}%
\bibitem [{\citenamefont {Grason}\ \emph {et~al.}(2025)\citenamefont {Grason},
  \citenamefont {Rogers},\ and\ \citenamefont {Hagan}}]{grason_teaching_2025}%
  \BibitemOpen
  \bibfield  {author} {\bibinfo {author} {\bibfnamefont {G.}~\bibnamefont
  {Grason}}, \bibinfo {author} {\bibfnamefont {W.~B.}\ \bibnamefont {Rogers}},\
  and\ \bibinfo {author} {\bibfnamefont {M.}~\bibnamefont {Hagan}},\ }\href
  {https://doi.org/10.1063/pt.e01d5a2a0a} {\bibfield  {journal} {\bibinfo
  {journal} {Physics Today}\ }\textbf {\bibinfo {volume} {78}},\ \bibinfo
  {pages} {24} (\bibinfo {year} {2025})}\BibitemShut {NoStop}%
\bibitem [{\citenamefont {Zandi}\ \emph {et~al.}(2020)\citenamefont {Zandi},
  \citenamefont {Dragnea}, \citenamefont {Travesset},\ and\ \citenamefont
  {Podgornik}}]{zandi-virusgrowth-2020}%
  \BibitemOpen
  \bibfield  {author} {\bibinfo {author} {\bibfnamefont {R.}~\bibnamefont
  {Zandi}}, \bibinfo {author} {\bibfnamefont {B.}~\bibnamefont {Dragnea}},
  \bibinfo {author} {\bibfnamefont {A.}~\bibnamefont {Travesset}},\ and\
  \bibinfo {author} {\bibfnamefont {R.}~\bibnamefont {Podgornik}},\ }\href
  {https://doi.org/https://doi.org/10.1016/j.physrep.2019.12.005} {\bibfield
  {journal} {\bibinfo  {journal} {Physics Reports}\ }\textbf {\bibinfo {volume}
  {847}},\ \bibinfo {pages} {1} (\bibinfo {year} {2020})}\BibitemShut {NoStop}%
\bibitem [{\citenamefont {Zlotnick}\ and\ \citenamefont
  {Mukhopadhyay}(2011)}]{zlotnick-virusassembly-2011}%
  \BibitemOpen
  \bibfield  {author} {\bibinfo {author} {\bibfnamefont {A.}~\bibnamefont
  {Zlotnick}}\ and\ \bibinfo {author} {\bibfnamefont {S.}~\bibnamefont
  {Mukhopadhyay}},\ }\href
  {https://doi.org/https://doi.org/10.1016/j.tim.2010.11.003} {\bibfield
  {journal} {\bibinfo  {journal} {Trends in Microbiology}\ }\textbf {\bibinfo
  {volume} {19}},\ \bibinfo {pages} {14} (\bibinfo {year} {2011})}\BibitemShut
  {NoStop}%
\bibitem [{\citenamefont {Prum}\ \emph {et~al.}(2009)\citenamefont {Prum},
  \citenamefont {Dufresne}, \citenamefont {Quinn},\ and\ \citenamefont
  {Waters}}]{prum-nanostructurefeathers-2009}%
  \BibitemOpen
  \bibfield  {author} {\bibinfo {author} {\bibfnamefont {R.~O.}\ \bibnamefont
  {Prum}}, \bibinfo {author} {\bibfnamefont {E.~R.}\ \bibnamefont {Dufresne}},
  \bibinfo {author} {\bibfnamefont {T.}~\bibnamefont {Quinn}},\ and\ \bibinfo
  {author} {\bibfnamefont {K.}~\bibnamefont {Waters}},\ }\href
  {https://doi.org/10.1098/rsif.2008.0466.focus} {\bibfield  {journal}
  {\bibinfo  {journal} {Journal of The Royal Society Interface}\ }\textbf
  {\bibinfo {volume} {6}},\ \bibinfo {pages} {S253} (\bibinfo {year} {2009})} \BibitemShut {NoStop}%
\bibitem [{\citenamefont {Dufresne}\ \emph {et~al.}(2009)\citenamefont
  {Dufresne}, \citenamefont {Noh}, \citenamefont {Saranathan}, \citenamefont
  {Mochrie}, \citenamefont {Cao},\ and\ \citenamefont
  {Prum}}]{dufresne-nanostructurefeatherassembly-2009}%
  \BibitemOpen
  \bibfield  {author} {\bibinfo {author} {\bibfnamefont {E.~R.}\ \bibnamefont
  {Dufresne}}, \bibinfo {author} {\bibfnamefont {H.}~\bibnamefont {Noh}},
  \bibinfo {author} {\bibfnamefont {V.}~\bibnamefont {Saranathan}}, \bibinfo
  {author} {\bibfnamefont {S.~G.~J.}\ \bibnamefont {Mochrie}}, \bibinfo
  {author} {\bibfnamefont {H.}~\bibnamefont {Cao}},\ and\ \bibinfo {author}
  {\bibfnamefont {R.~O.}\ \bibnamefont {Prum}},\ }\href
  {https://doi.org/10.1039/B902775K} {\bibfield  {journal} {\bibinfo  {journal}
  {Soft Matter}\ }\textbf {\bibinfo {volume} {5}},\ \bibinfo {pages} {1792}
  (\bibinfo {year} {2009})}\BibitemShut {NoStop}%
\bibitem [{\citenamefont {Saranathan}\ \emph {et~al.}(2012)\citenamefont
  {Saranathan}, \citenamefont {Forster}, \citenamefont {Noh}, \citenamefont
  {Liew}, \citenamefont {Mochrie}, \citenamefont {Cao}, \citenamefont
  {Dufresne},\ and\ \citenamefont {Prum}}]{saranathan-featherxrays-2012}%
  \BibitemOpen
  \bibfield  {author} {\bibinfo {author} {\bibfnamefont {V.}~\bibnamefont
  {Saranathan}}, \bibinfo {author} {\bibfnamefont {J.~D.}\ \bibnamefont
  {Forster}}, \bibinfo {author} {\bibfnamefont {H.}~\bibnamefont {Noh}},
  \bibinfo {author} {\bibfnamefont {S.-F.}\ \bibnamefont {Liew}}, \bibinfo
  {author} {\bibfnamefont {S.~G.~J.}\ \bibnamefont {Mochrie}}, \bibinfo
  {author} {\bibfnamefont {H.}~\bibnamefont {Cao}}, \bibinfo {author}
  {\bibfnamefont {E.~R.}\ \bibnamefont {Dufresne}},\ and\ \bibinfo {author}
  {\bibfnamefont {R.~O.}\ \bibnamefont {Prum}},\ }\href
  {https://doi.org/10.1098/rsif.2012.0191} {\bibfield  {journal} {\bibinfo
  {journal} {Journal of The Royal Society Interface}\ }\textbf {\bibinfo
  {volume} {9}},\ \bibinfo {pages} {2563} (\bibinfo {year} {2012})},\ \Eprint
  {https://arxiv.org/abs/https://royalsocietypublishing.org/rsif/article-pdf/9/75/2563/4760/rsif.2012.0191.pdf}
  {https://royalsocietypublishing.org/rsif/article-pdf/9/75/2563/4760/rsif.2012.0191.pdf}
  \BibitemShut {NoStop}%
\bibitem [{\citenamefont {Popp}\ and\ \citenamefont
  {Robinson}(2012)}]{popp-cellfilaments-2012}%
  \BibitemOpen
  \bibfield  {author} {\bibinfo {author} {\bibfnamefont {D.}~\bibnamefont
  {Popp}}\ and\ \bibinfo {author} {\bibfnamefont {R.~C.}\ \bibnamefont
  {Robinson}},\ }\href {https://doi.org/https://doi.org/10.1002/cm.21006}
  {\bibfield  {journal} {\bibinfo  {journal} {Cytoskeleton}\ }\textbf {\bibinfo
  {volume} {69}},\ \bibinfo {pages} {71} (\bibinfo {year} {2012})},\ \Eprint
  {https://arxiv.org/abs/https://onlinelibrary.wiley.com/doi/pdf/10.1002/cm.21006}
  {https://onlinelibrary.wiley.com/doi/pdf/10.1002/cm.21006} \BibitemShut
  {NoStop}%
\bibitem [{\citenamefont {WEISEL}(2007)}]{weisel-fibrin-2007}%
  \BibitemOpen
  \bibfield  {author} {\bibinfo {author} {\bibfnamefont {J.}~\bibnamefont
  {WEISEL}},\ }\href
  {https://doi.org/https://doi.org/10.1111/j.1538-7836.2007.02504.x} {\bibfield
   {journal} {\bibinfo  {journal} {Journal of Thrombosis and Haemostasis}\
  }\textbf {\bibinfo {volume} {5}},\ \bibinfo {pages} {116} (\bibinfo {year}
  {2007})}\BibitemShut {NoStop}%
\bibitem [{\citenamefont {Rosario}\ \emph {et~al.}(2025)\citenamefont
  {Rosario}, \citenamefont {McInally}, \citenamefont {Jelenkovic},
  \citenamefont {Goode},\ and\ \citenamefont
  {Kondev}}]{rosario-stereocilia-2025}%
  \BibitemOpen
  \bibfield  {author} {\bibinfo {author} {\bibfnamefont {A.}~\bibnamefont
  {Rosario}}, \bibinfo {author} {\bibfnamefont {S.~G.}\ \bibnamefont
  {McInally}}, \bibinfo {author} {\bibfnamefont {P.~R.}\ \bibnamefont
  {Jelenkovic}}, \bibinfo {author} {\bibfnamefont {B.~L.}\ \bibnamefont
  {Goode}},\ and\ \bibinfo {author} {\bibfnamefont {J.}~\bibnamefont
  {Kondev}},\ }\bibfield  {journal} {\bibinfo  {journal} {bioRxiv}\ }\href
  {https://doi.org/10.1101/2023.07.27.550898} {10.1101/2023.07.27.550898}
  (\bibinfo {year} {2025}),\ \Eprint
  {https://arxiv.org/abs/https://www.biorxiv.org/content/early/2025/12/21/2023.07.27.550898.full.pdf}
  {https://www.biorxiv.org/content/early/2025/12/21/2023.07.27.550898.full.pdf}
  \BibitemShut {NoStop}%
\bibitem [{\citenamefont {Krey}\ \emph {et~al.}(2023)\citenamefont {Krey},
  \citenamefont {Chatterjee}, \citenamefont {Halford}, \citenamefont
  {Cunningham}, \citenamefont {Perrin},\ and\ \citenamefont
  {Barr-Gillespie}}]{krey-stereocilia-2023}%
  \BibitemOpen
  \bibfield  {author} {\bibinfo {author} {\bibfnamefont {J.~F.}\ \bibnamefont
  {Krey}}, \bibinfo {author} {\bibfnamefont {P.}~\bibnamefont {Chatterjee}},
  \bibinfo {author} {\bibfnamefont {J.}~\bibnamefont {Halford}}, \bibinfo
  {author} {\bibfnamefont {C.~L.}\ \bibnamefont {Cunningham}}, \bibinfo
  {author} {\bibfnamefont {B.~J.}\ \bibnamefont {Perrin}},\ and\ \bibinfo
  {author} {\bibfnamefont {P.~G.}\ \bibnamefont {Barr-Gillespie}},\ }\href
  {https://doi.org/10.1371/journal.pbio.3001964} {\bibfield  {journal}
  {\bibinfo  {journal} {PLOS Biology}\ }\textbf {\bibinfo {volume} {21}},\
  \bibinfo {pages} {1} (\bibinfo {year} {2023})}\BibitemShut {NoStop}%
\bibitem [{\citenamefont {Hagan}\ and\ \citenamefont
  {Grason}(2021)}]{hagan_sla_2021}%
  \BibitemOpen
  \bibfield  {author} {\bibinfo {author} {\bibfnamefont {M.~F.}\ \bibnamefont
  {Hagan}}\ and\ \bibinfo {author} {\bibfnamefont {G.~M.}\ \bibnamefont
  {Grason}},\ }\href {https://doi.org/10.1103/RevModPhys.93.025008} {\bibfield
  {journal} {\bibinfo  {journal} {Rev. Mod. Phys.}\ }\textbf {\bibinfo {volume}
  {93}},\ \bibinfo {pages} {025008} (\bibinfo {year} {2021})}\BibitemShut
  {NoStop}%
\bibitem [{\citenamefont {Grason}(2016)}]{grason-perspective-2016}%
  \BibitemOpen
  \bibfield  {author} {\bibinfo {author} {\bibfnamefont {G.~M.}\ \bibnamefont
  {Grason}},\ }\href {https://doi.org/10.1063/1.4962629} {\bibfield  {journal}
  {\bibinfo  {journal} {The Journal of Chemical Physics}\ }\textbf {\bibinfo
  {volume} {145}},\ \bibinfo {pages} {110901} (\bibinfo {year} {2016})},\
  \Eprint
  {https://arxiv.org/abs/https://pubs.aip.org/aip/jcp/article-pdf/doi/10.1063/1.4962629/15517232/110901\_1\_online.pdf}
  {https://pubs.aip.org/aip/jcp/article-pdf/doi/10.1063/1.4962629/15517232/110901\_1\_online.pdf}
  \BibitemShut {NoStop}%
\bibitem [{\citenamefont {Schneider}\ and\ \citenamefont
  {Gompper}(2005)}]{schneider_shapes_2005}%
  \BibitemOpen
  \bibfield  {author} {\bibinfo {author} {\bibfnamefont {S.}~\bibnamefont
  {Schneider}}\ and\ \bibinfo {author} {\bibfnamefont {G.}~\bibnamefont
  {Gompper}},\ }\href {https://doi.org/10.1209/epl/i2004-10464-2} {\bibfield
  {journal} {\bibinfo  {journal} {Europhysics Letters (EPL)}\ }\textbf
  {\bibinfo {volume} {70}},\ \bibinfo {pages} {136} (\bibinfo {year}
  {2005})}\BibitemShut {NoStop}%
\bibitem [{\citenamefont {Armon}\ \emph {et~al.}(2014)\citenamefont {Armon},
  \citenamefont {Aharoni}, \citenamefont {Moshe},\ and\ \citenamefont
  {Sharon}}]{armon_shape_2014}%
  \BibitemOpen
  \bibfield  {author} {\bibinfo {author} {\bibfnamefont {S.}~\bibnamefont
  {Armon}}, \bibinfo {author} {\bibfnamefont {H.}~\bibnamefont {Aharoni}},
  \bibinfo {author} {\bibfnamefont {M.}~\bibnamefont {Moshe}},\ and\ \bibinfo
  {author} {\bibfnamefont {E.}~\bibnamefont {Sharon}},\ }\href
  {https://doi.org/10.1039/c3sm52313f} {\bibfield  {journal} {\bibinfo
  {journal} {Soft Matter}\ }\textbf {\bibinfo {volume} {10}},\ \bibinfo {pages}
  {2733} (\bibinfo {year} {2014})},\ \Eprint
  {https://arxiv.org/abs/https://pubs.rsc.org/sm/article-pdf/10/16/2733/3354555/c3sm52313f.pdf}
  {https://pubs.rsc.org/sm/article-pdf/10/16/2733/3354555/c3sm52313f.pdf}
  \BibitemShut {NoStop}%
\bibitem [{\citenamefont {Hall}\ \emph {et~al.}(2016)\citenamefont {Hall},
  \citenamefont {Bruss}, \citenamefont {Barone},\ and\ \citenamefont
  {Grason}}]{hall-chiralbundles-2016}%
  \BibitemOpen
  \bibfield  {author} {\bibinfo {author} {\bibfnamefont {D.~M.}\ \bibnamefont
  {Hall}}, \bibinfo {author} {\bibfnamefont {I.~R.}\ \bibnamefont {Bruss}},
  \bibinfo {author} {\bibfnamefont {J.~R.}\ \bibnamefont {Barone}},\ and\
  \bibinfo {author} {\bibfnamefont {G.~M.}\ \bibnamefont {Grason}},\ }\href
  {https://doi.org/10.1038/nmat4598} {\bibfield  {journal} {\bibinfo  {journal}
  {Nature Materials}\ }\textbf {\bibinfo {volume} {15}},\ \bibinfo {pages}
  {727} (\bibinfo {year} {2016})}\BibitemShut {NoStop}%
\bibitem [{\citenamefont {K\"ohler}\ \emph {et~al.}(2016)\citenamefont
  {K\"ohler}, \citenamefont {Backofen},\ and\ \citenamefont
  {Voigt}}]{kohler_stress_2016}%
  \BibitemOpen
  \bibfield  {author} {\bibinfo {author} {\bibfnamefont {C.}~\bibnamefont
  {K\"ohler}}, \bibinfo {author} {\bibfnamefont {R.}~\bibnamefont {Backofen}},\
  and\ \bibinfo {author} {\bibfnamefont {A.}~\bibnamefont {Voigt}},\ }\href
  {https://doi.org/10.1103/PhysRevLett.116.135502} {\bibfield  {journal}
  {\bibinfo  {journal} {Phys. Rev. Lett.}\ }\textbf {\bibinfo {volume} {116}},\
  \bibinfo {pages} {135502} (\bibinfo {year} {2016})}\BibitemShut {NoStop}%
\bibitem [{\citenamefont {Hall}\ \emph {et~al.}(2023)\citenamefont {Hall},
  \citenamefont {Stevens},\ and\ \citenamefont {Grason}}]{hall-wedges-2023}%
  \BibitemOpen
  \bibfield  {author} {\bibinfo {author} {\bibfnamefont {D.~M.}\ \bibnamefont
  {Hall}}, \bibinfo {author} {\bibfnamefont {M.~J.}\ \bibnamefont {Stevens}},\
  and\ \bibinfo {author} {\bibfnamefont {G.~M.}\ \bibnamefont {Grason}},\
  }\href {https://doi.org/10.1039/D2SM01371A} {\bibfield  {journal} {\bibinfo
  {journal} {Soft Matter}\ }\textbf {\bibinfo {volume} {19}},\ \bibinfo {pages}
  {858} (\bibinfo {year} {2023})}\BibitemShut {NoStop}%
\bibitem [{\citenamefont {Wang}\ and\ \citenamefont
  {Grason}(2025)}]{wang-polarwj-2025}%
  \BibitemOpen
  \bibfield  {author} {\bibinfo {author} {\bibfnamefont {M.}~\bibnamefont
  {Wang}}\ and\ \bibinfo {author} {\bibfnamefont {G.~M.}\ \bibnamefont
  {Grason}},\ }\href {https://doi.org/10.1039/D5SM00136F} {\bibfield  {journal}
  {\bibinfo  {journal} {Soft Matter}\ }\textbf {\bibinfo {volume} {21}},\
  \bibinfo {pages} {5423} (\bibinfo {year} {2025})}\BibitemShut {NoStop}%
\bibitem [{\citenamefont {Roy}\ \emph {et~al.}(2023)\citenamefont {Roy},
  \citenamefont {Terzi},\ and\ \citenamefont {Lenz}}]{leroy_collective_2023}%
  \BibitemOpen
  \bibfield  {author} {\bibinfo {author} {\bibfnamefont {H.~L.}\ \bibnamefont
  {Roy}}, \bibinfo {author} {\bibfnamefont {M.~M.}\ \bibnamefont {Terzi}},\
  and\ \bibinfo {author} {\bibfnamefont {M.}~\bibnamefont {Lenz}},\ }\href
  {https://doi.org/10.48550/arXiv.2308.04698} {\bibinfo {title} {Collective
  deformation modes promote fibrous self-assembly in protein-like particles}}
  (\bibinfo {year} {2023}),\ \Eprint {https://arxiv.org/abs/2308.04698}
  {arXiv:2308.04698 [cond-mat.soft]} \BibitemShut {NoStop}%
\bibitem [{\citenamefont {Bruss}\ and\ \citenamefont
  {Grason}(2012)}]{bruss_noneuclidean_2012}%
  \BibitemOpen
  \bibfield  {author} {\bibinfo {author} {\bibfnamefont {I.~R.}\ \bibnamefont
  {Bruss}}\ and\ \bibinfo {author} {\bibfnamefont {G.~M.}\ \bibnamefont
  {Grason}},\ }\href {https://doi.org/10.1073/pnas.1205606109} {\bibfield
  {journal} {\bibinfo  {journal} {Proceedings of the National Academy of
  Sciences}\ }\textbf {\bibinfo {volume} {109}},\ \bibinfo {pages} {10781}
  (\bibinfo {year} {2012})},\ \Eprint
  {https://arxiv.org/abs/https://www.pnas.org/doi/pdf/10.1073/pnas.1205606109}
  {https://www.pnas.org/doi/pdf/10.1073/pnas.1205606109} \BibitemShut {NoStop}%
\bibitem [{\citenamefont {Bruss}\ and\ \citenamefont
  {Grason}(2013)}]{bruss_topological_2013}%
  \BibitemOpen
  \bibfield  {author} {\bibinfo {author} {\bibfnamefont {I.~R.}\ \bibnamefont
  {Bruss}}\ and\ \bibinfo {author} {\bibfnamefont {G.~M.}\ \bibnamefont
  {Grason}},\ }\href {https://doi.org/10.1039/c3sm50672j} {\bibfield  {journal}
  {\bibinfo  {journal} {Soft Matter}\ }\textbf {\bibinfo {volume} {9}},\
  \bibinfo {pages} {8327} (\bibinfo {year} {2013})},\ \Eprint
  {https://arxiv.org/abs/https://pubs.rsc.org/sm/article-pdf/9/34/8327/3342734/c3sm50672j.pdf}
  {https://pubs.rsc.org/sm/article-pdf/9/34/8327/3342734/c3sm50672j.pdf}
  \BibitemShut {NoStop}%
\bibitem [{\citenamefont {Li}\ \emph {et~al.}(2019)\citenamefont {Li},
  \citenamefont {Zandi}, \citenamefont {Travesset},\ and\ \citenamefont
  {Grason}}]{li-crystalcaps-2019}%
  \BibitemOpen
  \bibfield  {author} {\bibinfo {author} {\bibfnamefont {S.}~\bibnamefont
  {Li}}, \bibinfo {author} {\bibfnamefont {R.}~\bibnamefont {Zandi}}, \bibinfo
  {author} {\bibfnamefont {A.}~\bibnamefont {Travesset}},\ and\ \bibinfo
  {author} {\bibfnamefont {G.~M.}\ \bibnamefont {Grason}},\ }\href
  {https://doi.org/10.1103/PhysRevLett.123.145501} {\bibfield  {journal}
  {\bibinfo  {journal} {Phys. Rev. Lett.}\ }\textbf {\bibinfo {volume} {123}},\
  \bibinfo {pages} {145501} (\bibinfo {year} {2019})}\BibitemShut {NoStop}%
\bibitem [{\citenamefont {Hackney}\ \emph {et~al.}(2023)\citenamefont
  {Hackney}, \citenamefont {Amey},\ and\ \citenamefont
  {Grason}}]{hackney-spins-2023}%
  \BibitemOpen
  \bibfield  {author} {\bibinfo {author} {\bibfnamefont {N.~W.}\ \bibnamefont
  {Hackney}}, \bibinfo {author} {\bibfnamefont {C.}~\bibnamefont {Amey}},\ and\
  \bibinfo {author} {\bibfnamefont {G.~M.}\ \bibnamefont {Grason}},\ }\href
  {https://doi.org/10.1103/PhysRevX.13.041010} {\bibfield  {journal} {\bibinfo
  {journal} {Phys. Rev. X}\ }\textbf {\bibinfo {volume} {13}},\ \bibinfo
  {pages} {041010} (\bibinfo {year} {2023})}\BibitemShut {NoStop}%
\bibitem [{\citenamefont {Aggeli}\ \emph {et~al.}(2001)\citenamefont {Aggeli},
  \citenamefont {Nyrkova}, \citenamefont {Bell}, \citenamefont {Harding},
  \citenamefont {Carrick}, \citenamefont {McLeish}, \citenamefont {Semenov},\
  and\ \citenamefont {Boden}}]{aggeli-chiralrods-2001}%
  \BibitemOpen
  \bibfield  {author} {\bibinfo {author} {\bibfnamefont {A.}~\bibnamefont
  {Aggeli}}, \bibinfo {author} {\bibfnamefont {I.~A.}\ \bibnamefont {Nyrkova}},
  \bibinfo {author} {\bibfnamefont {M.}~\bibnamefont {Bell}}, \bibinfo {author}
  {\bibfnamefont {R.}~\bibnamefont {Harding}}, \bibinfo {author} {\bibfnamefont
  {L.}~\bibnamefont {Carrick}}, \bibinfo {author} {\bibfnamefont {T.~C.~B.}\
  \bibnamefont {McLeish}}, \bibinfo {author} {\bibfnamefont {A.~N.}\
  \bibnamefont {Semenov}},\ and\ \bibinfo {author} {\bibfnamefont
  {N.}~\bibnamefont {Boden}},\ }\href {https://doi.org/10.1073/pnas.191250198}
  {\bibfield  {journal} {\bibinfo  {journal} {Proceedings of the National
  Academy of Sciences}\ }\textbf {\bibinfo {volume} {98}},\ \bibinfo {pages}
  {11857} (\bibinfo {year} {2001})},\ \Eprint
  {https://arxiv.org/abs/https://www.pnas.org/doi/pdf/10.1073/pnas.191250198}
  {https://www.pnas.org/doi/pdf/10.1073/pnas.191250198} \BibitemShut {NoStop}%
\bibitem [{\citenamefont {Grason}(2020)}]{grason-twistedbundles-2020}%
  \BibitemOpen
  \bibfield  {author} {\bibinfo {author} {\bibfnamefont {G.~M.}\ \bibnamefont
  {Grason}},\ }\href {https://doi.org/10.1039/C9SM01840A} {\bibfield  {journal}
  {\bibinfo  {journal} {Soft Matter}\ }\textbf {\bibinfo {volume} {16}},\
  \bibinfo {pages} {1102} (\bibinfo {year} {2020})}\BibitemShut {NoStop}%
\bibitem [{\citenamefont {Serafin}\ \emph
  {et~al.}(2021{\natexlab{a}})\citenamefont {Serafin}, \citenamefont {Lu},
  \citenamefont {Kotov}, \citenamefont {Sun},\ and\ \citenamefont
  {Mao}}]{serafin-polyhedra-2021}%
  \BibitemOpen
  \bibfield  {author} {\bibinfo {author} {\bibfnamefont {F.}~\bibnamefont
  {Serafin}}, \bibinfo {author} {\bibfnamefont {J.}~\bibnamefont {Lu}},
  \bibinfo {author} {\bibfnamefont {N.}~\bibnamefont {Kotov}}, \bibinfo
  {author} {\bibfnamefont {K.}~\bibnamefont {Sun}},\ and\ \bibinfo {author}
  {\bibfnamefont {X.}~\bibnamefont {Mao}},\ }\href
  {https://doi.org/10.1038/s41467-021-25139-9} {\bibfield  {journal} {\bibinfo
  {journal} {Nature Communications}\ }\textbf {\bibinfo {volume} {12}},\
  \bibinfo {pages} {4925} (\bibinfo {year} {2021}{\natexlab{a}})}\BibitemShut
  {NoStop}%
\bibitem [{\citenamefont {Berengut}\ \emph {et~al.}(2020)\citenamefont
  {Berengut}, \citenamefont {Wong}, \citenamefont {Berengut}, \citenamefont
  {Doye}, \citenamefont {Ouldridge},\ and\ \citenamefont
  {Lee}}]{berengut-polybricks-2020}%
  \BibitemOpen
  \bibfield  {author} {\bibinfo {author} {\bibfnamefont {J.~F.}\ \bibnamefont
  {Berengut}}, \bibinfo {author} {\bibfnamefont {C.~K.}\ \bibnamefont {Wong}},
  \bibinfo {author} {\bibfnamefont {J.~C.}\ \bibnamefont {Berengut}}, \bibinfo
  {author} {\bibfnamefont {J.~P.~K.}\ \bibnamefont {Doye}}, \bibinfo {author}
  {\bibfnamefont {T.~E.}\ \bibnamefont {Ouldridge}},\ and\ \bibinfo {author}
  {\bibfnamefont {L.~K.}\ \bibnamefont {Lee}},\ }\href
  {https://doi.org/10.1021/acsnano.0c07696} {\bibfield  {journal} {\bibinfo
  {journal} {ACS Nano}\ }\textbf {\bibinfo {volume} {14}},\ \bibinfo {pages}
  {17428} (\bibinfo {year} {2020})}\BibitemShut {NoStop}%
\bibitem [{\citenamefont {Meng}\ \emph {et~al.}(2014)\citenamefont {Meng},
  \citenamefont {Paulose}, \citenamefont {Nelson},\ and\ \citenamefont
  {Manoharan}}]{meng-curvedcrystals-2014}%
  \BibitemOpen
  \bibfield  {author} {\bibinfo {author} {\bibfnamefont {G.}~\bibnamefont
  {Meng}}, \bibinfo {author} {\bibfnamefont {J.}~\bibnamefont {Paulose}},
  \bibinfo {author} {\bibfnamefont {D.~R.}\ \bibnamefont {Nelson}},\ and\
  \bibinfo {author} {\bibfnamefont {V.~N.}\ \bibnamefont {Manoharan}},\ }\href
  {https://doi.org/10.1126/science.1244827} {\bibfield  {journal} {\bibinfo
  {journal} {Science}\ }\textbf {\bibinfo {volume} {343}},\ \bibinfo {pages}
  {634} (\bibinfo {year} {2014})},\ \Eprint
  {https://arxiv.org/abs/https://www.science.org/doi/pdf/10.1126/science.1244827}
  {https://www.science.org/doi/pdf/10.1126/science.1244827} \BibitemShut
  {NoStop}%
\bibitem [{\citenamefont {Zhang}\ \emph {et~al.}(2019)\citenamefont {Zhang},
  \citenamefont {Grossman}, \citenamefont {Danino},\ and\ \citenamefont
  {Sharon}}]{zhang-ribbons-2019}%
  \BibitemOpen
  \bibfield  {author} {\bibinfo {author} {\bibfnamefont {M.}~\bibnamefont
  {Zhang}}, \bibinfo {author} {\bibfnamefont {D.}~\bibnamefont {Grossman}},
  \bibinfo {author} {\bibfnamefont {D.}~\bibnamefont {Danino}},\ and\ \bibinfo
  {author} {\bibfnamefont {E.}~\bibnamefont {Sharon}},\ }\href
  {https://doi.org/10.1038/s41467-019-11473-6} {\bibfield  {journal} {\bibinfo
  {journal} {Nature Communications}\ }\textbf {\bibinfo {volume} {10}},\
  \bibinfo {pages} {3565} (\bibinfo {year} {2019})}\BibitemShut {NoStop}%
\bibitem [{\citenamefont {Serafin}\ \emph
  {et~al.}(2021{\natexlab{b}})\citenamefont {Serafin}, \citenamefont {Lu},
  \citenamefont {Kotov}, \citenamefont {Sun},\ and\ \citenamefont
  {Mao}}]{serafin_frustrated_2021}%
  \BibitemOpen
  \bibfield  {author} {\bibinfo {author} {\bibfnamefont {F.}~\bibnamefont
  {Serafin}}, \bibinfo {author} {\bibfnamefont {J.}~\bibnamefont {Lu}},
  \bibinfo {author} {\bibfnamefont {N.}~\bibnamefont {Kotov}}, \bibinfo
  {author} {\bibfnamefont {K.}~\bibnamefont {Sun}},\ and\ \bibinfo {author}
  {\bibfnamefont {X.}~\bibnamefont {Mao}},\ }\href
  {https://doi.org/10.1038/s41467-021-25139-9} {\bibfield  {journal} {\bibinfo
  {journal} {Nature Communications}\ }\textbf {\bibinfo {volume} {12}},\
  \bibinfo {pages} {4925} (\bibinfo {year} {2021}{\natexlab{b}})}\BibitemShut
  {NoStop}%
\bibitem [{\citenamefont {Meiri}\ and\ \citenamefont
  {Efrati}(2021)}]{meiri-gfa-2021}%
  \BibitemOpen
  \bibfield  {author} {\bibinfo {author} {\bibfnamefont {S.}~\bibnamefont
  {Meiri}}\ and\ \bibinfo {author} {\bibfnamefont {E.}~\bibnamefont {Efrati}},\
  }\href {https://doi.org/10.1103/PhysRevE.104.054601} {\bibfield  {journal}
  {\bibinfo  {journal} {Phys. Rev. E}\ }\textbf {\bibinfo {volume} {104}},\
  \bibinfo {pages} {054601} (\bibinfo {year} {2021})}\BibitemShut {NoStop}%
\bibitem [{\citenamefont {Hackney}\ and\ \citenamefont
  {Grason}(2025)}]{hackney-phasetransitions-2025}%
  \BibitemOpen
  \bibfield  {author} {\bibinfo {author} {\bibfnamefont {N.~W.}\ \bibnamefont
  {Hackney}}\ and\ \bibinfo {author} {\bibfnamefont {G.}~\bibnamefont
  {Grason}},\ }\href {https://doi.org/10.1103/bj18-bphb} {\bibfield  {journal}
  {\bibinfo  {journal} {Phys. Rev. E}\ }\textbf {\bibinfo {volume} {112}},\
  \bibinfo {pages} {065419} (\bibinfo {year} {2025})}\BibitemShut {NoStop}%
\bibitem [{\citenamefont {Ortiz-Tav\'arez}\ \emph {et~al.}(2025)\citenamefont
  {Ortiz-Tav\'arez}, \citenamefont {Yang}, \citenamefont {Kotov},\ and\
  \citenamefont {Mao}}]{ortiz-tavarez_statistical_2025}%
  \BibitemOpen
  \bibfield  {author} {\bibinfo {author} {\bibfnamefont {J.~M.}\ \bibnamefont
  {Ortiz-Tav\'arez}}, \bibinfo {author} {\bibfnamefont {Z.}~\bibnamefont
  {Yang}}, \bibinfo {author} {\bibfnamefont {N.}~\bibnamefont {Kotov}},\ and\
  \bibinfo {author} {\bibfnamefont {X.}~\bibnamefont {Mao}},\ }\href
  {https://doi.org/10.1103/PhysRevLett.134.147401} {\bibfield  {journal}
  {\bibinfo  {journal} {Phys. Rev. Lett.}\ }\textbf {\bibinfo {volume} {134}},\
  \bibinfo {pages} {147401} (\bibinfo {year} {2025})}\BibitemShut {NoStop}%
\bibitem [{\citenamefont {Spivack}\ \emph {et~al.}(2022)\citenamefont
  {Spivack}, \citenamefont {Hall},\ and\ \citenamefont
  {Grason}}]{Spivack-puzzlemers-2022}%
  \BibitemOpen
  \bibfield  {author} {\bibinfo {author} {\bibfnamefont {I.~R.}\ \bibnamefont
  {Spivack}}, \bibinfo {author} {\bibfnamefont {D.~M.}\ \bibnamefont {Hall}},\
  and\ \bibinfo {author} {\bibfnamefont {G.~M.}\ \bibnamefont {Grason}},\
  }\href {https://doi.org/10.1088/1367-2630/ac753e} {\bibfield  {journal}
  {\bibinfo  {journal} {New Journal of Physics}\ }\textbf {\bibinfo {volume}
  {24}},\ \bibinfo {pages} {063023} (\bibinfo {year} {2022})}\BibitemShut
  {NoStop}%
\bibitem [{\citenamefont {Wang}\ and\ \citenamefont
  {Grason}(2024)}]{wang-polybricks-2024}%
  \BibitemOpen
  \bibfield  {author} {\bibinfo {author} {\bibfnamefont {M.}~\bibnamefont
  {Wang}}\ and\ \bibinfo {author} {\bibfnamefont {G.}~\bibnamefont {Grason}},\
  }\href {https://doi.org/10.1103/PhysRevE.109.014608} {\bibfield  {journal}
  {\bibinfo  {journal} {Phys. Rev. E}\ }\textbf {\bibinfo {volume} {109}},\
  \bibinfo {pages} {014608} (\bibinfo {year} {2024})}\BibitemShut {NoStop}%
\bibitem [{\citenamefont {Lenz}\ and\ \citenamefont
  {Witten}(2017)}]{lenz_polygon_17}%
  \BibitemOpen
  \bibfield  {author} {\bibinfo {author} {\bibfnamefont {M.}~\bibnamefont
  {Lenz}}\ and\ \bibinfo {author} {\bibfnamefont {T.~A.}\ \bibnamefont
  {Witten}},\ }\href {https://doi.org/10.1038/nphys4184} {\bibfield  {journal}
  {\bibinfo  {journal} {Nature Physics}\ }\textbf {\bibinfo {volume} {13}},\
  \bibinfo {pages} {1100} (\bibinfo {year} {2017})}\BibitemShut {NoStop}%
\bibitem [{\citenamefont {Tyukodi}\ \emph {et~al.}(2022)\citenamefont
  {Tyukodi}, \citenamefont {Mohajerani}, \citenamefont {Hall}, \citenamefont
  {Grason},\ and\ \citenamefont {Hagan}}]{botond-tubules-2022}%
  \BibitemOpen
  \bibfield  {author} {\bibinfo {author} {\bibfnamefont {B.}~\bibnamefont
  {Tyukodi}}, \bibinfo {author} {\bibfnamefont {F.}~\bibnamefont {Mohajerani}},
  \bibinfo {author} {\bibfnamefont {D.~M.}\ \bibnamefont {Hall}}, \bibinfo
  {author} {\bibfnamefont {G.~M.}\ \bibnamefont {Grason}},\ and\ \bibinfo
  {author} {\bibfnamefont {M.~F.}\ \bibnamefont {Hagan}},\ }\href
  {https://doi.org/10.1021/acsnano.2c00865} {\bibfield  {journal} {\bibinfo
  {journal} {ACS Nano}\ }\textbf {\bibinfo {volume} {16}},\ \bibinfo {pages}
  {9077} (\bibinfo {year} {2022})}\BibitemShut {NoStop}%
\bibitem [{\citenamefont {Tanjeem}\ \emph
  {et~al.}(2022{\natexlab{a}})\citenamefont {Tanjeem}, \citenamefont {Hall},
  \citenamefont {Minnis}, \citenamefont {Hayward},\ and\ \citenamefont
  {Grason}}]{tanjeem-curvamers-2022}%
  \BibitemOpen
  \bibfield  {author} {\bibinfo {author} {\bibfnamefont {N.}~\bibnamefont
  {Tanjeem}}, \bibinfo {author} {\bibfnamefont {D.~M.}\ \bibnamefont {Hall}},
  \bibinfo {author} {\bibfnamefont {M.~B.}\ \bibnamefont {Minnis}}, \bibinfo
  {author} {\bibfnamefont {R.~C.}\ \bibnamefont {Hayward}},\ and\ \bibinfo
  {author} {\bibfnamefont {G.~M.}\ \bibnamefont {Grason}},\ }\href
  {https://doi.org/10.1103/PhysRevResearch.4.033035} {\bibfield  {journal}
  {\bibinfo  {journal} {Phys. Rev. Res.}\ }\textbf {\bibinfo {volume} {4}},\
  \bibinfo {pages} {033035} (\bibinfo {year} {2022}{\natexlab{a}})}\BibitemShut
  {NoStop}%
\bibitem [{\citenamefont {Sullivan}\ \emph {et~al.}(2024)\citenamefont
  {Sullivan}, \citenamefont {Hayward},\ and\ \citenamefont
  {Grason}}]{sullivan-curvamers-2024}%
  \BibitemOpen
  \bibfield  {author} {\bibinfo {author} {\bibfnamefont {K.~T.}\ \bibnamefont
  {Sullivan}}, \bibinfo {author} {\bibfnamefont {R.~C.}\ \bibnamefont
  {Hayward}},\ and\ \bibinfo {author} {\bibfnamefont {G.~M.}\ \bibnamefont
  {Grason}},\ }\href {https://doi.org/10.1103/PhysRevE.110.024602} {\bibfield
  {journal} {\bibinfo  {journal} {Phys. Rev. E}\ }\textbf {\bibinfo {volume}
  {110}},\ \bibinfo {pages} {024602} (\bibinfo {year} {2024})}\BibitemShut
  {NoStop}%
\bibitem [{\citenamefont {do~Carmo}(2016)}]{doCarmo2016Chapter4}%
  \BibitemOpen
  \bibfield  {author} {\bibinfo {author} {\bibfnamefont {M.~P.}\ \bibnamefont
  {do~Carmo}},\ }in\ \href@noop {} {\emph {\bibinfo {booktitle} {Differential
  Geometry of Curves and Surfaces}}}\ (\bibinfo  {publisher} {Dover
  Publications},\ \bibinfo {address} {Mineola, NY},\ \bibinfo {year} {2016})\
  \bibinfo {edition} {2nd}\ ed.,\ Chap.~\bibinfo {chapter} {4}\BibitemShut
  {NoStop}%
\bibitem [{\citenamefont {Aggarwal}\ \emph {et~al.}(2012)\citenamefont
  {Aggarwal}, \citenamefont {Rudnick}, \citenamefont {Bruinsma},\ and\
  \citenamefont {Klug}}]{aggarwal-viralshellelasticity-2012}%
  \BibitemOpen
  \bibfield  {author} {\bibinfo {author} {\bibfnamefont {A.}~\bibnamefont
  {Aggarwal}}, \bibinfo {author} {\bibfnamefont {J.}~\bibnamefont {Rudnick}},
  \bibinfo {author} {\bibfnamefont {R.~F.}\ \bibnamefont {Bruinsma}},\ and\
  \bibinfo {author} {\bibfnamefont {W.~S.}\ \bibnamefont {Klug}},\ }\href
  {https://doi.org/10.1103/PhysRevLett.109.148102} {\bibfield  {journal}
  {\bibinfo  {journal} {Phys. Rev. Lett.}\ }\textbf {\bibinfo {volume} {109}},\
  \bibinfo {pages} {148102} (\bibinfo {year} {2012})}\BibitemShut {NoStop}%
\bibitem [{\citenamefont {Singh}\ \emph {et~al.}(2020)\citenamefont {Singh},
  \citenamefont {Ko\ifmmode~\check{s}\else \v{s}\fi{}mrlj},\ and\ \citenamefont
  {Bruinsma}}]{singh-viralcapsidshells-2020}%
  \BibitemOpen
  \bibfield  {author} {\bibinfo {author} {\bibfnamefont {A.~R.}\ \bibnamefont
  {Singh}}, \bibinfo {author} {\bibfnamefont {A.}~\bibnamefont
  {Ko\ifmmode~\check{s}\else \v{s}\fi{}mrlj}},\ and\ \bibinfo {author}
  {\bibfnamefont {R.}~\bibnamefont {Bruinsma}},\ }\href
  {https://doi.org/10.1103/PhysRevLett.124.158101} {\bibfield  {journal}
  {\bibinfo  {journal} {Phys. Rev. Lett.}\ }\textbf {\bibinfo {volume} {124}},\
  \bibinfo {pages} {158101} (\bibinfo {year} {2020})}\BibitemShut {NoStop}%
\bibitem [{\citenamefont {Ko\ifmmode~\check{s}\else \v{s}\fi{}mrlj}\ and\
  \citenamefont {Nelson}(2017)}]{kosmrlj-sphericalshellstatmech-2017}%
  \BibitemOpen
  \bibfield  {author} {\bibinfo {author} {\bibfnamefont {A.}~\bibnamefont
  {Ko\ifmmode~\check{s}\else \v{s}\fi{}mrlj}}\ and\ \bibinfo {author}
  {\bibfnamefont {D.~R.}\ \bibnamefont {Nelson}},\ }\href
  {https://doi.org/10.1103/PhysRevX.7.011002} {\bibfield  {journal} {\bibinfo
  {journal} {Phys. Rev. X}\ }\textbf {\bibinfo {volume} {7}},\ \bibinfo {pages}
  {011002} (\bibinfo {year} {2017})}\BibitemShut {NoStop}%
\bibitem [{\citenamefont {Dervaux}\ \emph {et~al.}(2009)\citenamefont
  {Dervaux}, \citenamefont {Ciarletta},\ and\ \citenamefont {{Ben
  Amar}}}]{ciarlet-leafripples-2009}%
  \BibitemOpen
  \bibfield  {author} {\bibinfo {author} {\bibfnamefont {J.}~\bibnamefont
  {Dervaux}}, \bibinfo {author} {\bibfnamefont {P.}~\bibnamefont {Ciarletta}},\
  and\ \bibinfo {author} {\bibfnamefont {M.}~\bibnamefont {{Ben Amar}}},\
  }\href {https://doi.org/https://doi.org/10.1016/j.jmps.2008.11.011}
  {\bibfield  {journal} {\bibinfo  {journal} {Journal of the Mechanics and
  Physics of Solids}\ }\textbf {\bibinfo {volume} {57}},\ \bibinfo {pages}
  {458} (\bibinfo {year} {2009})}\BibitemShut {NoStop}%
\bibitem [{\citenamefont {Pezzulla}\ \emph {et~al.}(2017)\citenamefont
  {Pezzulla}, \citenamefont {Stoop}, \citenamefont {Jiang},\ and\ \citenamefont
  {Holmes}}]{holmes-noneuclideanshells-2017}%
  \BibitemOpen
  \bibfield  {author} {\bibinfo {author} {\bibfnamefont {M.}~\bibnamefont
  {Pezzulla}}, \bibinfo {author} {\bibfnamefont {N.}~\bibnamefont {Stoop}},
  \bibinfo {author} {\bibfnamefont {X.}~\bibnamefont {Jiang}},\ and\ \bibinfo
  {author} {\bibfnamefont {D.~P.}\ \bibnamefont {Holmes}},\ }\href
  {https://doi.org/10.1098/rspa.2017.0087} {\bibfield  {journal} {\bibinfo
  {journal} {Proceedings of the Royal Society A: Mathematical, Physical and
  Engineering Sciences}\ }\textbf {\bibinfo {volume} {473}},\ \bibinfo {pages}
  {20170087} (\bibinfo {year} {2017})},\ \Eprint
  {https://arxiv.org/abs/https://royalsocietypublishing.org/rspa/article-pdf/doi/10.1098/rspa.2017.0087/364358/rspa.2017.0087.pdf}
  {https://royalsocietypublishing.org/rspa/article-pdf/doi/10.1098/rspa.2017.0087/364358/rspa.2017.0087.pdf}
  \BibitemShut {NoStop}%
\bibitem [{\citenamefont {Pezzulla}\ \emph {et~al.}(2018)\citenamefont
  {Pezzulla}, \citenamefont {Stoop}, \citenamefont {Steranka}, \citenamefont
  {Bade},\ and\ \citenamefont {Holmes}}]{holmes-flytrapinstabilities-2018}%
  \BibitemOpen
  \bibfield  {author} {\bibinfo {author} {\bibfnamefont {M.}~\bibnamefont
  {Pezzulla}}, \bibinfo {author} {\bibfnamefont {N.}~\bibnamefont {Stoop}},
  \bibinfo {author} {\bibfnamefont {M.~P.}\ \bibnamefont {Steranka}}, \bibinfo
  {author} {\bibfnamefont {A.~J.}\ \bibnamefont {Bade}},\ and\ \bibinfo
  {author} {\bibfnamefont {D.~P.}\ \bibnamefont {Holmes}},\ }\href
  {https://doi.org/10.1103/PhysRevLett.120.048002} {\bibfield  {journal}
  {\bibinfo  {journal} {Phys. Rev. Lett.}\ }\textbf {\bibinfo {volume} {120}},\
  \bibinfo {pages} {048002} (\bibinfo {year} {2018})}\BibitemShut {NoStop}%
\bibitem [{\citenamefont {Stein-Montalvo}\ \emph {et~al.}(2019)\citenamefont
  {Stein-Montalvo}, \citenamefont {Costa}, \citenamefont {Pezzulla},\ and\
  \citenamefont {Holmes}}]{holmes-confinedswelling-2019}%
  \BibitemOpen
  \bibfield  {author} {\bibinfo {author} {\bibfnamefont {L.}~\bibnamefont
  {Stein-Montalvo}}, \bibinfo {author} {\bibfnamefont {P.}~\bibnamefont
  {Costa}}, \bibinfo {author} {\bibfnamefont {M.}~\bibnamefont {Pezzulla}},\
  and\ \bibinfo {author} {\bibfnamefont {D.~P.}\ \bibnamefont {Holmes}},\
  }\href {https://doi.org/10.1039/C8SM02035C} {\bibfield  {journal} {\bibinfo
  {journal} {Soft Matter}\ }\textbf {\bibinfo {volume} {15}},\ \bibinfo {pages}
  {1215} (\bibinfo {year} {2019})}\BibitemShut {NoStop}%
\bibitem [{\citenamefont {Lee}\ \emph {et~al.}(2024)\citenamefont {Lee},
  \citenamefont {Park},\ and\ \citenamefont
  {Holmes}}]{holmes-stimulishelltheory-2024}%
  \BibitemOpen
  \bibfield  {author} {\bibinfo {author} {\bibfnamefont {J.-H.}\ \bibnamefont
  {Lee}}, \bibinfo {author} {\bibfnamefont {H.~S.}\ \bibnamefont {Park}},\ and\
  \bibinfo {author} {\bibfnamefont {D.~P.}\ \bibnamefont {Holmes}},\ }\href
  {https://doi.org/10.1177/10812865231159676} {\bibfield  {journal} {\bibinfo
  {journal} {Mathematics and Mechanics of Solids}\ }\textbf {\bibinfo {volume}
  {29}},\ \bibinfo {pages} {1089} (\bibinfo {year} {2024})},\ \Eprint
  {https://arxiv.org/abs/https://doi.org/10.1177/10812865231159676}
  {https://doi.org/10.1177/10812865231159676} \BibitemShut {NoStop}%
\bibitem [{\citenamefont {Bende}\ \emph {et~al.}(2015)\citenamefont {Bende},
  \citenamefont {Evans}, \citenamefont {Innes-Gold}, \citenamefont {Marin},
  \citenamefont {Cohen}, \citenamefont {Hayward},\ and\ \citenamefont
  {Santangelo}}]{hayward-shellsnapping-2015}%
  \BibitemOpen
  \bibfield  {author} {\bibinfo {author} {\bibfnamefont {N.~P.}\ \bibnamefont
  {Bende}}, \bibinfo {author} {\bibfnamefont {A.~A.}\ \bibnamefont {Evans}},
  \bibinfo {author} {\bibfnamefont {S.}~\bibnamefont {Innes-Gold}}, \bibinfo
  {author} {\bibfnamefont {L.~A.}\ \bibnamefont {Marin}}, \bibinfo {author}
  {\bibfnamefont {I.}~\bibnamefont {Cohen}}, \bibinfo {author} {\bibfnamefont
  {R.~C.}\ \bibnamefont {Hayward}},\ and\ \bibinfo {author} {\bibfnamefont
  {C.~D.}\ \bibnamefont {Santangelo}},\ }\href
  {https://doi.org/10.1073/pnas.1509228112} {\bibfield  {journal} {\bibinfo
  {journal} {Proceedings of the National Academy of Sciences}\ }\textbf
  {\bibinfo {volume} {112}},\ \bibinfo {pages} {11175} (\bibinfo {year}
  {2015})},\ \Eprint
  {https://arxiv.org/abs/https://www.pnas.org/doi/pdf/10.1073/pnas.1509228112}
  {https://www.pnas.org/doi/pdf/10.1073/pnas.1509228112} \BibitemShut {NoStop}%
\bibitem [{\citenamefont {Paulsen}\ \emph {et~al.}(2016)\citenamefont
  {Paulsen}, \citenamefont {Hohlfeld}, \citenamefont {King}, \citenamefont
  {Huang}, \citenamefont {Qiu}, \citenamefont {Russell}, \citenamefont {Menon},
  \citenamefont {Vella},\ and\ \citenamefont
  {Davidovitch}}]{davidovitch-wrinkledsheets-2016}%
  \BibitemOpen
  \bibfield  {author} {\bibinfo {author} {\bibfnamefont {J.~D.}\ \bibnamefont
  {Paulsen}}, \bibinfo {author} {\bibfnamefont {E.}~\bibnamefont {Hohlfeld}},
  \bibinfo {author} {\bibfnamefont {H.}~\bibnamefont {King}}, \bibinfo {author}
  {\bibfnamefont {J.}~\bibnamefont {Huang}}, \bibinfo {author} {\bibfnamefont
  {Z.}~\bibnamefont {Qiu}}, \bibinfo {author} {\bibfnamefont {T.~P.}\
  \bibnamefont {Russell}}, \bibinfo {author} {\bibfnamefont {N.}~\bibnamefont
  {Menon}}, \bibinfo {author} {\bibfnamefont {D.}~\bibnamefont {Vella}},\ and\
  \bibinfo {author} {\bibfnamefont {B.}~\bibnamefont {Davidovitch}},\ }\href
  {https://doi.org/10.1073/pnas.1521520113} {\bibfield  {journal} {\bibinfo
  {journal} {Proceedings of the National Academy of Sciences}\ }\textbf
  {\bibinfo {volume} {113}},\ \bibinfo {pages} {1144} (\bibinfo {year}
  {2016})},\ \Eprint
  {https://arxiv.org/abs/https://www.pnas.org/doi/pdf/10.1073/pnas.1521520113}
  {https://www.pnas.org/doi/pdf/10.1073/pnas.1521520113} \BibitemShut {NoStop}%
\bibitem [{\citenamefont {Roback}\ \emph {et~al.}(2026)\citenamefont {Roback},
  \citenamefont {Moguel-Lehmer}, \citenamefont {Fransen}, \citenamefont
  {Santangelo},\ and\ \citenamefont
  {Hayward}}]{roback-dynamicalmodesnoneuclidean-2026}%
  \BibitemOpen
  \bibfield  {author} {\bibinfo {author} {\bibfnamefont {J.~C.}\ \bibnamefont
  {Roback}}, \bibinfo {author} {\bibfnamefont {C.~E.}\ \bibnamefont
  {Moguel-Lehmer}}, \bibinfo {author} {\bibfnamefont {K.~A.}\ \bibnamefont
  {Fransen}}, \bibinfo {author} {\bibfnamefont {C.~D.}\ \bibnamefont
  {Santangelo}},\ and\ \bibinfo {author} {\bibfnamefont {R.~C.}\ \bibnamefont
  {Hayward}},\ }\href {https://arxiv.org/abs/2605.09289} {\bibinfo {title}
  {Dynamical geometric modes in non-euclidean plates}} (\bibinfo {year}
  {2026}),\ \Eprint {https://arxiv.org/abs/2605.09289} {arXiv:2605.09289
  [cond-mat.soft]} \BibitemShut {NoStop}%
\bibitem [{\citenamefont {Efrati}\ \emph
  {et~al.}(2009{\natexlab{a}})\citenamefont {Efrati}, \citenamefont {Sharon},\
  and\ \citenamefont
  {Kupferman}}]{efrati-unconstrainednoneuclideanplates-2009}%
  \BibitemOpen
  \bibfield  {author} {\bibinfo {author} {\bibfnamefont {E.}~\bibnamefont
  {Efrati}}, \bibinfo {author} {\bibfnamefont {E.}~\bibnamefont {Sharon}},\
  and\ \bibinfo {author} {\bibfnamefont {R.}~\bibnamefont {Kupferman}},\ }\href
  {https://doi.org/https://doi.org/10.1016/j.jmps.2008.12.004} {\bibfield
  {journal} {\bibinfo  {journal} {Journal of the Mechanics and Physics of
  Solids}\ }\textbf {\bibinfo {volume} {57}},\ \bibinfo {pages} {762} (\bibinfo
  {year} {2009}{\natexlab{a}})}\BibitemShut {NoStop}%
\bibitem [{\citenamefont {Efrati}\ \emph
  {et~al.}(2009{\natexlab{b}})\citenamefont {Efrati}, \citenamefont {Sharon},\
  and\ \citenamefont {Kupferman}}]{efrati-noneuclideanplatebuckling-2009}%
  \BibitemOpen
  \bibfield  {author} {\bibinfo {author} {\bibfnamefont {E.}~\bibnamefont
  {Efrati}}, \bibinfo {author} {\bibfnamefont {E.}~\bibnamefont {Sharon}},\
  and\ \bibinfo {author} {\bibfnamefont {R.}~\bibnamefont {Kupferman}},\ }\href
  {https://doi.org/10.1103/PhysRevE.80.016602} {\bibfield  {journal} {\bibinfo
  {journal} {Phys. Rev. E}\ }\textbf {\bibinfo {volume} {80}},\ \bibinfo
  {pages} {016602} (\bibinfo {year} {2009}{\natexlab{b}})}\BibitemShut
  {NoStop}%
\bibitem [{\citenamefont {Sharon}\ and\ \citenamefont
  {Efrati}(2010)}]{efrati-noneuclideanplatetheory-2010}%
  \BibitemOpen
  \bibfield  {author} {\bibinfo {author} {\bibfnamefont {E.}~\bibnamefont
  {Sharon}}\ and\ \bibinfo {author} {\bibfnamefont {E.}~\bibnamefont
  {Efrati}},\ }\href {https://doi.org/10.1039/C0SM00479K} {\bibfield  {journal}
  {\bibinfo  {journal} {Soft Matter}\ }\textbf {\bibinfo {volume} {6}},\
  \bibinfo {pages} {5693} (\bibinfo {year} {2010})}\BibitemShut {NoStop}%
\bibitem [{\citenamefont {Davidovitch}\ \emph {et~al.}(2019)\citenamefont
  {Davidovitch}, \citenamefont {Sun},\ and\ \citenamefont
  {Grason}}]{davidovitch-confinedsheets-2016}%
  \BibitemOpen
  \bibfield  {author} {\bibinfo {author} {\bibfnamefont {B.}~\bibnamefont
  {Davidovitch}}, \bibinfo {author} {\bibfnamefont {Y.}~\bibnamefont {Sun}},\
  and\ \bibinfo {author} {\bibfnamefont {G.~M.}\ \bibnamefont {Grason}},\
  }\href {https://doi.org/10.1073/pnas.1815507116} {\bibfield  {journal}
  {\bibinfo  {journal} {Proceedings of the National Academy of Sciences}\
  }\textbf {\bibinfo {volume} {116}},\ \bibinfo {pages} {1483} (\bibinfo {year}
  {2019})},\ \Eprint
  {https://arxiv.org/abs/https://www.pnas.org/doi/pdf/10.1073/pnas.1815507116}
  {https://www.pnas.org/doi/pdf/10.1073/pnas.1815507116} \BibitemShut {NoStop}%
\bibitem [{\citenamefont {Moguel-Lehmer}\ and\ \citenamefont
  {Santangelo}(2025)}]{santangelo-frustratedribbon-2025}%
  \BibitemOpen
  \bibfield  {author} {\bibinfo {author} {\bibfnamefont {C.~E.}\ \bibnamefont
  {Moguel-Lehmer}}\ and\ \bibinfo {author} {\bibfnamefont {C.~D.}\ \bibnamefont
  {Santangelo}},\ }\href {https://doi.org/10.1103/q2dk-wf7t} {\bibfield
  {journal} {\bibinfo  {journal} {Phys. Rev. Lett.}\ }\textbf {\bibinfo
  {volume} {135}},\ \bibinfo {pages} {128201} (\bibinfo {year}
  {2025})}\BibitemShut {NoStop}%
\bibitem [{\citenamefont {Seung}\ and\ \citenamefont
  {Nelson}(1988)}]{seung-defectelasticity-1988}%
  \BibitemOpen
  \bibfield  {author} {\bibinfo {author} {\bibfnamefont {H.~S.}\ \bibnamefont
  {Seung}}\ and\ \bibinfo {author} {\bibfnamefont {D.~R.}\ \bibnamefont
  {Nelson}},\ }\href {https://doi.org/10.1103/PhysRevA.38.1005} {\bibfield
  {journal} {\bibinfo  {journal} {Phys. Rev. A}\ }\textbf {\bibinfo {volume}
  {38}},\ \bibinfo {pages} {1005} (\bibinfo {year} {1988})}\BibitemShut
  {NoStop}%
\bibitem [{\citenamefont {Lidmar}\ \emph {et~al.}(2003)\citenamefont {Lidmar},
  \citenamefont {Mirny},\ and\ \citenamefont
  {Nelson}}]{lidmar-viralshells-2003}%
  \BibitemOpen
  \bibfield  {author} {\bibinfo {author} {\bibfnamefont {J.}~\bibnamefont
  {Lidmar}}, \bibinfo {author} {\bibfnamefont {L.}~\bibnamefont {Mirny}},\ and\
  \bibinfo {author} {\bibfnamefont {D.~R.}\ \bibnamefont {Nelson}},\ }\href
  {https://doi.org/10.1103/PhysRevE.68.051910} {\bibfield  {journal} {\bibinfo
  {journal} {Phys. Rev. E}\ }\textbf {\bibinfo {volume} {68}},\ \bibinfo
  {pages} {051910} (\bibinfo {year} {2003})}\BibitemShut {NoStop}%
\bibitem [{\citenamefont {Hyde}\ \emph {et~al.}(1997)\citenamefont {Hyde},
  \citenamefont {Ninham}, \citenamefont {Andersson}, \citenamefont {Larsson},
  \citenamefont {Landh}, \citenamefont {Blum},\ and\ \citenamefont
  {Lidin}}]{hyde_beyond_1997}%
  \BibitemOpen
  \bibfield  {author} {\bibinfo {author} {\bibfnamefont {S.}~\bibnamefont
  {Hyde}}, \bibinfo {author} {\bibfnamefont {B.~W.}\ \bibnamefont {Ninham}},
  \bibinfo {author} {\bibfnamefont {S.}~\bibnamefont {Andersson}}, \bibinfo
  {author} {\bibfnamefont {K.}~\bibnamefont {Larsson}}, \bibinfo {author}
  {\bibfnamefont {T.}~\bibnamefont {Landh}}, \bibinfo {author} {\bibfnamefont
  {Z.}~\bibnamefont {Blum}},\ and\ \bibinfo {author} {\bibfnamefont
  {S.}~\bibnamefont {Lidin}},\ }in\ \href
  {https://doi.org/10.1016/B978-044481538-5/50005-8} {\emph {\bibinfo
  {booktitle} {The {Language} of {Shape}}}}\ (\bibinfo  {publisher}
  {Elsevier},\ \bibinfo {address} {Amsterdam},\ \bibinfo {year} {1997})\ pp.\
  \bibinfo {pages} {141--197}\BibitemShut {NoStop}%
\bibitem [{\citenamefont {Sethna}\ and\ \citenamefont
  {Kl\'eman}(1982)}]{sethna_spheric_1982}%
  \BibitemOpen
  \bibfield  {author} {\bibinfo {author} {\bibfnamefont {J.~P.}\ \bibnamefont
  {Sethna}}\ and\ \bibinfo {author} {\bibfnamefont {M.}~\bibnamefont
  {Kl\'eman}},\ }\href {https://doi.org/10.1103/PhysRevA.26.3037} {\bibfield
  {journal} {\bibinfo  {journal} {Phys. Rev. A}\ }\textbf {\bibinfo {volume}
  {26}},\ \bibinfo {pages} {3037(R)} (\bibinfo {year} {1982})}\BibitemShut
  {NoStop}%
\bibitem [{\citenamefont {DiDonna}\ and\ \citenamefont
  {Kamien}(2002)}]{didonna-bluephases-2002}%
  \BibitemOpen
  \bibfield  {author} {\bibinfo {author} {\bibfnamefont {B.~A.}\ \bibnamefont
  {DiDonna}}\ and\ \bibinfo {author} {\bibfnamefont {R.~D.}\ \bibnamefont
  {Kamien}},\ }\href {https://doi.org/10.1103/PhysRevLett.89.215504} {\bibfield
   {journal} {\bibinfo  {journal} {Phys. Rev. Lett.}\ }\textbf {\bibinfo
  {volume} {89}},\ \bibinfo {pages} {215504} (\bibinfo {year}
  {2002})}\BibitemShut {NoStop}%
\bibitem [{\citenamefont {DiDonna}\ and\ \citenamefont
  {Kamien}(2003)}]{didonna-bluephases-2003}%
  \BibitemOpen
  \bibfield  {author} {\bibinfo {author} {\bibfnamefont {B.~A.}\ \bibnamefont
  {DiDonna}}\ and\ \bibinfo {author} {\bibfnamefont {R.~D.}\ \bibnamefont
  {Kamien}},\ }\href {https://doi.org/10.1103/PhysRevE.68.041703} {\bibfield
  {journal} {\bibinfo  {journal} {Phys. Rev. E}\ }\textbf {\bibinfo {volume}
  {68}},\ \bibinfo {pages} {041703} (\bibinfo {year} {2003})}\BibitemShut
  {NoStop}%
\bibitem [{\citenamefont {Kamien}\ \emph {et~al.}(2023)\citenamefont {Kamien},
  \citenamefont {Nastishin},\ and\ \citenamefont
  {Pansu}}]{kamien-focalconics-2023}%
  \BibitemOpen
  \bibfield  {author} {\bibinfo {author} {\bibfnamefont {R.~D.}\ \bibnamefont
  {Kamien}}, \bibinfo {author} {\bibfnamefont {Y.}~\bibnamefont {Nastishin}},\
  and\ \bibinfo {author} {\bibfnamefont {B.}~\bibnamefont {Pansu}},\ }\href
  {https://doi.org/10.1073/pnas.2311957120} {\bibfield  {journal} {\bibinfo
  {journal} {Proceedings of the National Academy of Sciences}\ }\textbf
  {\bibinfo {volume} {120}},\ \bibinfo {pages} {e2311957120} (\bibinfo {year}
  {2023})},\ \Eprint
  {https://arxiv.org/abs/https://www.pnas.org/doi/pdf/10.1073/pnas.2311957120}
  {https://www.pnas.org/doi/pdf/10.1073/pnas.2311957120} \BibitemShut {NoStop}%
\bibitem [{\citenamefont {Ciarlet}\ and\ \citenamefont
  {Paumier}(1986)}]{ciarlet-mvkshells-1986}%
  \BibitemOpen
  \bibfield  {author} {\bibinfo {author} {\bibfnamefont {P.~G.}\ \bibnamefont
  {Ciarlet}}\ and\ \bibinfo {author} {\bibfnamefont {J.~C.}\ \bibnamefont
  {Paumier}},\ }\href {https://doi.org/10.1007/BF00272623} {\bibfield
  {journal} {\bibinfo  {journal} {Computational Mechanics}\ }\textbf {\bibinfo
  {volume} {1}},\ \bibinfo {pages} {177} (\bibinfo {year} {1986})}\BibitemShut
  {NoStop}%
\bibitem [{\citenamefont {Seffen}(2006)}]{seffen-bistableshells-2006}%
  \BibitemOpen
  \bibfield  {author} {\bibinfo {author} {\bibfnamefont {K.}~\bibnamefont
  {Seffen}},\ }\href {https://doi.org/10.1098/rspa.2006.1750} {\bibfield
  {journal} {\bibinfo  {journal} {Proceedings of the Royal Society A:
  Mathematical, Physical and Engineering Sciences}\ }\textbf {\bibinfo {volume}
  {463}},\ \bibinfo {pages} {67} (\bibinfo {year} {2006})},\ \Eprint
  {https://arxiv.org/abs/https://royalsocietypublishing.org/rspa/article-pdf/463/2077/67/695090/rspa.2006.1750.pdf}
  {https://royalsocietypublishing.org/rspa/article-pdf/463/2077/67/695090/rspa.2006.1750.pdf}
  \BibitemShut {NoStop}%
\bibitem [{\citenamefont {Landau}\ and\ \citenamefont
  {Lifshitz}(1986)}]{LandauLifshitzElasticityChapter2}%
  \BibitemOpen
  \bibfield  {author} {\bibinfo {author} {\bibfnamefont {L.~D.}\ \bibnamefont
  {Landau}}\ and\ \bibinfo {author} {\bibfnamefont {E.~M.}\ \bibnamefont
  {Lifshitz}},\ }in\ \href@noop {} {\emph {\bibinfo {booktitle} {Theory of
  Elasticity}}},\ \bibinfo {series} {Course of Theoretical Physics},
  Vol.~\bibinfo {volume} {7}\ (\bibinfo  {publisher} {Butterworth-Heinemann},\
  \bibinfo {address} {Oxford},\ \bibinfo {year} {1986})\ \bibinfo {edition}
  {3rd}\ ed.,\ Chap.~\bibinfo {chapter} {2}, pp.\ \bibinfo {pages}
  {38--86}\BibitemShut {NoStop}%
\bibitem [{Note1()}]{Note1}%
  \BibitemOpen
  \bibinfo {note} {For a shell composed of a homogeneous elastic material of
  thickness $t$ with Young's modulus $E$ and Poisson ratio $\nu $, bending
  moduli are given by $B_\parallel = Et^3/12(1-\nu ^2)$, $B_\perp =
  B_{\parallel }\nu $ and the 2D Young's modulus is $Y = Et$.}\BibitemShut
  {Stop}%
\bibitem [{\citenamefont {Stukowski}(2009)}]{ovito}%
  \BibitemOpen
  \bibfield  {author} {\bibinfo {author} {\bibfnamefont {A.}~\bibnamefont
  {Stukowski}},\ }\href {https://doi.org/10.1088/0965-0393/18/1/015012}
  {\bibfield  {journal} {\bibinfo  {journal} {Modelling and Simulation in
  Materials Science and Engineering}\ }\textbf {\bibinfo {volume} {18}},\
  \bibinfo {pages} {015012} (\bibinfo {year} {2009})}\BibitemShut {NoStop}%
\bibitem [{\citenamefont {Thompson}\ \emph {et~al.}(2022)\citenamefont
  {Thompson}, \citenamefont {Aktulga}, \citenamefont {Berger}, \citenamefont
  {Bolintineanu}, \citenamefont {Brown}, \citenamefont {Crozier}, \citenamefont
  {in~'t Veld}, \citenamefont {Kohlmeyer}, \citenamefont {Moore}, \citenamefont
  {Nguyen}, \citenamefont {Shan}, \citenamefont {Stevens}, \citenamefont
  {Tranchida}, \citenamefont {Trott},\ and\ \citenamefont {Plimpton}}]{LAMMPS}%
  \BibitemOpen
  \bibfield  {author} {\bibinfo {author} {\bibfnamefont {A.~P.}\ \bibnamefont
  {Thompson}}, \bibinfo {author} {\bibfnamefont {H.~M.}\ \bibnamefont
  {Aktulga}}, \bibinfo {author} {\bibfnamefont {R.}~\bibnamefont {Berger}},
  \bibinfo {author} {\bibfnamefont {D.~S.}\ \bibnamefont {Bolintineanu}},
  \bibinfo {author} {\bibfnamefont {W.~M.}\ \bibnamefont {Brown}}, \bibinfo
  {author} {\bibfnamefont {P.~S.}\ \bibnamefont {Crozier}}, \bibinfo {author}
  {\bibfnamefont {P.~J.}\ \bibnamefont {in~'t Veld}}, \bibinfo {author}
  {\bibfnamefont {A.}~\bibnamefont {Kohlmeyer}}, \bibinfo {author}
  {\bibfnamefont {S.~G.}\ \bibnamefont {Moore}}, \bibinfo {author}
  {\bibfnamefont {T.~D.}\ \bibnamefont {Nguyen}}, \bibinfo {author}
  {\bibfnamefont {R.}~\bibnamefont {Shan}}, \bibinfo {author} {\bibfnamefont
  {M.~J.}\ \bibnamefont {Stevens}}, \bibinfo {author} {\bibfnamefont
  {J.}~\bibnamefont {Tranchida}}, \bibinfo {author} {\bibfnamefont
  {C.}~\bibnamefont {Trott}},\ and\ \bibinfo {author} {\bibfnamefont {S.~J.}\
  \bibnamefont {Plimpton}},\ }\href {https://doi.org/10.1016/j.cpc.2021.108171}
  {\bibfield  {journal} {\bibinfo  {journal} {Comp. Phys. Comm.}\ }\textbf
  {\bibinfo {volume} {271}},\ \bibinfo {pages} {108171} (\bibinfo {year}
  {2022})}\BibitemShut {NoStop}%
\bibitem [{\citenamefont {Ghafouri}\ and\ \citenamefont
  {Bruinsma}(2005)}]{ghafouri-ribbons-2005}%
  \BibitemOpen
  \bibfield  {author} {\bibinfo {author} {\bibfnamefont {R.}~\bibnamefont
  {Ghafouri}}\ and\ \bibinfo {author} {\bibfnamefont {R.}~\bibnamefont
  {Bruinsma}},\ }\href {https://doi.org/10.1103/PhysRevLett.94.138101}
  {\bibfield  {journal} {\bibinfo  {journal} {Phys. Rev. Lett.}\ }\textbf
  {\bibinfo {volume} {94}},\ \bibinfo {pages} {138101} (\bibinfo {year}
  {2005})}\BibitemShut {NoStop}%
\bibitem [{\citenamefont {Tanjeem}\ \emph
  {et~al.}(2022{\natexlab{b}})\citenamefont {Tanjeem}, \citenamefont {Minnis},
  \citenamefont {Hayward},\ and\ \citenamefont
  {Shields~IV}}]{tanjeem-shapechanging-2022}%
  \BibitemOpen
  \bibfield  {author} {\bibinfo {author} {\bibfnamefont {N.}~\bibnamefont
  {Tanjeem}}, \bibinfo {author} {\bibfnamefont {M.~B.}\ \bibnamefont {Minnis}},
  \bibinfo {author} {\bibfnamefont {R.~C.}\ \bibnamefont {Hayward}},\ and\
  \bibinfo {author} {\bibfnamefont {C.~W.}\ \bibnamefont {Shields~IV}},\ }\href
  {https://doi.org/https://doi.org/10.1002/adma.202105758} {\bibfield
  {journal} {\bibinfo  {journal} {Advanced Materials}\ }\textbf {\bibinfo
  {volume} {34}},\ \bibinfo {pages} {2105758} (\bibinfo {year}
  {2022}{\natexlab{b}})},\ \Eprint
  {https://arxiv.org/abs/https://onlinelibrary.wiley.com/doi/pdf/10.1002/adma.202105758}
  {https://onlinelibrary.wiley.com/doi/pdf/10.1002/adma.202105758} \BibitemShut
  {NoStop}%
\bibitem [{\citenamefont {Kuenstler}\ \emph {et~al.}(2020)\citenamefont
  {Kuenstler}, \citenamefont {Lahikainen}, \citenamefont {Zhou}, \citenamefont
  {Xu}, \citenamefont {Priimagi},\ and\ \citenamefont
  {Hayward}}]{kuenstler-curvedhydrogels-2020}%
  \BibitemOpen
  \bibfield  {author} {\bibinfo {author} {\bibfnamefont {A.~S.}\ \bibnamefont
  {Kuenstler}}, \bibinfo {author} {\bibfnamefont {M.}~\bibnamefont
  {Lahikainen}}, \bibinfo {author} {\bibfnamefont {H.}~\bibnamefont {Zhou}},
  \bibinfo {author} {\bibfnamefont {W.}~\bibnamefont {Xu}}, \bibinfo {author}
  {\bibfnamefont {A.}~\bibnamefont {Priimagi}},\ and\ \bibinfo {author}
  {\bibfnamefont {R.~C.}\ \bibnamefont {Hayward}},\ }\href
  {https://doi.org/10.1021/acsmacrolett.0c00469} {\bibfield  {journal}
  {\bibinfo  {journal} {ACS Macro Letters}\ }\textbf {\bibinfo {volume} {9}},\
  \bibinfo {pages} {1172} (\bibinfo {year} {2020})},\ \bibinfo {note} {pMID:
  32864191},\ \Eprint
  {https://arxiv.org/abs/https://doi.org/10.1021/acsmacrolett.0c00469}
  {https://doi.org/10.1021/acsmacrolett.0c00469} \BibitemShut {NoStop}%
\bibitem [{\citenamefont {Jeon}\ and\ \citenamefont
  {Hayward}(2020)}]{jeon-trilayergels-2020}%
  \BibitemOpen
  \bibfield  {author} {\bibinfo {author} {\bibfnamefont {S.-J.}\ \bibnamefont
  {Jeon}}\ and\ \bibinfo {author} {\bibfnamefont {R.~C.}\ \bibnamefont
  {Hayward}},\ }\href {https://doi.org/10.1039/C9SM01922G} {\bibfield
  {journal} {\bibinfo  {journal} {Soft Matter}\ }\textbf {\bibinfo {volume}
  {16}},\ \bibinfo {pages} {688} (\bibinfo {year} {2020})}\BibitemShut
  {NoStop}%
\bibitem [{\citenamefont {Na}\ \emph {et~al.}(2016)\citenamefont {Na},
  \citenamefont {Bende}, \citenamefont {Bae}, \citenamefont {Santangelo},\ and\
  \citenamefont {Hayward}}]{na-greyscaleshapes-2016}%
  \BibitemOpen
  \bibfield  {author} {\bibinfo {author} {\bibfnamefont {J.-H.}\ \bibnamefont
  {Na}}, \bibinfo {author} {\bibfnamefont {N.~P.}\ \bibnamefont {Bende}},
  \bibinfo {author} {\bibfnamefont {J.}~\bibnamefont {Bae}}, \bibinfo {author}
  {\bibfnamefont {C.~D.}\ \bibnamefont {Santangelo}},\ and\ \bibinfo {author}
  {\bibfnamefont {R.~C.}\ \bibnamefont {Hayward}},\ }\href
  {https://doi.org/10.1039/C6SM00714G} {\bibfield  {journal} {\bibinfo
  {journal} {Soft Matter}\ }\textbf {\bibinfo {volume} {12}},\ \bibinfo {pages}
  {4985} (\bibinfo {year} {2016})}\BibitemShut {NoStop}%
\bibitem [{\citenamefont {Jana}\ \emph {et~al.}(2017)\citenamefont {Jana},
  \citenamefont {de~Frutos}, \citenamefont {Davidson},\ and\ \citenamefont
  {Abacassis}}]{Jana_NanoPlatlets_2017}%
  \BibitemOpen
  \bibfield  {author} {\bibinfo {author} {\bibfnamefont {S.}~\bibnamefont
  {Jana}}, \bibinfo {author} {\bibfnamefont {M.}~\bibnamefont {de~Frutos}},
  \bibinfo {author} {\bibfnamefont {P.}~\bibnamefont {Davidson}},\ and\
  \bibinfo {author} {\bibfnamefont {B.}~\bibnamefont {Abacassis}},\ }\href
  {https://doi.org/10.1126/sciadv.1701483} {\bibfield  {journal} {\bibinfo
  {journal} {Science Advances}\ }\textbf {\bibinfo {volume} {3}},\ \bibinfo
  {pages} {e1701483} (\bibinfo {year} {2017})},\ \Eprint
  {https://arxiv.org/abs/https://www.science.org/doi/pdf/10.1126/sciadv.1701483}
  {https://www.science.org/doi/pdf/10.1126/sciadv.1701483} \BibitemShut
  {NoStop}%
\bibitem [{\citenamefont {Guillemeney}\ \emph {et~al.}(2022)\citenamefont
  {Guillemeney}, \citenamefont {Lermusiaux}, \citenamefont {Landaburu},
  \citenamefont {Wagnon},\ and\ \citenamefont
  {Abécassis}}]{guillemeney_curvature_2022}%
  \BibitemOpen
  \bibfield  {author} {\bibinfo {author} {\bibfnamefont {L.}~\bibnamefont
  {Guillemeney}}, \bibinfo {author} {\bibfnamefont {L.}~\bibnamefont
  {Lermusiaux}}, \bibinfo {author} {\bibfnamefont {G.}~\bibnamefont
  {Landaburu}}, \bibinfo {author} {\bibfnamefont {B.}~\bibnamefont {Wagnon}},\
  and\ \bibinfo {author} {\bibfnamefont {B.}~\bibnamefont {Abécassis}},\
  }\href {https://doi.org/10.1038/s42004-021-00621-z} {\bibfield  {journal}
  {\bibinfo  {journal} {Communications Chemistry}\ }\textbf {\bibinfo {volume}
  {5}},\ \bibinfo {pages} {7} (\bibinfo {year} {2022})}\BibitemShut {NoStop}%
\bibitem [{\citenamefont {Monego}\ \emph {et~al.}(2024)\citenamefont {Monego},
  \citenamefont {Dutta}, \citenamefont {Grossman}, \citenamefont {Krapez},
  \citenamefont {Bauer}, \citenamefont {Hubley}, \citenamefont {Margueritat},
  \citenamefont {Mahler}, \citenamefont {Widmer-Cooper},\ and\ \citenamefont
  {Abécassis}}]{abecassis-nanoplatelet_curvature-2024}%
  \BibitemOpen
  \bibfield  {author} {\bibinfo {author} {\bibfnamefont {D.}~\bibnamefont
  {Monego}}, \bibinfo {author} {\bibfnamefont {S.}~\bibnamefont {Dutta}},
  \bibinfo {author} {\bibfnamefont {D.}~\bibnamefont {Grossman}}, \bibinfo
  {author} {\bibfnamefont {M.}~\bibnamefont {Krapez}}, \bibinfo {author}
  {\bibfnamefont {P.}~\bibnamefont {Bauer}}, \bibinfo {author} {\bibfnamefont
  {A.}~\bibnamefont {Hubley}}, \bibinfo {author} {\bibfnamefont
  {J.}~\bibnamefont {Margueritat}}, \bibinfo {author} {\bibfnamefont
  {B.}~\bibnamefont {Mahler}}, \bibinfo {author} {\bibfnamefont
  {A.}~\bibnamefont {Widmer-Cooper}},\ and\ \bibinfo {author} {\bibfnamefont
  {B.}~\bibnamefont {Abécassis}},\ }\href
  {https://doi.org/10.1073/pnas.2316299121} {\bibfield  {journal} {\bibinfo
  {journal} {Proceedings of the National Academy of Sciences}\ }\textbf
  {\bibinfo {volume} {121}},\ \bibinfo {pages} {e2316299121} (\bibinfo {year}
  {2024})},\ \Eprint
  {https://arxiv.org/abs/https://www.pnas.org/doi/pdf/10.1073/pnas.2316299121}
  {https://www.pnas.org/doi/pdf/10.1073/pnas.2316299121} \BibitemShut {NoStop}%
\bibitem [{\citenamefont {Boucenna}\ \emph {et~al.}(2025)\citenamefont
  {Boucenna}, \citenamefont {Carn},\ and\ \citenamefont
  {Mourchid}}]{mourchid-thermoresponsive_nanoplatelets-2025}%
  \BibitemOpen
  \bibfield  {author} {\bibinfo {author} {\bibfnamefont {I.}~\bibnamefont
  {Boucenna}}, \bibinfo {author} {\bibfnamefont {F.}~\bibnamefont {Carn}},\
  and\ \bibinfo {author} {\bibfnamefont {A.}~\bibnamefont {Mourchid}},\ }\href
  {https://doi.org/10.1021/acs.langmuir.5c03354} {\bibfield  {journal}
  {\bibinfo  {journal} {Langmuir}\ }\textbf {\bibinfo {volume} {41}},\ \bibinfo
  {pages} {27837} (\bibinfo {year} {2025})},\ \Eprint
  {https://arxiv.org/abs/https://pubs.acs.org/langd5/article-pdf/41/41/27837/42137622/la5c03354.pdf}
  {https://pubs.acs.org/langd5/article-pdf/41/41/27837/42137622/la5c03354.pdf}
  \BibitemShut {NoStop}%
\bibitem [{\citenamefont {Ithurria}\ and\ \citenamefont
  {Dubertret}(2008)}]{cadmium-nanoplatelets-2008}%
  \BibitemOpen
  \bibfield  {author} {\bibinfo {author} {\bibfnamefont {S.}~\bibnamefont
  {Ithurria}}\ and\ \bibinfo {author} {\bibfnamefont {B.}~\bibnamefont
  {Dubertret}},\ }\href {https://doi.org/10.1021/ja807724e} {\bibfield
  {journal} {\bibinfo  {journal} {Journal of the American Chemical Society}\
  }\textbf {\bibinfo {volume} {130}},\ \bibinfo {pages} {16504} (\bibinfo
  {year} {2008})},\ \Eprint
  {https://arxiv.org/abs/https://pubs.acs.org/jacsat/article-pdf/130/49/16504/8395480/ja807724e.pdf}
  {https://pubs.acs.org/jacsat/article-pdf/130/49/16504/8395480/ja807724e.pdf}
  \BibitemShut {NoStop}%
\bibitem [{\citenamefont {Dietz}\ \emph {et~al.}(2009)\citenamefont {Dietz},
  \citenamefont {Douglas},\ and\ \citenamefont {Shih}}]{dietz-foldingdna-2009}%
  \BibitemOpen
  \bibfield  {author} {\bibinfo {author} {\bibfnamefont {H.}~\bibnamefont
  {Dietz}}, \bibinfo {author} {\bibfnamefont {S.~M.}\ \bibnamefont {Douglas}},\
  and\ \bibinfo {author} {\bibfnamefont {W.~M.}\ \bibnamefont {Shih}},\ }\href
  {https://doi.org/10.1126/science.1174251} {\bibfield  {journal} {\bibinfo
  {journal} {Science}\ }\textbf {\bibinfo {volume} {325}},\ \bibinfo {pages}
  {725} (\bibinfo {year} {2009})},\ \Eprint
  {https://arxiv.org/abs/https://www.science.org/doi/pdf/10.1126/science.1174251}
  {https://www.science.org/doi/pdf/10.1126/science.1174251} \BibitemShut
  {NoStop}%
\bibitem [{\citenamefont {Han}\ \emph {et~al.}(2011)\citenamefont {Han},
  \citenamefont {Pal}, \citenamefont {Nangreave}, \citenamefont {Deng},
  \citenamefont {Liu},\ and\ \citenamefont {Yan}}]{Han-dnacurvature-2011}%
  \BibitemOpen
  \bibfield  {author} {\bibinfo {author} {\bibfnamefont {D.}~\bibnamefont
  {Han}}, \bibinfo {author} {\bibfnamefont {S.}~\bibnamefont {Pal}}, \bibinfo
  {author} {\bibfnamefont {J.}~\bibnamefont {Nangreave}}, \bibinfo {author}
  {\bibfnamefont {Z.}~\bibnamefont {Deng}}, \bibinfo {author} {\bibfnamefont
  {Y.}~\bibnamefont {Liu}},\ and\ \bibinfo {author} {\bibfnamefont
  {H.}~\bibnamefont {Yan}},\ }\href {https://doi.org/10.1126/science.1202998}
  {\bibfield  {journal} {\bibinfo  {journal} {Science}\ }\textbf {\bibinfo
  {volume} {332}},\ \bibinfo {pages} {342} (\bibinfo {year} {2011})},\ \Eprint
  {https://arxiv.org/abs/https://www.science.org/doi/pdf/10.1126/science.1202998}
  {https://www.science.org/doi/pdf/10.1126/science.1202998} \BibitemShut
  {NoStop}%
\bibitem [{\citenamefont {Cheng}\ \emph {et~al.}(2012)\citenamefont {Cheng},
  \citenamefont {Aggarwal},\ and\ \citenamefont
  {Stevens}}]{Cheng_SoftMatter_2012}%
  \BibitemOpen
  \bibfield  {author} {\bibinfo {author} {\bibfnamefont {S.}~\bibnamefont
  {Cheng}}, \bibinfo {author} {\bibfnamefont {A.}~\bibnamefont {Aggarwal}},\
  and\ \bibinfo {author} {\bibfnamefont {M.~J.}\ \bibnamefont {Stevens}},\
  }\href {https://doi.org/10.1039/C2SM25068C} {\bibfield  {journal} {\bibinfo
  {journal} {Soft Matter}\ }\textbf {\bibinfo {volume} {8}},\ \bibinfo {pages}
  {5666} (\bibinfo {year} {2012})}\BibitemShut {NoStop}%
\bibitem [{\citenamefont {Videbæk}\ \emph {et~al.}(2024)\citenamefont
  {Videbæk}, \citenamefont {Hayakawa}, \citenamefont {Grason}, \citenamefont
  {Hagan}, \citenamefont {Fraden},\ and\ \citenamefont
  {Rogers}}]{videbaekk-economical-2024}%
  \BibitemOpen
  \bibfield  {author} {\bibinfo {author} {\bibfnamefont {T.~E.}\ \bibnamefont
  {Videbæk}}, \bibinfo {author} {\bibfnamefont {D.}~\bibnamefont {Hayakawa}},
  \bibinfo {author} {\bibfnamefont {G.~M.}\ \bibnamefont {Grason}}, \bibinfo
  {author} {\bibfnamefont {M.~F.}\ \bibnamefont {Hagan}}, \bibinfo {author}
  {\bibfnamefont {S.}~\bibnamefont {Fraden}},\ and\ \bibinfo {author}
  {\bibfnamefont {W.~B.}\ \bibnamefont {Rogers}},\ }\href
  {https://doi.org/10.1126/sciadv.ado5979} {\bibfield  {journal} {\bibinfo
  {journal} {Science Advances}\ }\textbf {\bibinfo {volume} {10}},\ \bibinfo
  {pages} {eado5979} (\bibinfo {year} {2024})},\ \Eprint
  {https://arxiv.org/abs/https://www.science.org/doi/pdf/10.1126/sciadv.ado5979}
  {https://www.science.org/doi/pdf/10.1126/sciadv.ado5979} \BibitemShut
  {NoStop}%
\end{thebibliography}%

\end{document}